\documentclass[aps,eqsecnum,preprint,floats,epsf,epsfig,nofootinbib,letter]{revtex4}
\usepackage{epsfig,color}
\usepackage{multirow}
\usepackage{subfigure}

\begin{document}
\def\be{\begin{eqnarray}}
\def\en{\end{eqnarray}}
\def\non{\nonumber}
\def\B{{\cal B}}
\def\ov{\overline}
\def\CP{{\it CP}~}
\def\la{\langle}
\def\ra{\rangle}
\def\pr{{\sl Phys. Rev.}~}
\def\prl{{\sl Phys. Rev. Lett.}~}
\def\pl{{\sl Phys. Lett.}~}
\def\np{{\sl Nucl. Phys.}~}
\def\zp{{\sl Z. Phys.}~}
\def\lsim{ {\ \lower-1.2pt\vbox{\hbox{\rlap{$<$}\lower5pt\vbox{\hbox{$\sim$}
}}}\ } }
\def\gsim{ {\ \lower-1.2pt\vbox{\hbox{\rlap{$>$}\lower5pt\vbox{\hbox{$\sim$}
}}}\ } }


\vskip 1.5 cm

\centerline{\large\bf Charmed Baryon Physics Circa 2021}

\vskip 1.5 cm

\bigskip
\bigskip
\centerline{\bf Hai-Yang Cheng}
\medskip
\centerline{Institute of Physics, Academia Sinica}
\centerline{Taipei, Taiwan 115, Republic of China}
\medskip

\bigskip
\bigskip
\bigskip
\small

This is an update of the two articles (Cheng, 2009; Cheng, 2015) in which charmed baryon physics around 2007 and 2015, respectively, were reviewed.  In this review
we emphasize the experimental progress and the theoretical development since 2015.

\pagebreak

\tableofcontents

\newpage

\section{Introduction}
Charm baryon spectroscopy provides an excellent ground for studying the dynamics of light quarks in the environment of a heavy quark or two heavy quarks.  In the past two decades,
many new excited charmed baryon states have been
discovered by BaBar, Belle, CLEO and LHCb. Both $B$ decays and the $e^+e^-\to c\bar c$ continuum provide very rich sources of charmed baryons. Many efforts have been made to identify  the quantum numbers of these new states and understand their properties. In 2017 LHCb has explored the charmed baryon sector of the $\Omega_c$ and observed five narrow excited $\Omega_c$ states \cite{LHCb:Omegac}. In 2020 LHCb has announced the observation of  three new $\Xi_c$ excited states in the $\Lambda_c^+K^-$ spectrum \cite{LHCb:Xic2923}. Both have
triggered a flood of interest in attempting to identify their spin-parity quantum numbers. In 2017 LHCb observed a new resonance in the $\Lambda_c^+K^-\pi^+\pi^+$ mass spectrum with a mass of $3621.40\pm0.78$ MeV \cite{LHCb:Xiccpp}, which was consistent with expectations for the doubly charmed baryon $\Xi_{cc}^{++}$ baryon and significantly larger than the mass of 3519 MeV measured by SELEX for $\Xi_{cc}^+$ \cite{Selex02}. This opens a new chapter on doubly heavy baryons.

Lifetimes of the heavy baryons are commonly analyzed within the framework of heavy quark expansion (HQE). In this approach, the predicted lifetime pattern for charmed baryons, namely,
$\tau(\Xi_c^+)>\tau(\Lambda_c^+)>\tau(\Xi_c^0)>\tau(\Omega_c^0)$, is in agreement with experiments listed in the 2018 version of the Particle Data Group (PDG) \cite{PDG2018}. This lifetime hierarchy in which the $\Omega_c$ is shortest-lived among the four charmed baryons owing to its large constructive Pauli interference has been known to the community for a long time. However, in 2018 LHCb has reported a new measurement of the $\Omega_c^0$ lifetime \cite{LHCb:tauOmegac}, which is nearly four times larger than the current world-average value of $\tau(\Omega_c^0)$  inferred from fixed target experiments. This indicates that
the $\Omega_c^0$, which is naively expected to be shortest-lived in the charmed baryon system, now lives longer than the $\Lambda_c^+$. The observation of a huge jump of the $\Omega_c^0$ lifetime is beyond imagination.
This very striking experimental result calls for a better theoretical understanding of charmed baryon lifetimes based on HQE. After the first observation of the $\Xi_{cc}^{++}$, LHCb has presented the first lifetime measurement of this doubly charm baryon  \cite{LHCb:tauXiccpp}.
It is thus imperative to study the lifetimes of doubly heavy baryons.

Theoretical interest in hadronic weak decays of charmed baryons peaked around the early 1990s and then faded away. To date, we still do not have a good phenomenological model, not mentioning the quantum chromodynamics (QCD)-inspired approach which has been applied successfully to heavy meson decays, to describe the complicated physics of baryon decays.
Nevertheless, there were several major breakthroughs in recent charmed-baryon experiments in regard to the hadronic weak decays of antitriplet charmed baryons $\Lambda_c^+$ and $\Xi_c^{+,0}$. First of all, the absolute branching fraction of $\Lambda_c^+\to pK^-\pi^+$, which is a benchmark for nearly all other branching fractions of the $\Lambda_c^+$, has been measured by Belle \cite{Zupanc} and BESIII \cite{BES:pKpi} independently with much smaller uncertainties.
Second, in 2015 BESIII has measured the absolute branching fractions for more than a dozen of  decay modes directly for the first time \cite{BES:pKpi}.  Not only the central values are substantially different from the PDG ones (versions before 2016), but also the uncertainties are significantly improved.

Likewise, for the $\Xi_c^{+,0}$ systems, Belle has recently reported the first measurements of the absolute branching fractions of $\Xi_c^0\to \Xi^-\pi^+$, $\Xi_c^+\to \Xi^-\pi^+\pi^+$ and
$\Xi_c^+\to pK^-\pi^+$ \cite{Belle:Xic0,Belle:Xic+}. There is a special class of hadronic decays of charmed baryons that
can be theoretically studied in a reliable way, namely, heavy-flavor-conserving
nonleptonic decays such as $\Xi_Q\to\Lambda_Q\pi$. In these decays, it is naively expected that only
the light quarks inside the heavy baryon will participate in weak
interactions and the heavy quark behaves as a ``spectator".
The charm-flavor-conserving decay $\Xi_c^0\to\Lambda_c^+\pi^-$ first advocated and studied in 1992 \cite{Cheng:HFC} was finally measured by the LHCb this year with a surprisingly large branching fraction of $(0.55\pm0.02\pm0.18)\%$ \cite{LHCb:HFC}.
The first observation of $\Xi_{cc}^{++}\to  \Xi_c^+\pi^+$ with a significance of 5.9 standard devitations has been reported by the LHCb \cite{LHCb:XiccDecay}. Motivated by these experimental progresses, there exist growing theoretical activities in the study of weak decays of singly and doubly charm baryons.

For exclusive semileptonic decays of charmed baryons,
measurement of the absolute branching fraction was first made by BESIII in 2015 for $\Lambda_c^+\to\Lambda e^+\nu_e$ \cite{BESIII:LamcSL} and subsequently  for $\Lambda_c^+\to\Lambda \mu^+\nu_\mu$ \cite{BESIII:LamcSLmu}. Very recently, both Belle \cite{Belle:Xic0semi} and ALICE \cite{ALICE:XicSL} have reported the measurements of the branching fractions of the semileptonic $\Xi_c^0\to \Xi^-\ell^+\nu_\ell$ decays. Although the electromagnetic decay rates of $S$-wave charmed
baryons such as $\Xi'^0_c\to\Xi_c^0\gamma$, $\Xi'^+_c\to
\Xi^+_c\gamma$ and $\Omega_c^{*0}\to\Omega_c^0\gamma$ have not been measured so far, Belle has recently presented a successful search for the electromagnetic decays $\Xi_c(2790,2815)\to \Xi_c\gamma$ with the neutral orbitally excited $\Xi_c$ baryons \cite{Belle:rad}.

This review is essentially an update of Refs. \cite{Cheng:2009} and \cite{Cheng:2015} in which charmed baryon physics around 2007 and 2015, respectively, were described. In this review we shall emphasize the experimental progress and achievement since 2015 and the theoretical development. Some other review articles on charmed baryons can be found in
\cite{Korner94,Bigireview,Roberts,Klempt,Crede,Bphysics,Bigi,Li:2021iwf,Groote:2021pxt}.

\section{Spectroscopy}
\subsection{Singly charmed baryons}
A singly charmed baryon is composed of a charmed quark and
two light quarks. Each
light quark is a triplet of flavor SU(3). There are two different SU(3)
multiplets of charmed baryons: a symmetric sextet ${\bf 6}_F$ and an
antisymmetric antitriplet ${\bf \bar 3}_F$. The $\Lambda_c^+$,
$\Xi_c^+$ and $\Xi_c^0$ form a ${\bf \bar 3}_F$ representation and
they all decay weakly. The $\Omega_c^0$, $\Xi'^+_c$, $\Xi'^0_c$
and $\Sigma_c^{++,+,0}$ form a ${\bf 6}_F$ representation; among them,
only $\Omega_c^0$ decays weakly.

In the quark model, the orbital angular momentum of the light
diquark can be decomposed into ${\bf L}_\ell={\bf L}_\rho+{\bf
L}_\lambda$, where ${\bf L}_\rho$ is the orbital angular momentum
between the two light quarks, and ${\bf L}_\lambda$ is the orbital
angular momentum between the diquark and the charmed quark.
Although the separate spin angular momentum
$S_\ell$ and orbital angular momentum $L_\ell$ of the light
degrees of freedom are not well defined, they are included for
guidance from the quark model. In the heavy quark limit, the spin
of the charmed quark ${\bf S}_c$, and the total angular momentum of the
two light quarks ${\bf J}_\ell={\bf S}_\ell+{\bf L}_\ell$, are separately conserved. The total angular momentum is given by ${\bf J}={\bf S}_c+{\bf J}_\ell$.
It is convenient to use $S_\ell$, $L_\ell$, and $J_\ell$ to enumerate the spectrum of states.
Denoting the quantum
numbers $L_\rho$ and $L_\lambda$ as the eigenvalues of ${\bf L}_\rho^2$ and
${\bf L}_\lambda^2$, respectively, the $\rho$-orbital momentum $L_\rho$ describes relative
orbital excitations of the two light quarks, and the $\lambda$-orbital
momentum $L_\lambda$ describes orbital excitations of the center of mass of the two light quarks relative to the heavy quark. The $P$-wave heavy baryon can be in either the
$(L_\rho=0,L_\lambda=1)$ $\lambda$-state or the $(L_\rho=1,L_\lambda=0)$ $\rho$-state.
The next orbitally excited states are the positive-parity
excitations with $L_\rho+L_\lambda=2$.  There exist multiplets
with the
symmetric orbital wave function, corresponding to
$L_\lambda=2,L_\rho=0$ and $L_\lambda=0,L_\rho=2$.
The orbital $L_\lambda=L_\rho=1$ states
are antisymmetric under the interchange of two light quarks.


\begin{table}[!]
\caption{Mass spectra and widths (in units of MeV) of
charmed baryons. Experimental values are taken from the 2021 update of the Particle
Data Group (PDG) \cite{PDG}. Note that the states $\Xi_c(2939)^0$ and $\Xi_c(2965)^0$ observed by the LHCb \cite{LHCb:Xic2923} have been identified with $\Xi_c(2930)^0$ and $\Xi_c(2970)^0$, respectively, by the PDG.}
\label{tab:spectrum}
\begin{center}
\footnotesize{
\begin{tabular}{cc  c c c} \hline \hline
~~~~State~~ & ~~$J^P$~~  &
~~~~~~~~~Mass~~~~~~~~~ & ~~~~Width~~~~ &~Decay modes~\\
\hline
 $\Lambda_c^+$ & ${1\over 2}^+$  & $2286.46\pm0.14$ & & weak  \\
 $\Lambda_c(2595)^+$ & ${1\over 2}^-$  & $2592.25\pm0.28$ &
 $2.6\pm0.6$ & $\Lambda_c\pi\pi,\Sigma_c\pi$ \\
 $\Lambda_c(2625)^+$ & ${3\over 2}^-$  & $2628.11\pm0.19$ &
 $<0.97$ & $\Lambda_c\pi\pi,\Sigma_c\pi$ \\
 $\Lambda_c(2765)^+$ & $?^?$ & $2766.6\pm2.4$ & $50$ & $\Sigma_c\pi,\Lambda_c\pi\pi$ \\
 $\Lambda_c(2860)^+$ & ${3\over 2}^+$  & $2856.1^{+2.3}_{-6.0}$ & $68^{+12}_{-22}$
 & $\Sigma_c^{(*)}\pi,\Lambda_c\pi\pi,D^0p,D^+n$ \\
 $\Lambda_c(2880)^+$ & ${5\over 2}^+$ & $2881.63\pm0.24$ & $5.6^{+0.8}_{-0.6}$
 & $\Sigma_c^{(*)}\pi,\Lambda_c\pi\pi,D^0p,D^+n$  \\
 $\Lambda_c(2940)^+$ & ${3\over 2}^-$ & $2939.6^{+1.3}_{-1.5}$ & $20^{+6}_{-5}$ &
 $\Sigma_c^{(*)}\pi,\Lambda_c\pi\pi,D^0p,D^+n$  \\
 $\Sigma_c(2455)^{++}$ & ${1\over 2}^+$  & $2453.97\pm0.14$ &
 $1.89^{+0.09}_{-0.18}$ & $\Lambda_c\pi$ \\
 $\Sigma_c(2455)^{+}$ & ${1\over 2}^+$  & $2452.9\pm0.4$ &
 $<4.6$ & $\Lambda_c\pi$\\
 $\Sigma_c(2455)^{0}$ & ${1\over 2}^+$  & $2453.75\pm0.14$
 & $1.83^{+0.11}_{-0.19}$ & $\Lambda_c\pi$ \\
 $\Sigma_c(2520)^{++}$ & ${3\over 2}^+$ & $2518.41^{+0.21}_{-0.19}$
 & $14.78^{+0.30}_{-0.40}$ & $\Lambda_c\pi$\\
 $\Sigma_c(2520)^{+}$ & ${3\over 2}^+$  & $2517.5\pm2.3$
 & $<17$ & $\Lambda_c\pi$ \\
 $\Sigma_c(2520)^{0}$ & ${3\over 2}^+$  & $2518.48\pm0.20$
 & $15.3^{+0.4}_{-0.5}$ & $\Lambda_c\pi$ \\
 $\Sigma_c(2800)^{++}$ & $?^?$ & $2801^{+4}_{-6}$ & $75^{+22}_{-17}$ &
 $\Lambda_c\pi,\Sigma_c^{(*)}\pi,\Lambda_c\pi\pi$ \\
 $\Sigma_c(2800)^{+}$ & $?^?$  & $2792^{+14}_{-~5}$ & $62^{+60}_{-40}$ &
 $\Lambda_c\pi,\Sigma_c^{(*)}\pi,\Lambda_c\pi\pi$ \\
 $\Sigma_c(2800)^{0}$ & $?^?$ & $2806^{+5}_{-7}$ & $72^{+22}_{-15}$ &
 $\Lambda_c\pi,\Sigma_c^{(*)}\pi,\Lambda_c\pi\pi$\\
 $\Xi_c^+$ & ${1\over 2}^+$  & $2467.71\pm0.23$ & & weak \\
 $\Xi_c^0$ & ${1\over 2}^+$  & $2470.44\pm0.28$ & & weak \\
 $\Xi'^+_c$ & ${1\over 2}^+$ & $2578.2\pm0.5$ & & $\Xi_c\gamma$ \\
 $\Xi'^0_c$ & ${1\over 2}^+$  & $2578.7\pm0.5$ & & $\Xi_c\gamma$ \\
 $\Xi_c(2645)^+$ & ${3\over 2}^+$  & $2645.10\pm0.30$ & $2.14\pm0.19$ & $\Xi_c\pi$ \\
 $\Xi_c(2645)^0$ & ${3\over 2}^+$ & $2646.16\pm0.25$ & $2.35\pm0.22$ & $\Xi_c\pi$ \\
 $\Xi_c(2790)^+$ & ${1\over 2}^-$ & $2791.9\pm0.5$ & $8.9\pm1.0$ & $\Xi'_c\pi,\Xi_c\pi$\\
 $\Xi_c(2790)^0$ & ${1\over 2}^-$ & $2793.9\pm0.5$ & $10.0\pm1.1$ & $\Xi'_c\pi,\Xi_c\pi$ \\
 $\Xi_c(2815)^+$ & ${3\over 2}^-$  & $2816.51\pm0.25$ & $2.43\pm0.26$ & $\Xi^*_c\pi,\Xi_c\pi\pi,\Xi_c'\pi$ \\
 $\Xi_c(2815)^0$ & ${3\over 2}^-$ & $2819.79\pm0.30$ & $2.54\pm0.25$ & $\Xi^*_c\pi,\Xi_c\pi\pi,\Xi_c'\pi$ \\
 $\Xi_c(2923)^0$ & $?^?$  & $2923.04\pm0.35$ & $7.1\pm2.0$ & $\Lambda_c \ov K$ \\
 $\Xi_c(2930)^+$ & $?^?$  & $2942\pm5$ & $15\pm9$ & $\Lambda_c \ov K$ \\
 $\Xi_c(2930)^0$ & $?^?$  & $2938.55\pm0.30$ & $10.2\pm1.4$ & $\Lambda_c \ov K$ \\
 $\Xi_c(2970)^+$ & $?^?$  & $2964.3\pm1.5$ & $20.9^{+2.4}_{-3.5}$
 & $\Sigma_c \ov K,\Lambda_c \ov K\pi,\Xi_c\pi\pi$  \\
 $\Xi_c(2970)^0$ & $?^?$ & $2967.1\pm1.7$ & $$
 & ~$\Sigma_c \ov K,\Lambda_c \ov K\pi,\Xi_c\pi\pi$~ \\
 $\Xi_c(3055)^+$  & $?^?$ & $3055.9\pm0.4$ & $7.8\pm1.9$ &
 $\Sigma_c \ov K,\Lambda_c \ov K\pi,D\Lambda$  \\
 $\Xi_c(3080)^+$ & $?^?$  & $3077.2\pm0.4$ & $3.6\pm1.1$ &
 $\Sigma_c \ov K,\Lambda_c \ov K\pi,D\Lambda$  \\
 $\Xi_c(3080)^0$  & $?^?$ & $3079.9\pm1.4$ & $5.6\pm2.2$
 & $\Sigma_c \ov K,\Lambda_c \ov K\pi,D\Lambda$ \\
 $\Omega_c^0$ & ${1\over 2}^+$  & $2695.2\pm1.7$ & & weak \\
 $\Omega_c(2770)^0$ & ${3\over 2}^+$  & $2765.9\pm2.0$ & & $\Omega_c\gamma$ \\
 $\Omega_c(3000)^0$ & $?^?$  & $3000.41\pm0.22$ & $4.5\pm0.7$ & $\Xi_c\ov K$ \\
 $\Omega_c(3050)^0$ & $?^?$  & $3050.20\pm0.13$ & $<1.2$ & $\Xi_c\ov K$ \\
 $\Omega_c(3065)^0$ & $?^?$ & $3065.46\pm0.28$ & $3.5\pm0.4$ & $\Xi_c \ov K$ \\
 $\Omega_c(3090)^0$ & $?^?$ & $3090.0\pm0.5$ & $8.7\pm1.3$ & $\Xi'_c\ov K,\Xi_c \ov K$ \\
 $\Omega_c(3120)^0$ & $?^?$  & $3119.1\pm1.0$ & $<2.6$ & $\Xi'_c\ov K,\Xi_c\ov K$ \\
 \hline \hline
\end{tabular}}
\end{center}
\end{table}

Singly charmed baryon spectroscopy provides an ideal place for studying
the dynamics of the light quarks in the environment of a heavy
quark. The observed mass spectra and decay widths of singly charmed baryons are
summarized in Table \ref{tab:spectrum} with the data taken from the Particle Data Group \cite{PDG}.
By now, the $J^P={1\over 2}^+,\frac12^-,\frac32^+,\frac32^-$ and $\frac52^+$ antitriplet states $\Lambda_c,\Xi_c$ and
$J^P={1\over 2}^+,{3\over 2}^+$ sextet states $\Omega_c,\Xi'_c,\Sigma_c$
are  established (see Table \ref{tab:3and6} below for details). Notice that except for the parity of the lightest
$\Lambda_c^+$ and the heavier ones $\Lambda_c(2880)^+$ \cite{Belle:Lamc2880,LHCb:Lambdac2880}, $\Lambda_c(2860)^+$ \cite{LHCb:Lambdac2880} and $\Lambda_c^+(2940)$ \cite{LHCb:Lambdac2880} in the $\Lambda_c^+$ sector,
and $\Xi_c(2970)^+$ in the $\Xi_c$ sector \cite{Belle:Xic2970}, none of the other $J^P$ quantum numbers shown in
Table \ref{tab:spectrum} has been measured. One has to rely on the
quark model to determine the $J^P$ assignments.

The charmed baryon spectroscopy has been studied extensively in many
various models.  It appears that  the spectroscopy is reasonably well described by the model based on the relativitsic heavy quark-light diquark model advocated by Ebert, Faustov and Galkin (EFG) \cite{Ebert:2011}. Indeed, the
quantum numbers $J^P=\frac52^+$ of $\Lambda_c(2880)$ have been correctly predicted in the model based on the diquark idea \cite{Selem} even before its discovery in the Belle experiment~\cite{Belle:Lamc2880}.
Based on the heavy quark-light diquark model, EFG have constructed the Regge trajectories of heavy baryons for orbital and radial excitations; all available experimental data on heavy baryons fit nicely to linear Regge trajectories, namely, the trajectories in the $(J,M^2)$ and $(n_r,M^2)$ planes for orbitally and radially excited heavy baryons, respectively:
\begin{eqnarray}
J=\alpha M^2+\alpha_0, \qquad n_r=\beta M^2+\beta_0,
\end{eqnarray}
where $J$ is the baryon spin, $M$ is the baryon mass, $n_r$ is the radial excitation quantum number, $\alpha$, $\beta$ are the slopes and $\alpha_0$, $\beta_0$ are the intercepts. The Regge trajectories can be plotted for charmed baryons with natural $(P=(-1)^{J-1/2})$ and unnatural $(P=(-1)^{J+1/2})$ parities.
Various Regge trajectories in the $(J^P,M^2)$ plane for $\Omega_c,\Lambda_c,\Xi_c,\Xi'_c$ and $\Sigma_c$ states will be shown below.  We have proposed in \cite{Cheng:2015}  to employ the predictions of the spin-parity quantum numbers of charmed baryons and their masses in \cite{Ebert:2011} as a theoretical benchmark, where the linearity, parallelism and equidistance of the Regge trajectories were verified in their calculations.

\vskip 0.3cm
In the following, we discuss some of the excited charmed baryon
states:
\subsubsection{$\Lambda_c$ states}

$\Lambda_c(2765)^+/\Sigma_c(2765)^+$ is a broad state first seen in the $\Lambda_c^+\pi^+\pi^-$ decay by CLEO~\cite{CLEO:Lamc2880} and later reported by Belle in the $\Sigma_c^{++/0}\pi^\mp$ final state \cite{Belle:Lamc2880}.
However, it is still not known whether it is
$\Lambda_c^+$ or $\Sigma_c^+$ and whether the large width might be due
to overlapping states.  In the quark-diquark model, it has also been proposed to be either the first radial ($2S$)
excitation of the $\Lambda_c$ with $J^P=\frac12^+$ containing the light scalar diquark  or the first orbital
excitation ($1P$) of the $\Sigma_c$ with $J^P=\frac32^-$ containing the light axial-vector diquark \cite{Ebert:2007}. Recently, Belle has determined the isospin of $\Lambda_c(2765)^+/\Sigma_c(2765)^+$ by searching for possible isospin partners in the decays to
$\Sigma_c^{++/0}\pi^0\to \Lambda_c^+\pi^\pm\pi^0$ \cite{Belle:Lamc2765}. The isospin is determined to be $I=0$, suggesting that the name should go with $\Lambda_c(2765)^+$. Therefore, we shall assign the quantum numbers $\frac12^+(2S)$ to $\Lambda_c(2765)^+$.

\begin{figure}[t]
\begin{center}
\includegraphics[width=80mm]{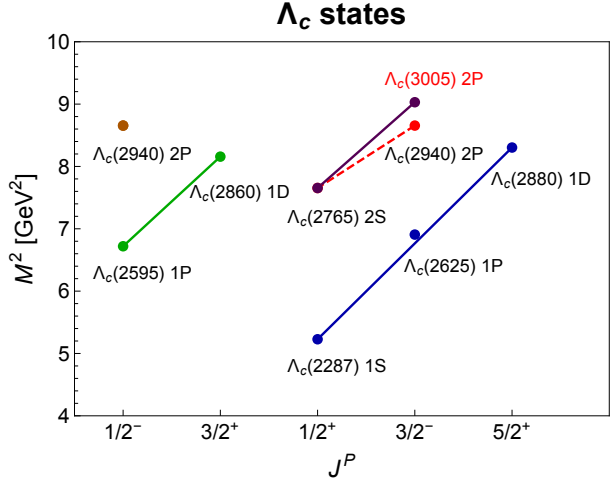}
\caption{Regge trajectories of the $\Lambda_c$ states in the $(J^P,M^2)$ plane with natural $(1/2^+,3/2^-,5/2^+)$ and unnatural $(1/2^-,3/2^+)$ parities.  The yet detected state is labeled in red.  The red dashed line shows a discordant identification of $\Lambda_c(2940)$ as a $3/2^-(2P)$ state.}
\label{fig:Lambdac}
\end{center}
\end{figure}

The state $\Lambda_c(2880)^+$, first observed by CLEO~\cite{CLEO:Lamc2880} in the $\Lambda_c^+\pi^+\pi^-$ decay, was also seen by
BaBar in the $D^0p$ spectrum~\cite{BaBar:Lamc2940}.  Belle studied the
experimental constraint on its $J^P$ quantum numbers~\cite{Belle:Lamc2880} and found that $J^P=\frac52^+$ was favored by the angular analysis of $\Lambda_c(2880)^+\to\Sigma_c^{0,++}\pi^\pm$ decays. The mass, width and quantum numbers of $\Lambda_c(2880)$ were later confirmed by LHCb~\cite{LHCb:Lambdac2880}. The $\frac12^+(1S)$ $\Lambda_c$, $\frac32^-(1P)$ $\Lambda_c(2625)$ and $\frac52^+(1D)$ $\Lambda_c(2880)$ states form a Regge trajectory as depicted in Fig.~\ref{fig:Lambdac}. The new resonance $\Lambda_c(2860)^+$ observed by LHCb, as manifested in the near-threshold enhancement in the $D^0p$ amplitude through an amplitude analysis of the $\Lambda_b^0\to D^0p\pi^-$ decay, has $J^P=\frac 32^+$ with mass and width shown in Table~\ref{tab:spectrum}~\cite{LHCb:Lambdac2880}.
It forms another Regge trajectory with $\frac12^-(1P)$ $\Lambda_c(2595)$.
It is worth mentioning that the existence of this new state $\Lambda_c(2860)^+$ was noticed  before the LHCb experiment~\cite{Chen:Sigmac,Lu:2016ctt,Chen:Dwave}.
We see from Fig.~\ref{fig:Lambdac} that both trajectories are parallel to each other nicely.

The highest state $\Lambda_c(2940)^+$ was first discovered by BaBar in
the $D^0p$ decay mode~\cite{BaBar:Lamc2940} and  confirmed by
Belle in the $\Sigma_c^0\pi^+$ and $\Sigma_c^{++}\pi^-$ decays, which
subsequently decay into $\Lambda_c^+\pi^+\pi^-$~\cite{Belle:Lamc2880}. Its spin-parity assignment is quite diverse (see \cite{Cheng:2015} for a review). The constraints on its spin and parity were recently studied by LHCb~\cite{LHCb:Lambdac2880}. The most likely assignment was found to be $J^P=\frac32^-$. However,
if we draw a Regge trajectory connecting $\Lambda_c(2940)$ and $\Lambda_c(2765)$ with $\frac12^+(2S)$, we shall see that this Regge line is not parallel to the other two Regge trajectories. If we use the quark-diquark model prediction of $\Lambda_c(3005)$ for the $\frac32^-(2P)$ state \cite{Ebert:2011}, the trajectories satisfy the parallelism nicely. Hence, we suggest that the quantum numbers of $\Lambda_c(2940)^+$ are most likely $\frac12^-(2P)$.
Indeed, LHCb has cautiously stated that ``The most likely spin-parity assignment for $\Lambda_c(2940)$ is $J^P=\frac32^-$ but the other solutions with spin 1/2 to 7/2 cannot be excluded.'' In order to clarify this issue, it is thus important to search for the $\Lambda_c^+$ state with a mass of order 3005 MeV and verify its quantum numbers as $\frac32^-(2P)$.

In the quark-model calculations the mass of $\Lambda_c(3/2^-,2P)$ lies in the range of $3000\sim 3040$ MeV \cite{Ebert:2011,Chen:Sigmac,Capstick:1985xss,Chen:2015lpa}, while $\Lambda_c(1/2^-,2P)$ is slightly lighter by not more than 25 MeV. If $\Lambda_c(2940)$ is identified with the state $\Lambda_c(3/2^-,2P)$, there will be a low mass puzzle for $\Lambda_c(2940)$. It was argued recently in \cite{Luo:2019qkm} that the mass of $\Lambda_c(3/2^-,2P)$ can be lowered down to be consistent with the experimental data of $\Lambda_c(2940)$ by introducing the $D^*N$ coupled channel contribution, in analogue to the states, for example, $\Lambda(1405)$ and $D^*_{s0}(2317)$ whose masses  are typically 100 MeV lower than the expectation from the quark model calculations due to the coupled channel effects from the nearby $\bar KN$ and $DK$ thresholds, respectively. In this scenario, the mass of $\Lambda_c(1/2^-,2P)$ is higher than that of $\Lambda_c(3/2^-,2P)$ by around 40 MeV; that is, it leads to an interesting mass inverted relation in the charmed baryon system.

As we shall see below,  the Regge trajectories do respect the parallelism satisfactorily not only for the antitriplet $\Lambda_c$ and $\Xi_c$ states but also for the sextet $\Omega_c$, $\Xi'_c$ and $\Sigma_c$ ones. Therefore, we still believe that the parallelism of Regge trajectories strongly suggests that
$\Lambda(3/2^-,2P)$ should not be identified with $\Lambda_c(2940)$, irrespective of its small mass puzzle. At any rate, we need to wait for the experimental clarification for this issue.

\subsubsection{$\Xi_c$ states}

\begin{figure}[t]
\begin{center}
\vspace{10pt}
\includegraphics[width=70mm]{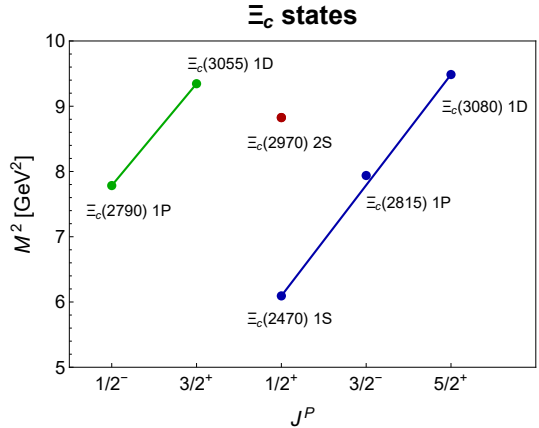}
\caption{Regge trajectories of the $\Xi_c$ states in the $(J^P,M^2)$ plane  with natural $(1/2^+,3/2^-,5/2^+)$ and unnatural $(1/2^-,3/2^+)$ parities.}
\label{fig:Xic}
\end{center}
\end{figure}

Another example showing the usefulness of the Regge phenomenology in the $J^P$ assignment of charmed baryons is the $\Xi_c$ states.
The Regge analysis suggests $3/2^+(1D)$ for $\Xi_c(3055)$ and $5/2^+(1D)$ for $\Xi_c(3080)$~\cite{Ebert:2011} (see also discussions in~\cite{Chen:Lambdac2880}). The $\Xi_c(2470)$, $\Xi_c(2815)$ and $\Xi_c(3080)$ states form a $\frac12^+$ Regge trajectory, while  $\Xi_c(2790)$ and $\Xi_c(3055)$ form a $\frac12^-$ one (see Fig.~\ref{fig:Xic}). They are parallel to each other nicely.

The state $\Xi_c(2970)$ has been observed in different decay modes:  $\Lambda_c^+\bar K\pi$, ${\Xi'}_c^+\pi^-$ and $\Xi_c(2645)^+\pi^-$ by Belle \cite{Belle:Xic2006,Belle:Xic2008,Belle:Xic2016} and $\Sigma_c(2455)^0K_S^0$ by BaBar \cite{BaBar:Xic2008}. For the neutral one, we quote the 2020 version of PDG \cite{PDG} for its mass and width
\footnote{The new state $\Xi_c(2965)^0$ discovered by the LHCb \cite{LHCb:Xic2923} has been identified with $\Xi_c(2970)^0$ by the PDG \cite{PDG} due to their mass closeness. As will be discussed in Sec. II.A.4 below, they are mostly likely two different states.}
\be \label{eq:Xic2970}
\Xi_c(2970)^0:\qquad M=2970.9^{+0.4}_{-0.6}\,{\rm MeV}, \qquad \Gamma=28.1^{+3.4}_{-4.0}\,{\rm MeV}.
\en
Belle has recently made an effort to measure the spin and parity of $\Xi_c(2970)$ for the first time \cite{Belle:Xic2970}. The spin was determined by testing possible spin hypothesis of
$\Xi_c(2970)$ with the angular analysis of the decay $\Xi_c(2970)^+\to\Xi_c(2645)^0\pi^+\to\Xi_c^+\pi^-\pi^+$. Likewise, its parity was established
from the ratio of branching fractions of the two decays, $\Xi_c(2970)^+\to\Xi_c(2645)^0\pi^+$ and $\Xi_c(2970)^+\to{\Xi'_c}^0\pi^+$. The analysis favors the spin-parity $J^P=1/2^+$ with the spin of the light-quark degrees of freedom $S_\ell=0$. Indeed, an assignment of a 2$S$ state with $J^P=1/2^+$ is expected in most of the theoretical studies, see e.g. \cite{Ebert:2007,Ebert:2011,Chen:LambdaQ,Cheng:2007HHChPT,Cheng:2015HHChPT}.

\subsubsection{$\Omega_c$ states}
For a long time, only two ground states had been observed  for the $\Omega_c$ baryons: $\frac12^+$ $\Omega_c^0$ and $\frac32^+$ $\Omega_c(2770)^0$. The latter was seen by BaBar in the electromagnetic decay
$\Omega_c(2770)\to\Omega_c\gamma$ \cite{BaBar:Omegacst}. The mass difference between $\Omega_c^*$ and $\Omega_c$ is too small for any strong decay to occur.
LHCb has started to explore this sector in 2017 and observed five new, narrow excited $\Omega_c$ states decaying into $\Xi_c^+K^-$: $\Omega_c(3000)$, $\Omega_c(3050)$, $\Omega_c(3065)$, $\Omega_c(3090)$ and $\Omega_c(3120)$ \cite{LHCb:Omegac}. Except $\Omega_c(3120)$, the first four states were also confirmed by Belle later \cite{Belle:Omegac}. The observation of excited $\Omega_c^0$ states has triggered a flood of interest in attempting to identify their spin-parity quantum numbers \cite{Agaev,Cheng:Omegac,Chen:Pwave,Karliner:2017,Yang,Zhao,Wangwei,Huang:Omegac,Wangzg,Wang:2017xam,Padmanath,Chen:Omegac,Zhao:Omegac,Aliev:Omegac,Kim,Agaev_2,Santopinto,Faustov,Yang:2021lce,Jia:2020vek}.

\begin{table}[t]
\caption{The $P$-wave $\Omega_c$ baryons in terms of the notations $\Omega_{cJ_\ell}(J^P)$, $\tilde \Omega_{cJ_\ell}(J^P)$ and $[{\bf 6}_F,J_\ell,S_\ell, \rho/\lambda]$,
where $J_\ell$ is the total angular momentum of
the two light quarks \cite{Cheng:2007HHChPT,Zhu}. } \label{tab:pwave}
\begin{center}
\begin{tabular}{|c|cccc c|} \hline
~~~~~State~~~~~ & SU(3) & ~~$S_\ell$~~ & ~~$L_\ell(L_\rho,L_\lambda)$~~&
~~$J_\ell^{P_\ell}$~~ & ~~$[{\bf 6}_F,J_\ell,S_\ell, \rho/\lambda]$~~ \\
 \hline
 $\Omega_{c0}({1\over 2}^{-})$ & ${\bf 6}_F$ & 1 & 1\,(0,1) & $0^-$ & $[{\bf 6}_F,0,1,\lambda]$
 \\
 $\Omega_{c1}({1\over 2}^{-},{3\over 2}^{-})$ & ${\bf 6}_F$ & 1 & 1\,(0,1) & $1^-$ & $[{\bf 6}_F,1,1,\lambda]$
 \\
 $\Omega_{c2}({3\over 2}^{-},{5\over 2}^{-})$ & ${\bf 6}_F$ & 1 & 1\,(0,1) & $2^-$ &  $[{\bf 6}_F,2,1,\lambda]$
 \\
 $\tilde \Omega_{c1}({1\over 2}^{-},{3\over 2}^{-})$ & ${\bf 6}_F$ & 0 & 1\,(1,0) & $1^-$ & $[{\bf 6}_F,1,0,\rho]$
 \\
 \hline
\end{tabular}
\end{center}
\end{table}

Assuming that the spin of the two light quarks $S_\ell$ is 1 for the sextet baryon in the quark model, there are five first $P$-wave orbitally excited $\Omega_c$ states, namely $\Omega_{c0}({1\over 2}^-)$,  $\Omega_{c1}({1\over 2}^-,{3\over 2}^-)$ and $\Omega_{c2}({3\over 2}^-,{5\over 2}^-)$  with $J_\ell$ being the total angular momentum of the two light quarks \cite{Cheng:2007HHChPT,Zhu} (see Table \ref{tab:pwave}). This has led many authors \cite{Karliner:2017,Zhao,Wangwei,Wangzg,Padmanath,Chen:Omegac,Jia:2020vek} to
advocate that the observed five $\Omega_c$ resonances can be assigned to the five orbitally excited $1P(1/2^-,3/2^-,5/2^-)$ states (see Table \ref{tab:JPOmegac}).

However, it is very likely that not all the observed $\Omega_c$ baryons can be interpreted as the $P$-wave orbitally excited states. The argument goes as follows. In the presence of the spin-orbit interaction ${\bf S}_c\cdot{\bf L}$ and the tensor interaction, states with the same $J^P$ but different $J_\ell$ will mix together \cite{Ebert:2011}.
Following \cite{Chen:Sigmac,Chen:Omegac}, we write
\begin{eqnarray} \label{eq:1/2mixing}
\left(\begin{array}{c} (1P,1/2^-)_l \\ (1P,1/2^-)_h
\end{array}\right)=\left(
\begin{array}{cc} \cos\theta_1 & -\sin\theta_1 \\
\sin\theta_1 & \cos\theta_1 \end{array}\right)
\left(\begin{array}{c} \Omega_{c0}(1/2^-) \\ \Omega_{c1}(1/2^-)
\end{array}\right),
\end{eqnarray}
and
\begin{eqnarray} \label{eq:3/2mixing}
\left(\begin{array}{c} (1P,3/2^-)_h \\ (1P,3/2^-)_l
\end{array}\right)=\left(
\begin{array}{cc} \cos\theta_2 & -\sin\theta_2 \\
\sin\theta_2 & \cos\theta_2 \end{array}\right)
\left(\begin{array}{c} \Omega_{c1}(3/2^-) \\ \Omega_{c2}(3/2^-)
\end{array}\right).
\end{eqnarray}
Since the width of $\Omega_{c0}(\frac12^-)$ is estimated to be of order 410 MeV in heavy hadron chiral perturbation theory \cite{Cheng:Omegac},
$\Omega_c(3000)$ cannot be a pure $\Omega_{c0}(\frac12^-)$ state due to a very broad width expected for the $s$-wave transition. Nevertheless, it can be identified with $\Omega_{c1}(\frac12^-)$ since its decay into $\Xi_c K$ is prohibited in the heavy quark limit but could be allowed when heavy quark symmetry is broken. This means
the mixing angles $\theta_1$ and $\theta_2$ defined in Eqs.~(\ref{eq:1/2mixing}) and (\ref{eq:3/2mixing}) are constrained by the data to be around $96^\circ$ and $160^\circ$, respectively \cite{Cheng:Omegac}. Obviously, the other state $(1P,1/2^-)_h$ will be too broad to be observed. Hence, it was conclude in \cite{Cheng:Omegac} that only one of the  $(1P,1/2^-)$ states can be identified with the observed narrow $\Omega_c$ baryon. The state $\Omega_c(3000)$ is narrow in this scenario because it is primarily a $\Omega_{c1}(\frac12^-)$ state with a very small component of $\Omega_{c0}(\frac12^-)$.

\begin{table}[t]
\caption{Proposed spin-parity quantum numbers for the excited $\Omega_c$ baryons.} \label{tab:JPOmegac}
\begin{center}
\begin{tabular}{l ccc c c } \hline\hline
  & ~$\Omega_c(3000)^0$~ & ~~$\Omega_c(3050)^0$~~ & ~$\Omega_c(3065)^0$~ & ~$\Omega_c(3090)^0$~ & ~$\Omega_c(3120)^0$~ \\
 \hline
Agaev {\it et al.} \cite{Agaev_2} & $1/2^-$ & $3/2^-$ & $1/2^+$ & $1/2^+$ & $3/2^+$ \\
Aliev {\it et al.} \cite{Aliev:Omegac}  & $1/2^-$ & $$ & $3/2^-$ & $$ & $$ \\
B. Chen, X. Liu \cite{Chen:Omegac} & $1/2^-$ & $3/2^-$ & $5/2^-$ & $1/2^-$ & $3/2^-$ \\
H. Chen {\it et al.}  \cite{Chen:Dwave} & $1/2^-$ & $1/2^-$ & $1/2^-$ or $1/2^+$ & & $3/2^+$ \\
Cheng, Chiang \cite{Cheng:Omegac} & $1/2^-$ & $3/2^-$ & $5/2^-$ & $1/2^+$ & $3/2^+$ \\
Faustov, Galkin \cite{Faustov}  & $3/2^-$ & $5/2^-$ & $3/2^-$ & $1/2^+$ & $3/2^+$ \\
Huang {\it et al.} \cite{Huang:Omegac}  & $$ & $$ & $$ &  & $1/2^-$ \\
Jia {\it et al.} \cite{Jia:2020vek} & $1/2^-$ & $1/2^-$ & $3/2^-$ & $3/2^-$ & $5/2^-$ \\
Karliner, Rosner \cite{Karliner:2017}: (i) & $1/2^-$ & $1/2^-$ & $3/2^-$ & $3/2^-$ & $5/2^-$ \\
~~~~~~~~(ii) & $3/2^-$ & $3/2^-$ & $5/2^-$ & $1/2^+$ & $3/2^+$ \\
Padmanath {\it et al.} \cite{Padmanath}  & $1/2^-$ & $1/2^-$ & $3/2^-$ & $3/2^-$ & $5/2^-$ \\
Santopinto {\it et al.} \cite{Santopinto} & $1/2^-$ & $3/2^-$ & $1/2^-$ & $3/2^-$ & $5/2^-$ \\
K. Wang {\it et al.}  \cite{Zhao} & $1/2^-$ & $3/2^-$ & $3/2^-$ & $5/2^-$ & $1/2^+$ or $3/2^+$ \\
W. Wang, R.L. Zhu  \cite{Wangwei} & $1/2^-$ & $1/2^-$ & $3/2^-$ & $3/2^-$ & $5/2^-$ \\
Z. Wang \cite{Wangzg} & $1/2^-$ & $1/2^-$ & $3/2^-$ & $3/2^-$ & $5/2^-$ \\
Z. Wang {\it et al.} \cite{Wang:2017xam} & $1/2^-$ & $$ & $$ & $3/2^-$ or $1/2^+$ & $3/2^+$ \\
Yang, H. Chen \cite{Yang:2021lce} &  $1/2^-$ or $3/2^-$ & $1/2^-$ & $3/2^-$ & $3/2^-$ & $5/2^-$ \\
Z. Zhao {\it et al.} \cite{Zhao:Omegac}: (i) & $1/2^+$  & $5/2^+$  & $3/2^-$  & $3/2^-$  & $5/2^+$   \\
~~~~~~~~(ii) & $3/2^+$ & $7/2^+$ &  $5/2^-$  &  $5/2^-$ &  $7/2^+$  \\
 \hline\hline
\end{tabular}
\end{center}
\end{table}

Very recently, LHCb has reported the re-observation of the four sates  $\Omega_c(3000)^0$, $\Omega_c(3050)^0$, $\Omega_c(3065)^0$ and $\Omega_c(3090)^0$ produced in the $\Xi_c^+K^-$ mass projection of the $\Omega_b^-\to \Xi_c^+K^-\pi^-$ decays, while the $\Omega_c(3120)^0$ was not observed \cite{LHCb:excitedOmegac}. The updated results are
\be
\Omega_c(3000)^0:&& \quad M=2999.2\pm0.9\pm0.9^{+0.19}_{-0.22}\,{\rm MeV}, \quad \Gamma=4.8\pm2.1\pm2.5\,{\rm MeV}, \non \\
\Omega_c(3050)^0:&& \quad M=3050.1\pm0.3\pm0.2^{+0.19}_{-0.22}\,{\rm MeV}, \quad \Gamma< 1.6\,{\rm MeV}, \non \\
\Omega_c(3065)^0:&& \quad M=3065.9\pm0.4\pm0.4^{+0.19}_{-0.22}\,{\rm MeV}, \quad \Gamma=1.7\pm1.0\pm0.5\,{\rm MeV}, \non \\
\Omega_c(3090)^0:&& \quad M=3091.0\pm1.1\pm1.0^{+0.19}_{-0.22}\,{\rm MeV}, \quad \Gamma=7.4\pm3.1\pm2.8\,{\rm MeV},
\en
where the natural width of the $\Omega_c(3050)^0$ is consistent with zero, therefore an upper limit is set.
They are consistent with the previous results measured in prompt production (see Table \ref{tab:spectrum}). Moreover, LHCb has performed the spin analysis based on the distribution of the helicity angle in the decay chain $\Omega_b^-\to \Omega_c^{**0}\pi^-$ followed by $\Omega_c^{**0}\to \Xi_c^+K^-$. The results of the spin analysis are shown in Table \ref{tab:Omegaspin} in which the significance of the rejection of the $J$ hypothesis are listed. Significance values of excluding the $J=1/2$ spin hypothesis for $\Omega_c(3050)^0$ and $\Omega_c(3065)^0$ are $2.2\,\sigma$ and $3.6\,\sigma$, respectively. The combined hypothesis of the four states in the order $J=1/2, 1/2, 3/2, 3/2$ as proposed in \cite{Jia:2020vek,Karliner:2017,Wangwei,Wangzg,Padmanath} is rejected by $3.5\,\sigma$. This is consistent with the  observation made in \cite{Cheng:Omegac} that only one of the  $(1P,1/2^-)$ states can be identified with the observed narrow $\Omega_c$ baryon, namely, the $\Omega_c(3000)^0$ state (see also \cite{Zhao}).

\begin{table}[t]
\caption{Shown are the significance values of excluding the $J$ spin hypothesis for the excited $\Omega_c^0$ baryons in the LHCb experiment \cite{LHCb:excitedOmegac}.} \label{tab:Omegaspin}
\begin{center}
\begin{tabular}{l ccc  } \hline\hline
State & ~~$J=1/2$~~ &  ~~$J=3/2$~~ & ~~$J=5/2$~~ \\
\hline
$\Omega_c(3000)^0$ & $0.5\,\sigma$ & $0.8\,\sigma$ & $0.4\,\sigma$ \\
$\Omega_c(3050)^0$ & $2.2\,\sigma$ & $0.1\,\sigma$ & $1.2\,\sigma$ \\
$\Omega_c(3065)^0$ & $3.6\,\sigma$ & $0.6\,\sigma$ & $1.2\,\sigma$ \\
$\Omega_c(3090)^0$ & $0.3\,\sigma$ & $0.8\,\sigma$ & $0.5\,\sigma$ \\
 \hline\hline
\end{tabular}
\end{center}
\end{table}

LHCb pointed out that the spin assignment of the four observed peaks is consistent with $\lambda$-mode excitations with quantum numbers $J=1/2,3/2,3/2$ and $5/2$ and the non-observation of the $\Omega_c(3120)^0$ baryon would be consistent with the state being either one of the $2S$ doublet or a $\rho$-mode $P$-wave excitation \cite{LHCb:excitedOmegac}.
It is interesting to notice that among various models listed in Table \ref{tab:JPOmegac}, only the predicted $J^P$ quantum numbers in Ref. \cite{Zhao} are consistent with the above-mentioned scenario. In the following, we shall use this plausible spin-parity assignment
\be \label{eq:JPOmega}
&&J^P[\Omega_c(3000)^0]=1/2^-, \qquad J^P[\Omega_c(3050)^0]=3/2^-, \non \\ && J^P[\Omega_c(3065)^0]=3/2^-, \qquad J^P[\Omega_c(3090)^0]=5/2^-.
\en

Since the mass gap between the observed excited $\Omega_c$ states and the ground $\Omega_c(2700)$ state ranges from 300 to 400 MeV, it is conceivable that a light quark-antiquark pair is excited from the QCD vacuum. Indeed, baryon excitation by pulling out a $q\bar q$ pair may need less energy than that by the traditional orbital excitation. Hence, the excited $\Omega_c$ states may be composed of the compact pentaquarks \cite{Wang:penta,C.Wang:penta,Kim:penta,An:penta,Yang} or molecular pentaquark states. In the molecular picture, $\Omega_c$ baryons are described as the meson-baryon molecular states such as $\Xi D, \Xi_c K$ and $\Xi'_c K$ \cite{Huang:molcu,Debastiani:molcu,Debastiani:1710,Y.Huang:molcu,Chen:molcu,Montana:molcu,Huang:Omegac}.

\subsubsection{$\Xi'_c$ states}
Besides the ground-state $1/2^+$ $\Xi_c(2578)$ and $3/2^+$ $\Xi_c(2645)$ baryons, a broader resonance  $\Xi_c(2930)$ at 2930 MeV was seen by BaBar \cite{BaBar:Xic2930} and subsequently by Belle \cite{Belle:Xic2930} in $B^-\to K^-\Lambda_c^+\bar\Lambda_c^-$ decays. In a recent LHCb experiment, three new resonances $\Xi_c(2923)^0$, $\Xi_c(2939)^0$ and $\Xi_c(2965)^0$ were observed in the $\Lambda_c^+K^-$ spectrum \cite{LHCb:Xic2923}
\be
\Xi_c(2923)^0:&& \quad M=2923.04\pm0.25\pm0.24\,{\rm MeV}, \qquad \Gamma=7.1\pm0.8\pm1.8\,{\rm MeV}, \non \\
\Xi_c(2939)^0:&& \quad M=2938.55\pm0.21\pm0.22\,{\rm MeV}, \qquad \Gamma=10.2\pm0.8\pm1.1\,{\rm MeV}, \non \\
\Xi_c(2965)^0:&& \quad M=2964.88\pm0.26\pm0.20\,{\rm MeV}, \qquad \Gamma=14.1\pm0.9\pm1.3\,{\rm MeV}.
\en
As claimed by the LHCb, the lack of any $\Xi_c(2930)^0$ signal indicates that the broad bump observed in $B^-\to K^-\Lambda_c^+\bar\Lambda_c^-$ decays \cite{BaBar:Xic2930,Belle:Xic2930} might be due to the overlap of two narrower states, such as $\Xi_c(2923)^0$ and $\Xi_c(2939)^0$.
As for the $\Xi_c(2965)^0$ baryon, it is in the vicinity of the known $\Xi_c(2970)$ baryon
and has been identified with $\Xi_c(2970)^0$ by the PDG \cite{PDG} due to their mass closeness.  However, the natural width and mass of the $\Xi_c(2965)^0$ resonance  differ significantly from those of the $\Xi_c(2970)^0$ baryon given in Eq. (\ref{eq:Xic2970}).
Hence, it was advocated by the LHCb that they are two different states and
further studies are required to establish whether the $\Xi_c(2965)^0$ is indeed a different baryon \cite{LHCb:Xic2923}. We shall see shortly that the equal spacing rule together the spin analysis of $\Omega_c(3090)$ by the LHCb suggests that the spin-parity quantum numbers of $\Xi_c(2965)$ are preferred to be $5/2^-$. Since the $J^P$ quantum numbers of $\Xi_c(2970)$ have been recently measured by Belle to be $1/2^+$ \cite{Belle:Xic2970}, this is another argument against the identification of $\Xi_c(2965)$ with $\Xi_c(2970)$.

\begin{figure}[t]
\begin{center}
\vspace{10pt}
\includegraphics[width=70mm]{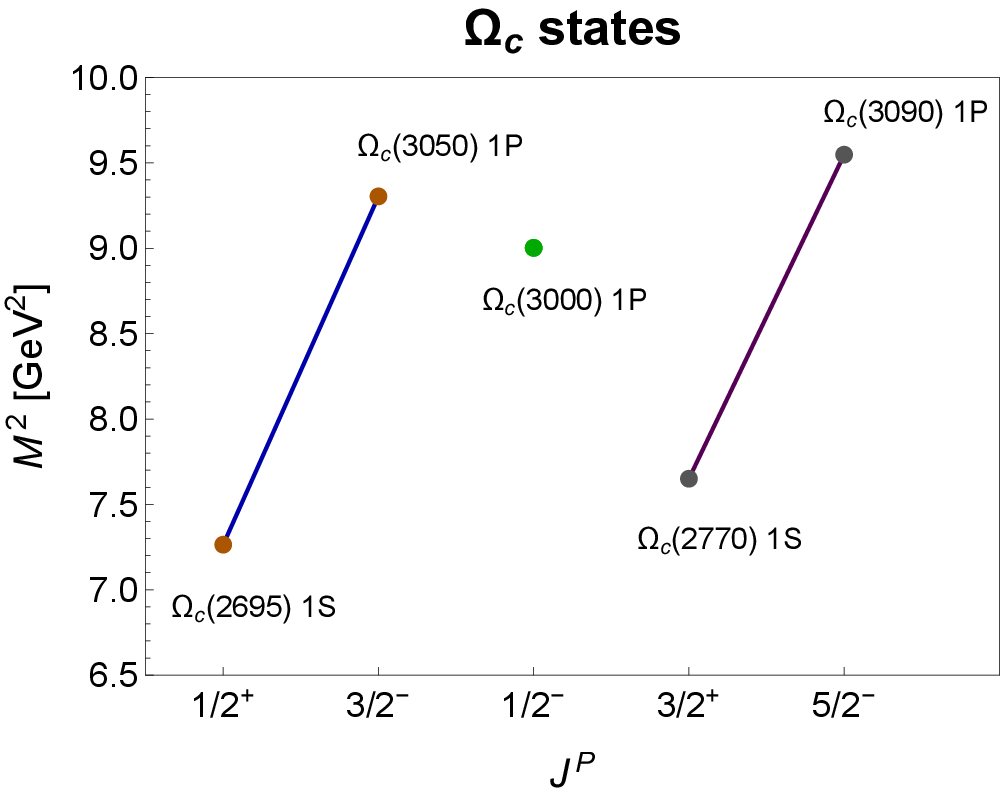}\hspace{0.8cm}
\includegraphics[width=70mm]{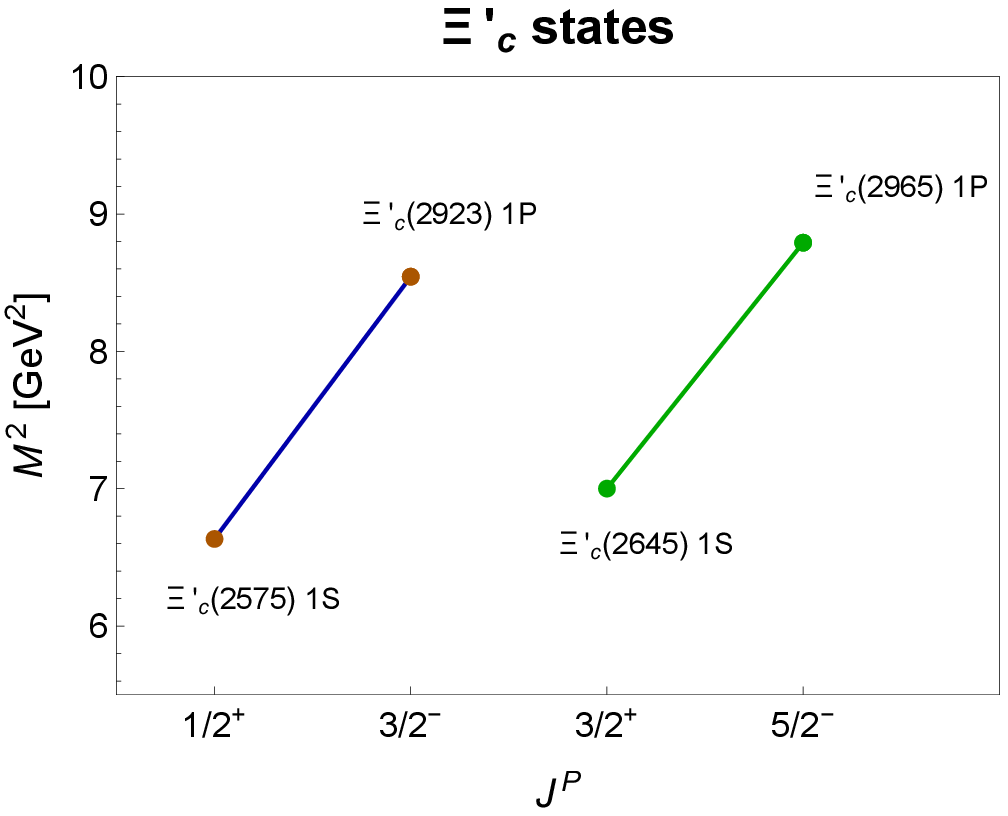}
\caption{Regge trajectories of the excited $\Omega_c$ and $\Xi'_c$ states in the $(J^P,M^2)$ plane with natural $(1/2^+,3/2^-)$ and unnatural $(1/2^-,3/2^+,5/2^-)$ parities. As there are two $3/2^-$ states in each charmed baryon family, we assign the lighter ones to the Regge trajectories, namely, $\Omega_c(3050)$ and $\Xi'_c(2923)$.
}
\label{fig:Omegac,Xicp}
\end{center}
\end{figure}

We note that the equal spacing rule applies well to the sextet of the $J^P=3/2^+$ charmed baryon ground states:
\be
M[\Omega_c(2770)^0]-M[\Xi_c(2645)^0]\simeq M[\Xi_c(2645)^0]-M[\Sigma_c(2520)^0]\simeq 125\,{\rm MeV}.
\en
The same rule also holds for the excited $\Xi'_c$ baryons within a precision of a few MeV:
\be \label{eq:spacingrule}
&& M[\Omega_c(3050)^0]-M[\Xi_c(2923)^0]\simeq M[\Xi_c(2923)^0]-M[\Sigma_c(2800)^0]\simeq 125\,{\rm MeV}, \non \\
&& M[\Omega_c(3065)^0]-M[\Xi_c(2939)^0]\simeq 125\,{\rm MeV}, \\
&& M[\Omega_c(3090)^0]-M[\Xi_c(2965)^0]\simeq 125\,{\rm MeV}. \non
\en
The universal behavior of mass gaps in the singly heavy baryon system was recently analyzed and justified within the framework of a constituent quark model \cite{Chen:massgap}.
Hence, we shall assume that $\Xi_c(2923)$ and $\Omega_c(3050)$ belong to the same ${\bf 6}_F$ multiplet and likewise for $\Xi_c(2939)$ and $\Omega_c(3065)$ as well as $\Xi_c(2965)$ and $\Omega_c(3090)$.
From the spin-parity quantum numbers for the excited $\Omega_c$ baryons given in Eq. (\ref{eq:JPOmega}), it follows that
\be \label{eq:JPof Xic}
J^P[\Xi'_c(2923)^0]=3/2^-, \qquad J^P[\Xi'_c(2939)^0]=3/2^-, \qquad J^P[\Xi'_c(2965)^0]=5/2^-.
\en
The $J^P$ quantum numbers of the newly discovered excited $\Xi'_c$ baryons have been extensively studied in the literature \cite{Yang:2021lce,Bijker:Xic,Agaev:Xic,Zhu:Xic,Wang:Xic,Lu:Xic,Yang:Xic}.
Among them, the predicted $J^P$ in \cite{Wang:Xic} are consistent with the above assignments. Indeed, the predicted spin-parity quantum numbers of excited $\Omega_c$ states based on the approach of \cite{Wang:Xic} also agree with Eq. (\ref{eq:JPOmega}) \cite{Zhao}.

In Fig. \ref{fig:Omegac,Xicp} we show the Regge trajectories in the $(J^P,M^2)$ plane for excited $\Omega_c$ and $\Xi'_c$ states. We will discuss their implications in Sec. II.A.6\,.

\subsubsection{$\Sigma_c$ states}
An inspection of Table \ref{tab:spectrum} for the mass spectra and widths of charmed baryons, it is obvious that the natural widths of the singly-charged $\Sigma_c(2455)^+$ and $\Sigma_c(2520)^+$ have not been measured. Also their masses are less accurately measured than their doubly-charged and neutral analogs. Using the experimental value of $\Gamma(\Sigma_c(2455)^{++})$ as input to fix the magnitude of the coupling constant $g_2$ within the framework of heavy hadron chiral perturbation theory (HHChPT), we have predicted
$\Gamma(\Sigma_c(2455)^{+})=2.3^{+0.1}_{-0.2}$ MeV and $\Gamma(\Sigma_c(2520)^{+})=15.2^{+0.6}_{-1.3}$ MeV \cite{Cheng:2015HHChPT}. A new measurement of the masses and widths by Belle \cite{Belle:Sigmac2021} yields
\be
M(\Sigma_c(2455)^+)-M(\Lambda_c^+)&=&166.17\pm0.05^{+0.16}_{-0.07}\,{\rm MeV}, \non\\
M(\Sigma_c(2520)^+)-M(\Lambda_c^+)&=&230.9\pm0.5^{+0.5}_{-0.1}\,{\rm MeV},
\en
and
\be
\Gamma(\Sigma_c(2455)^{+})=2.3\pm0.3\pm0.3\,{\rm MeV}, \qquad
\Gamma(\Sigma_c(2520)^{+})=17.2^{+2.3+3.1}_{-2.1-0.7}\,{\rm MeV}.
\en
The measured partial widths are thus in agreement with theory, especially for $\Sigma_c(2455)^{+}$.

\begin{figure}[t]
\begin{center}
\includegraphics[width=70mm]{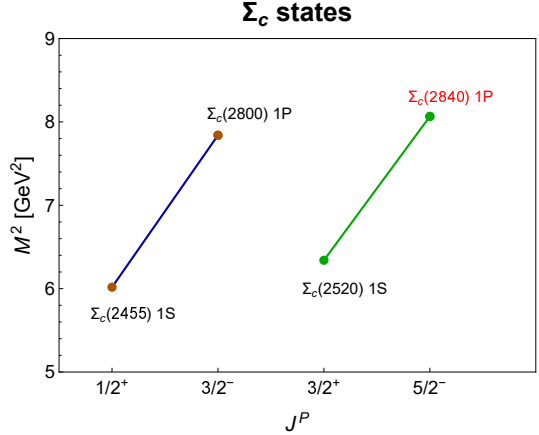}
\caption{Regge trajectories of the $\Sigma_c$ states in the $(J^P,M^2)$ plane  with natural $(1/2^+,3/2^-)$ and unnatural $(3/2^+,5/2^-)$ parities.  The yet detected state is labeled in red.}
\label{fig:Sigmac}
\end{center}
\end{figure}

The highest isotriplet charmed baryons, $\Sigma_c(2800)^{++,+,0}$,
which decay to $\Lambda_c^+\pi$, were first measured by Belle
\cite{Belle:Sigc2800} with widths of the order of 70 MeV. The possible quark states are $\Sigma_{c0}({1\over 2}^-)$,  $\Sigma_{c1}({1\over 2}^-,{3\over 2}^-)$, $\tilde\Sigma_{c1}({1\over 2}^-,{3\over 2}^-)$, and $\Sigma_{c2}({3\over 2}^-,{5\over 2}^-)$, in analog to $\Omega_c^{**}$ (see Table \ref{tab:pwave}).
The states $\Sigma_{c1}$ and $\tilde\Sigma_{c1}$ are ruled out
because their decays to $\Lambda_c^+\pi$ are prohibited in the heavy quark limit.  Now the
$\Sigma_{c2}({3\over 2}^-,{5\over 2}^-)$ baryon decays primarily into the
$\Lambda_c\pi$ system in a $D$-wave, whereas $\Sigma_{c0}({1\over 2}^-)$  decays into $\Lambda_c\pi$ in an $S$-wave. Because HHChPT implies a very broad $\Sigma_{c0}$ with a width of the order of 885 MeV \cite{Cheng:2015HHChPT}, this  $P$-wave state is also unlikely. Therefore, $\Sigma_c(2800)^{++,+,0}$ are likely to
be either $\Sigma_{c2}({3\over 2}^-)$ or $\Sigma_{c2}({5\over 2}^-)$ or a mixture of the two. In the quark-diquark model \cite{Ebert:2011}, both of them have very close masses compatible with experiment.  Given that for light strange baryons, the first orbital excitation of the $\Sigma$ also has the quantum numbers $J^P=3/2^-$, we will advocate a $\Sigma_{c2}(3/2^-)$ state for $\Sigma_c(2800)$.

Just like the $\Omega_c^{**}$, in principle we can have a $J^P={1\over 2}^-$ $\Sigma_c^{**}$ resonance with the width comparable to $\Sigma_c(2800)$ or even smaller. This is possible if we consider the mixing of $\Sigma_{c0}({1\over 2}^-)$ and $\Sigma_{c1}({1\over 2}^-)$ so that the physical $\Sigma_c({1\over 2}^-)$ state with smaller width is primarily the component of  $\Sigma_{c1}({1\over 2}^-)$. The other $\Sigma_c({1\over 2}^-)$ state governed by $\Sigma_{c0}({1\over 2}^-)$ will be very broad.

According to the equal spacing rule in Eq. (\ref{eq:spacingrule}), $\Omega_c(3050)^0$, $\Xi'_c(2923)^0$ and $\Sigma_c(2800)$ form a $3/2^-$ sextet multiplet. Likewise,  $\Omega_c(3065)^0$, $\Xi'_c(2939)^0$ and $\Sigma_c(2815)$ form another $3/2^-$ sextet multiplet, while $\Omega_c(3090)^0$, $\Xi'_c(2965)^0$ and $\Sigma_c(2840)$ form a $5/2^-$ sextet multiplet provided that the yet-detected states $3/2^-$ $\Sigma_c(2815)$ and $5/2^-$ $\Sigma_c(2840)$ are introduced. Notice that BaBar has observed an excited $\Sigma_c^0$ state in the decay $B^-\to\Sigma_c(2846)^0\bar p\to\Lambda_c^+\pi^-\bar p$ with a mass of $2846\pm8\pm10$~MeV and a width of $86^{+33}_{-22}$~MeV~\cite{BaBar:Sigmac2846}. It remains to check if the state $\Sigma_c(2846)$ has $J^P=5/2^-$.

\vskip 0.4 cm
\subsubsection{Regge trajectories}

Various Regge trajectories in the $(J^P,M^2)$ plane for $\Omega_c,\Lambda_c,\Xi_c,\Xi'_c$ and $\Sigma_c$ states are depicted in Figs.~\ref{fig:Lambdac} to \ref{fig:Sigmac}. In the phenomenology of Regge trajectories, the Regge slopes are usually assumed to be the same for all the baryon multiplets. This ansatz leads to the parallelism among trajectories with natural or unnatural parities, and the parallelism
between natural and unnatural parities.
Empirically, this is nicely supported by the Regge trajectories of
the antitriplet $\Lambda_c$ and $\Xi_c$ states. We see that their Regge trajectories for the orbital excitations of $\frac12^-$ and $\frac32^-$ are parallel to each other, as shown in Figs.~\ref{fig:Lambdac} and \ref{fig:Xic}. Based on this nice property of parallelism, we have argued that the quantum numbers of $\Lambda_c(2940)^+$ are most likely $\frac12^-(2P)$ rather than $\frac32^-$ favored by LHCb \cite{LHCb:Lambdac2880}.

Previously it was found in \cite{Cheng:Omegac} that the Regge trajectories for the sextet $\Omega_c$, $\Xi'_c$ and $\Sigma_c$ states did not respect the parallelism satisfactorily: the slope of the Regge trajectory for the orbital excitation of $\frac12^+$ is slightly larger than that of the $\frac32^+$ one (see Figs. 1, 5 and 6 in \cite{Cheng:Omegac}). This is now overcome in Figs. \ref{fig:Omegac,Xicp} and \ref{fig:Sigmac}. The key point is that the $5/2^-$ states which were previously identified with $\Omega_c(3066)$, $\Xi'_c(2921)$ and $\Sigma_c(2790)$ are now designated to $\Omega_c(3090)$, $\Xi'_c(2965)$ and $\Sigma_c(2840)$, respectively.
As a result,  the Regge trajectories for the orbital excitations of $\frac12^+$ and $\frac32^+$ will be parallel to each other. Therefore, the parallelism of the Regge trajectories is very useful for identifying the $5/2^-$ states.

\vskip 0.2 cm
\subsubsection{Antitriplet and sextet states}
The antitriplet and sextet states of charmed baryons are listed in Table \ref{tab:3and6}. To date, the following states are established: the $J^P={1\over 2}^+$, ${1\over
2}^-$, and ${3\over 2}^-$ ${\bf \bar 3}_F$ states, ($\Lambda_c^+$, $\Xi_c^+,\Xi_c^0)$,
($\Lambda_c(2595)^+$, $\Xi_c(2790)^+,\Xi_c(2790)^0)$, and ($\Lambda_c(2625)^+$, $\Xi_c(2815)^+,\Xi_c(2815)^0)$, respectively, and the
$J^P={1\over 2}^+$ and ${3\over 2}^+$ ${\bf 6}_F$ states,
($\Omega_c,\Xi'_c,\Sigma_c$) and ($\Omega_c^*,\Xi'^*_c,\Sigma_c^*$), respectively. The mass difference $m_{\Xi_c}-m_{\Lambda_c}$ in the antitriplet states clearly lies between 180 and 200 MeV. From Table \ref{tab:3and6}, we see that $\Xi_c(3080)$ and $\Lambda_c(2880)$ form a nice $J^P=5/2^+$ antitriplet and theor mass difference  is consistent with that observed in other antitriplets. Likewise, $\Xi_c(2970)$ and $\Lambda_c(2765)$ form an $1/2^+(2S)$ antitriplet.

\begin{table}[t]
\caption{Antitriplet and sextet states of charmed baryons.
Mass differences $\Delta m_{\Xi_c\Lambda_c}\equiv m_{\Xi_c}-m_{\Lambda_c}$, $\Delta m_{\Xi'_c\Sigma_c}\equiv m_{\Xi'_c}-m_{\Sigma_c}$, $\Delta m_{\Omega_c\Xi'_c}\equiv m_{\Omega_c}-m_{\Xi'_c}$ are all in units of MeV. } \label{tab:3and6}
\begin{center}
\begin{tabular}{c ccc } \hline\hline
  & $J^P(nL)$ & States & Mass difference  \\
 \hline
 ~~${\bf \bar 3}_F$~~ & ~~${1\over 2}^+(1S)$~~ &  $\Lambda_c(2287)^+$, $\Xi_c(2470)^+,\Xi_c(2470)^0$ & ~~$\Delta m_{\Xi_c\Lambda_c}=183$ ~~  \\
 & ~~${1\over 2}^-(1P)$~~ &  $\Lambda_c(2595)^+$, $\Xi_c(2790)^+,\Xi_c(2790)^0$ & $\Delta m_{\Xi_c\Lambda_c}=200$  \\
 & ~~${3\over 2}^-(1P)$~~ &  $\Lambda_c(2625)^+$, $\Xi_c(2815)^+,\Xi_c(2815)^0$ & $\Delta m_{\Xi_c\Lambda_c}=190$  \\
 & ~~${1\over 2}^+(2S)$~~ &  $\Lambda_c(2765)^+$, $\Xi_c(2970)^+,\Xi_c(2970)^0$ & $\Delta m_{\Xi_c\Lambda_c}=200$  \\
 & ~~${3\over 2}^+(1D)$~~ &  $\Lambda_c(2860)^+$, $\Xi_c(3055)^+,\Xi_c(3055)^0$ & $\Delta m_{\Xi_c\Lambda_c}=200$  \\
 & ~~${5\over 2}^+(1D)$~~ &  $\Lambda_c(2880)^+$, $\Xi_c(3080)^+,\Xi_c(3080)^0$ & $\Delta m_{\Xi_c\Lambda_c}=197$  \\
 \hline
 ~~${\bf 6}_F$~~ & ~~${1\over 2}^+(1S)$~~ &  $\Omega_c(2695)^0$, $\Xi'_c(2575)^{+,0},\Sigma_c(2455)^{++,+,0}$ & ~~~~$\Delta  m_{\Omega_c\Xi'_c}=119$, $\Delta m_{\Xi'_c\Sigma_c}=124$~~  \\
 & ~~~${3\over 2}^+(1S)$~~~ &  $\Omega_c(2770)^0$, $\Xi'_c(2645)^{+,0},\Sigma_c(2520)^{++,+,0}$ & ~~~~ $\Delta m_{\Omega_c\Xi'_c}=120$, $\Delta m_{\Xi'_c\Sigma_c}=128$~~ \\
 \hline\hline
\end{tabular}
\end{center}
\end{table}

\begin{table}[t]
\caption{Proposed assignments of quantum numbers for the sextet states of charmed baryons.
 The yet detected states are labeled in red bold face.} \label{tab:more6}
\begin{center}
\begin{tabular}{c l cc } \hline\hline
  & ~~$J^P(nL)$ & States & Mass differences  \\
 \hline
 ~~${\bf 6}_F$~~
 & ~~${3\over 2}^-(1P)_\ell$~ &  $\Omega_c(3050)^0$, ${\Xi'_c(2923)^{+,0}},\Sigma_c(2800)^{++,+,0}$ & ~~~~$\Delta m_{\Omega_c\Xi'_c}=127$, $\Delta m_{\Xi'_c\Sigma_c}=123$~~  \\
 & ~~${3\over 2}^-(1P)_h$~~ &  ~$\Omega_c(3065)^0$, ${\Xi'_c(2939)}^{+,0}$, ${\bf \color{red} \Sigma_c(2815)^{++,+,0}}$ & ~~~~$\Delta m_{\Omega_c\Xi'_c}=127$, $\Delta m_{\Xi'_c\Sigma_c}=125$~~  \\
 & ~~${5\over 2}^-(1P)$~~ &  ~$\Omega_c(3090)^0$, ${\Xi'_c(2965)^{+,0}}, {\bf\color{red} \Sigma_c(2840)^{++,+,0} }$ & ~~~~$\Delta m_{\Omega_c\Xi'_c}=125$, $\Delta m_{\Xi'_c\Sigma_c}=125$~~  \\
 \hline\hline
\end{tabular}
\end{center}
\end{table}

In Table \ref{tab:more6} we show the proposed sextet states classified according to $J^P={3/ 2}^-(1P)_\ell$, ${3/2}^-(1P)_h$ and ${5/ 2}^-(1P)$. Note that the two yet-detected $\Sigma_c(2815)$ and $\Sigma_c(2840)$ states are inferred from the equal spacing rule  discussed in passing. They need experimental confirmation.

\subsection{Doubly charmed baryons}
Evidence of doubly charmed baryon states had been reported by SELEX
in $\Xi_{cc}(3519)^+\to\Lambda_c^+K^-\pi^+$ \cite{Selex02}.
Further observation of $\Xi_{cc}^+\to pD^+K^-$ was also announced by SELEX
\cite{Selex04}. However, none of the doubly charmed states
discovered by SELEX has been confirmed by FOCUS \cite{FOCUS:dc}, BaBar
\cite{BaBar:dc}, Belle \cite{Belle:dc} and LHCb \cite{LHCb:dc}, although $10^6$
$\Lambda_c^+$ events were produced in $B$ factories, for example, versus 1630
$\Lambda_c^+$ events observed at SELEX.

The doubly charmed baryons $\Xi_{cc}^{(*)++},\Xi_{cc}^{(*)+},\Omega_{cc}^{(*)+}$ with the quark contents $ccu,ccd,ccs$ form an SU(3) triplet. They
have been studied extensively using many different approaches: the quark model, light quark--heavy diquark model, QCD sum rules, and lattice simulation. The tabulated doubly charmed baryon masses calculated in various models can be found in \cite{Guo,Karliner:2014,Karliner:2018,Li:2019,Soto:2020pfa}.

\begin{figure}[t]
\begin{center}
\includegraphics[height=52mm]{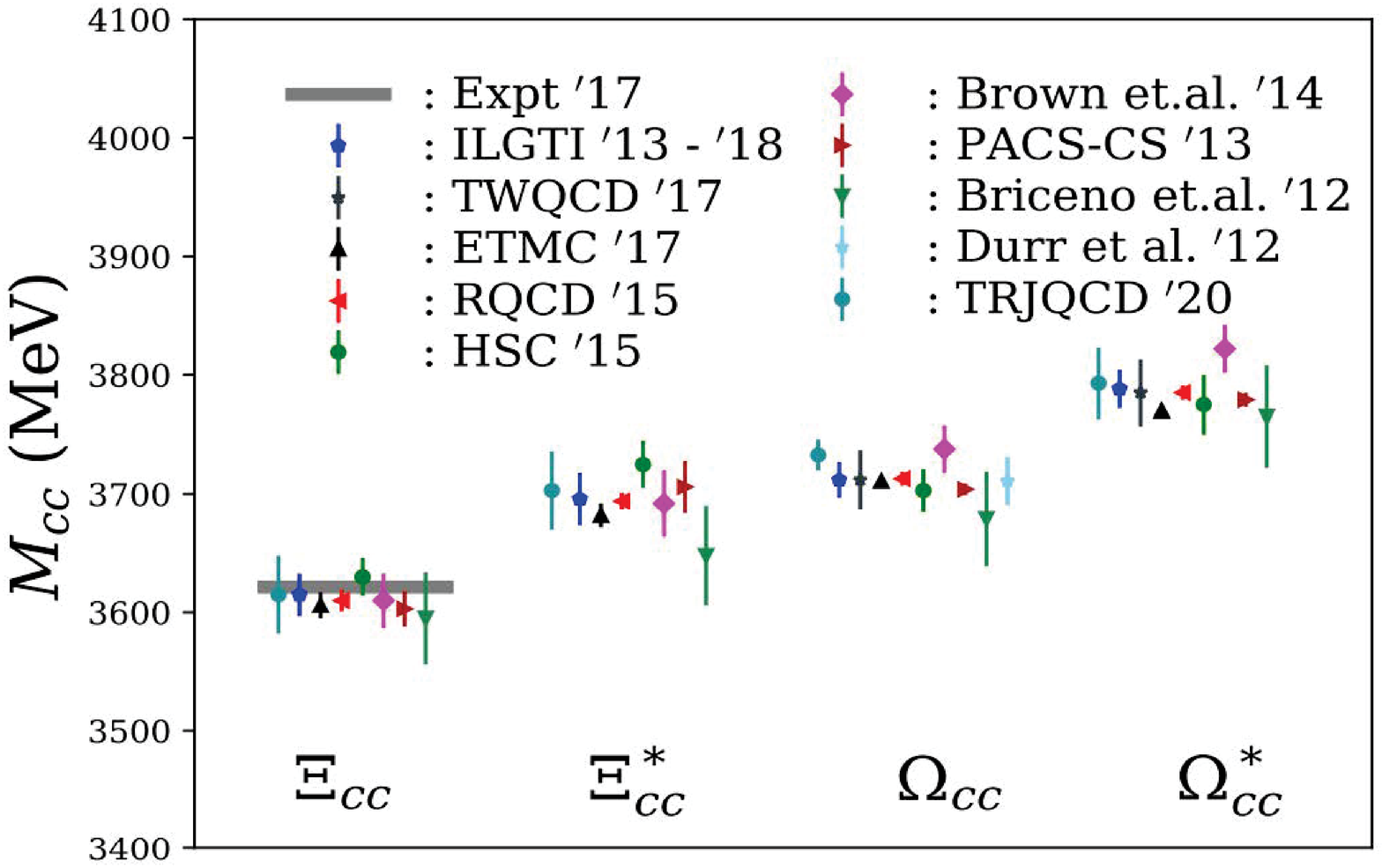}
\caption{Doubly charmed low-lying baryon spectra taken from \cite{Padmanath:2021}. } \label{fig:doubly}
\end{center}
\end{figure}

The results of recent lattice studies of doubly charmed baryon spectra by several different lattice groups are summarized  in Fig. \ref{fig:doubly} (see also \cite{Bahtiyar:2020uuj} and references therein).
The various lattice results are consistent with each other and they fall into the ranges
\begin{eqnarray} \label{eq:dcbmass}
M(\Xi_{cc})=3.54\sim 3.68\, {\rm GeV}, &\qquad&  M(\Xi^*_{cc})=3.62\sim 3.75\,{\rm GeV}, \nonumber \\
M(\Omega_{cc})=3.62\sim 3.76\, {\rm GeV}, &\qquad&  M(\Omega^*_{cc})=3.69\sim 3.88\,{\rm GeV}.
\end{eqnarray}
For comparison, the updated mass of $\Xi_{cc}^{++}$ measured by the LHCb is given by \cite{LHCb:Xiccmass}
\be
M(\Xi_{cc}^{++})=3621.55\pm0.23\pm0.30\,{\rm MeV}.
\en

The hyperfine mass splittings of doubly charmed baryons are related to that of charmed mesons. They satisfy the well-known mass relations
\begin{eqnarray} \label{eq:massrel}
m_{\Xi^*_{cc}}-m_{\Xi_{cc}}={3\over 4}(m_{D^*}-m_D), \qquad
m_{\Omega^*_{cc}}-m_{\Omega_{cc}}={3\over 4}(m_{D^*_s}-m_{D_s}),
\end{eqnarray}
which have been derived in various contents, such as HQET \cite{Savage:baryon},
pNRQCD (potential NRQCD) \cite{Brambilla,Fleming} and the quark model \cite{Lewis,Ebert:2005}.

In 2017 it was first pointed out in \cite{Yu:Xiccpp} that $\Xi_{cc}^{++}\to\Lambda_c^+ K^-\pi^+\pi^+$ and $\Xi_c^+\pi^+$ were the most promising modes to search for doubly heavy charmed baryons with branching fractions of order $10\%$. Soon afterwards,
the LHCb collaboration observed a resonance in the $\Lambda_c^+K^-\pi^+\pi^+$ mass spectrum at a mass of $3621.40\pm0.78$ MeV \cite{LHCb:Xiccpp}, which is consistent with expectations for the doubly charmed baryon $\Xi_{cc}^{++}$ baryon and is significantly larger than the mass of 3519 MeV measured by SELEX for $\Xi_{cc}^+$ \cite{Selex02}.

\section{Lifetimes}

\subsection{Singly charmed baryons}

\subsubsection{Lifetime measurements}

Among singly charmed baryons, the antitriplet states $\Lambda_c^+,~\Xi_c^+,~\Xi_c^0$, and the $\Omega_c^0$ baryon in the sextet decay weakly. According to the 2004 version of PDG \cite{PDG2004}, the world averages of their lifetimes in 2004 were given by
  \begin{eqnarray} \label{eq:exptlifetime}
&& \tau(\Lambda^+_c)= (2.00\pm0.06)\times 10^{-13}s, \qquad
\tau(\Xi^+_c)= (4.42\pm0.26)\times 10^{-13}s,
  \nonumber \\
&& \tau(\Xi^0_c)= (1.12^{+0.13}_{-0.10})\times 10^{-13}s, \qquad\quad~
\tau(\Omega^0_c)= (0.69\pm0.12)\times 10^{-13}s.
  \end{eqnarray}
The charmed baryon lifetime measurements were dominated by the FOCUS experiments performed during the period of 2001-2003 \cite{FOCUS:Lambdac,FOCUS:Xicp,FOCUS:Xic0,FOCUS:Omegac}.
The above-mentioned world averages of charmed baryon lifetimes remained stable from 2004 to 2018. They led to the lifetime pattern
\be
\tau(\Xi_c^+)>\tau(\Lambda_c^+)>\tau(\Xi_c^0)>\tau(\Omega_c^0).
\en
It is quite different from the bottom baryon case where the lifetime hierarchy reads \cite{PDG}
\be
\tau(\Omega_b^-)>\tau(\Xi_b^-)>\tau(\Xi_b^0)\simeq\tau(\Lambda_b^0).
\en

\begin{table}[t]
\caption{Evolution of the charmed baryon lifetimes measured in units of $10^{-13}s$.}
\label{tab:expt_lifetimes}
\begin{center}
\begin{tabular}{l c c c c} \hline \hline
 & $\tau(\Xi_c^+)$ & $\tau(\Lambda_c^+)$ & $\tau(\Xi_c^0)$ & $\tau(\Omega_c^0)$ \\
\hline
PDG (2004-2018) \cite{PDG2018} & $4.42\pm0.26$ & $2.00\pm0.06$ & $1.12^{+0.13}_{-0.10}$ & $0.69\pm0.12$\\
LHCb (2018) \cite{LHCb:tauOmegac} & & & & ~~$2.68\pm0.26$~~ \\
LHCb (2019) \cite{LHCb:2019ldj} & $4.57\pm0.06$ & $2.035\pm0.022$ & $1.545\pm0.026$ & \\
PDG (2020) \cite{PDG} & ~~$4.56\pm0.05$~~ & ~~$2.024\pm0.031$~~ & ~~$1.53\pm 0.06$~~ & ~~$2.68\pm0.26$~~ \\
LHCb (2021) \cite{LHCb:tauOmegac_2} & & & $1.480\pm0.032$ & $2.765\pm0.141$ \\
WA (2021) & ~~$4.56\pm0.05$~~ & ~~$2.024\pm0.031$~~ & ~~$1.520\pm 0.020$~~ & ~~$2.745\pm0.124$~~ \\
\hline \hline
\end{tabular}
\end{center}
\end{table}

\begin{figure}[t]
\begin{center}
\includegraphics[height=45mm]{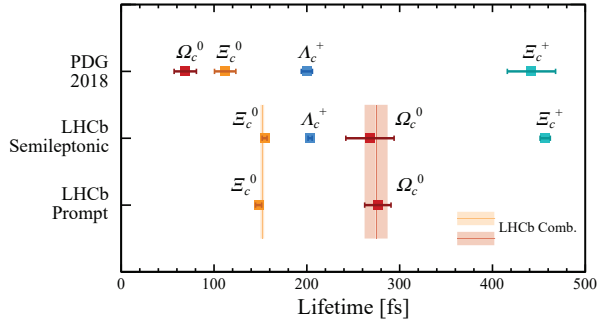}
\caption{Graph \cite{LHCb:tauOmegac_2} showing previous world-average values of the charmed baryon baryon lifetimes from the PDG \cite{PDG2018} and the LHCb measurements of the $\Omega_c^0$ and $\Xi_c^0$ lifetimes obtained from semileptonic beauty-hadron decays and prompt signals. The combined LHCb results are shown in coloured bands.}
\label{fig:charmbarylife}
\end{center}
\end{figure}

However, the situation was dramatically changed in 2018 when LHCb reported a new measurement of the charmed baryon $\Omega_c^0$ lifetime using semileptonic $b$-hadron decays \cite{LHCb:tauOmegac}.  More precisely, LHCb found $\tau(\Omega_c^0)=(2.68\pm0.24\pm0.10\pm0.02)\times 10^{-13}s$ , using the semileptonic decay $\Omega_b^-\to\Omega_c^0\mu^-\bar \nu_\mu X$ followed by $\Omega_c^0\to pK^- K^-\pi^+$. This value is nearly four times larger than the 2018 world-average value of $\tau(\Omega_c^0)=(0.69\pm0.12)\times 10^{-13}s$ \cite{PDG2018} extracted from fixed target experiments. As a result, a new lifetime pattern was emerged
\be
\tau(\Xi_c^+)>{\tau(\Omega_c^0)}>\tau(\Lambda_c^+)>\tau(\Xi_c^0).
\en
In 2019, LHCb reported precision measurements of the $\Lambda_c^+$, $\Xi_c^+$ and $\Xi_c^0$ lifetimes \cite{LHCb:2019ldj} as displayed in Table \ref{tab:expt_lifetimes} and Fig. \ref{fig:charmbarylife}. The $\Xi_c^0$ baryon lifetime is approximately 3.3 standard deviations larger than the world average value.

Finally, LHCb \cite{LHCb:tauOmegac_2} has reported in 2021 a new measurement using promptly produced $\Omega_c^0$ and $\Xi_c^0$ baryons with 5.4 fb$^{-1}$ of the LHCb data in which $\Omega_c^0$ and $\Xi_c^0$ were reconstructed through their decays to $pK^-K^+\pi^+$. The results are
\be
\tau(\Omega_c^0) &=& (2.765\pm0.134\pm0.044\pm0.007)\times 10^{-13}s, \non \\
\tau(\Xi_c^0) &=& (1.480\pm0.023\pm0.022\pm0.002)\times 10^{-13}s.
\en
Hence, the previous LHCb measurement of $\tau(\Omega_c^0)$ is confirmed and its precision is improved by a factor of 2.
The LHCb observation of a huge jump of the $\Omega_c^0$ baryon lifetime in 2018 is very shocking and striking from both experimental and theoretical points of view.  This is the first time in the history of particle physics that the lifetime of a hadron measured in a new experiment was so drastically different from the old one. The LHCb  has collected 978 events of the $b$-tagged $\Omega_c^0$ decays which is about five times larger than those accumulated by all predecessors FOCUS, WA89 and E687 in fixed target experiments \cite{LHCb:tauOmegac}. As stressed in \cite{Bigi}, the lifetime value measured is so large that could have been easily measured much earlier than 2018 by experiments at $e^+e^-$ whose resolution is about $150\times 10^{-15}s$ typically. Since CLEO-c and Belle have both observed $\Omega_c^0$ and measured its mass, they should/could have measured quite easily the lifetime value measured by the LHCb.

On the theoretical side, heavy quark expansion (HQE) in powers of $1/m_Q$ is the standard theoretical framework for analyzing the lifetimes of bottom and charmed hadrons. Lifetime differences arise from the spectator effects such as weak $W$-exchange and Pauli interference. Heavy quark expansion to the order of $1/m_b^3$ works very well for $B$ mesons and bottom baryons (see e.g. \cite{Cheng:2018}).
It turns out that the observed lifetime hierarchy $\tau(\Xi_c^+)>\tau(\Lambda_c^+)>\tau(\Xi_c^0)>\tau(\Omega_c^0)$ settled for a long time until 2018 can be understood at least qualitatively if not quantitatively in the OPE approach up to $1/m_c^3$ expansion.  Among the four charmed baryons, $\Omega_c^0$ is expected to be shortest-lived as it receives largest constructive Pauli interference in both nonleptonic and semileptonic decays. Presumably $\Lambda_c^+$ should live longer than $\Omega_c^0$.
Therefore, the LHCb observation of a much longer $\tau(\Omega_c^0)$ is beyond imagination. Does it mean that HQE fails to apply to the charmed baryon system? What is wrong with HQE for the charm system? We will come to address this issue later.

\subsubsection{Heavy quark expansion}
It was realized since the late 1970s and 1980s that  the lifetime differences of singly heavy hadrons containing a heavy quark $Q$ arise mainly from
the spectator effects like $W$-exchange and Pauli interference
due to the identical quarks produced in heavy quark decay and in the wave function of the
heavy hadron \cite{Bigi92,BS93}. The spectator
effects were expressed in 1980s
in terms of local four-quark operators by relating the total widths to
the imaginary part of certain forward scattering amplitudes
\cite{Bilic,Guberina:1986,SV}.
With the advent of heavy quark effective theory (HQET), it was recognized
in early 1990s that nonperturbative corrections to the parton picture
can be systematically expanded in powers of $1/m_Q$ \cite{Bigi92,BS93}.
Within the QCD-based framework, namely the heavy
quark expansion (HQE), which is a generalization of the operator product
expansion (OPE) in $1/m_Q$ in the Minkowski space,  some
phenomenological assumptions in 1980s acquired a firm theoretical footing in 1990s and nonperturbative effects can be systematically studied (for a review, see \cite{Lenz:2014}).
The observed lifetime hierarchy
$\tau(\Xi_c^+)>\tau(\Lambda_c^+)>\tau(\Xi_c^0)
>\tau(\Omega_c^0)$ can be understood at least qualitatively if not quantitatively in the OPE approach up to $1/m_c^3$ expansion. However, as we shall see below, this is not the whole story and a big surprise will come out by the end of the story.

On the basis of the OPE approach, the inclusive rate of the charmed baryon $\B_c$ is schematically
represented by
 \begin{eqnarray}
 \Gamma({\cal B}_c\to f) = {G_F^2m_c^5\over
192\pi^3}V_{\rm CKM}\left(A_0+{A_2\over m_c^2}+{A_3\over
m_c^3}+{\cal O}\left({1\over m_c^4}\right)\right),
 \end{eqnarray}
where $V_{\rm CKM}$ is the relevant Cabibbo-Kobayashi-Maskawa matrix element.
The $A_0$ term comes from the $c$ quark decay and is common to all
charmed hadrons. There are no linear $1/m_Q$ corrections to the
inclusive decay rate owing to the lack of gauge-invariant
dimension-four operators \cite{Chay,Bigi92}, a consequence known
as Luke's theorem \cite{Luke}. Nonperturbative corrections start
at order $1/m_Q^2$ and are model-independent. Spectator
effects in inclusive decays due to the Pauli interference and
$W$-exchange contributions account for the $1/m_c^3$ corrections and
they have two noteworthy features: First, the estimate of the spectator
effects is model dependent; the hadronic four-quark matrix
elements are usually evaluated by assuming the factorization
approximation for mesons and the quark model for baryons. Second,
there is a two-body phase-space enhancement factor of $16\pi^2$
for spectator effects relative to the three-body phase space for
heavy quark decay. This implies that spectator effects, which are of the
order of $1/m_c^3$, are comparable to and even exceed the $1/m_c^2$
terms.

\begin{figure}[t]
\begin{center}
\includegraphics[height=25mm]{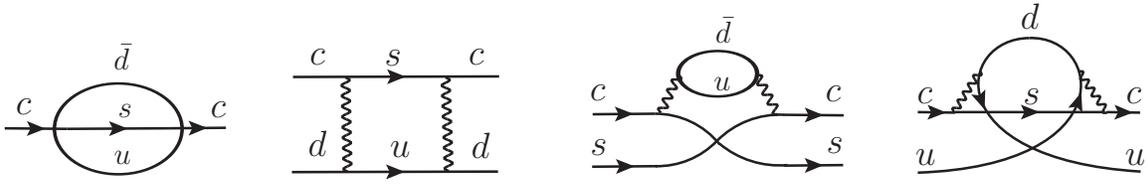}
\caption{Contributions to nonleptonic decay rates of charmed
baryons from four-quark operators: (a) $c$-quark decay, (b)
$W$-exchange, (c) constructive Pauli interference and (d)
destructive Pauli interference.} \label{fig:fourquarkNL}
\end{center}
\end{figure}

The total width of the charmed baryon ${\cal B}_c$
generally receives contributions from inclusive nonleptonic and semileptonic
decays: $\Gamma({\cal B}_c)=\Gamma_{\rm NL}({\cal
B}_c)+\Gamma_{\rm SL}({\cal B}_c)$. The nonleptonic contribution
can be further decomposed into
 \begin{eqnarray}
 \Gamma_{\rm NL}({\cal B}_c)=\Gamma^{\rm dec}({\cal B}_c)+\Gamma^{\rm ann}
 ({\cal B}_c)+\Gamma^{\rm
 int}_+({\cal B}_c)+\Gamma^{\rm int}_-({\cal B}_c),
 \end{eqnarray}
corresponding to the $c$-quark decay, $W$-exchange
contribution, constructive and destructive Pauli interference, respectively, as depicted in Fig. \ref{fig:fourquarkNL}. The
inclusive decay rate is known to be governed by the
imaginary part of an effective nonlocal forward transition
operator $T$
\begin{eqnarray}  \label{eq:NLrate}
\Gamma(\B_{Q}) &=& {G_F^2m_Q^5\over 192\pi^3}\,V_{\rm CKM}\,{1\over 2m_{\B_{Q}}}
\Bigg\{ c_{3,Q}\langle \B_{Q}|\bar QQ|\B_{Q}\rangle+ {c_{5,Q}\over m_Q^2} \langle\B_{Q}|\bar Q \sigma\cdot GQ|\B_{Q}\rangle \nonumber \\ &+& {c_{6,Q} \over m_Q^3} \langle \B_{Q}|T_6|\B_{Q}\rangle+{c_{7,Q} \over m_Q^4} \langle \B_{Q}|T_7|\B_{Q}\rangle+\cdots\Bigg\},
\end{eqnarray}
where the dimension-6 operator $T_6$ consists of the four-quark operators $(\bar Q\Gamma q)(\bar q\Gamma Q)$ with $\Gamma$ representing a combination of the Lorentz and color matrices, while a subset of dimension-7 $T_7$ is governed by the four-quark operators containing derivative insertions \cite{Gabbiani:2003pq,Gabbiani:2004tp}.
Therefore, $\Gamma^{\rm dec}$ corresponds to the
imaginary part of Fig. \ref{fig:fourquarkNL}(a) sandwiched between
the same ${\cal B}_c$ states. At the Cabibbo-allowed level,
$\Gamma^{\rm dec}$ represents the decay rate of $c\to su\bar d$,
and $\Gamma^{\rm ann}$ denotes the contribution from the
$W$-exchange diagram $cd\to us\to cd$ (Fig. \ref{fig:fourquarkNL}(b)). The interference $\Gamma^{\rm
int}_+$ ($\Gamma^{\rm int}_-$) in Fig. \ref{fig:fourquarkNL}(c) (Fig. \ref{fig:fourquarkNL}(d)) arises from
constructive (destructive) interference between the $s$ ($u$) quark produced
in $c$-quark decay and the spectator $s$ ($u$) quark in the
charmed baryon ${\cal B}_c$. Note that the constructive Pauli
interference is unique to the charmed baryon sector, as it does not
occur in the bottom sector. From the quark content of the
charmed baryons, it is clear that
at the Cabibbo-allowed level, destructive interference occurs
in $\Lambda_c^+$ and $\Xi_c^+$ decays, whereas $\Xi_c^+,\Xi_c^0$ and
$\Omega_c^0$ can have constructive interference $\Gamma^{\rm int}_+$.
Because $\Omega_c^0$
contains two $s$ quarks, it is natural to expect that $\Gamma^{\rm
int}_+(\Omega_c^0)\gg \Gamma^{\rm int}_+(\Xi_c)$. The $W$-exchange contribution
occurs  only for $\Xi_c^0$ and $\Lambda_c^+$ at the same
Cabibbo-allowed level. In the HQE approach, the
above-mentioned spectator effects can be described in terms of the
matrix elements of local four-quark operators.

\begin{table}[t]
\caption{Various contributions to the decay rates (in units of
$10^{-12}$ GeV) of singly charmed baryons in HQE to order $1/m_c^3$ \cite{Cheng:2018}.   Experimental values of charmed baryon lifetimes are taken from 2018 PDG \cite{PDG2018}.}
\label{tab:lifetime3}
\begin{center}
\begin{tabular}{c c c c c l l  l  l} \hline \hline
 & $\Gamma^{\rm dec}$ & $\Gamma^{\rm ann}$ & $\Gamma^{\rm int}_-$ &
$\Gamma^{\rm int}_+$ & ~ $\Gamma^{\rm semi}$ & ~$\Gamma^{\rm tot}$ &
~$\tau(10^{-13}s)$~ & ~ $\tau_{\rm expt}(10^{-13}s)$ \\
\hline
 ~$\Lambda_c^+$ & ~0.886~ & ~1.479~ & ~$-0.400$~ & 0.042 & ~0.215~ &
~2.221~ & ~  2.96~ &  ~$2.00\pm 0.06$~   \\
 ~$\Xi_c^+$ & 0.886 & 0.085 & $-0.431$ & 0.882 & ~0.726 &
~2.148~ & ~ 3.06~ & ~$4.42\pm0.26$  \\
 ~$\Xi_c^0$ & 0.886 & 1.591 & & 0.882 & ~0.726 &
~4.084 & ~ 1.61 & ~$1.12^{+0.13}_{-0.10}$ \\
 ~$\Omega_c^0$ & 1.019 & 0.515 & & 2.974 & ~1.901 &
~6.409 & ~ 1.03  & ~$0.69\pm 0.12$  \\
\hline \hline
\end{tabular}
\end{center}
\end{table}

\vskip 0.4cm

The inclusive
nonleptonic rates of charmed baryons in the valence quark
approximation and in the limit $m_s/m_c=0$ are expressed as \cite{Cheng:1997}:
 \begin{eqnarray} \label{eq:lifetimes}
 \Gamma_{\rm NL}(\Lambda_c^+) &=& \Gamma^{\rm
 dec}(\Lambda_c^+)+\cos\theta_C^2\Gamma^{\rm ann}+\Gamma^{\rm
 int}_-+\sin\theta_C^2\Gamma^{\rm int}_+,  \nonumber \\
 \Gamma_{\rm NL}(\Xi_c^+) &=& \Gamma^{\rm
 dec}(\Xi_c^+)+\sin\theta_C^2\Gamma^{\rm ann}+\Gamma^{\rm
 int}_-+\cos\theta_C^2\Gamma^{\rm int}_+,  \nonumber \\
 \Gamma_{\rm NL}(\Xi_c^0) &=& \Gamma^{\rm
 dec}(\Xi_c^0)+\Gamma^{\rm ann}+\Gamma^{\rm int}_+,  \nonumber \\
 \Gamma_{\rm NL}(\Omega_c^0) &=& \Gamma^{\rm
 dec}(\Omega_c^0)+6\sin\theta_C^2\Gamma^{\rm ann}+{10\over 3}\cos\theta_C^2\Gamma^{\rm int}_+,
 \end{eqnarray}
where $\theta_C$ is the Cabibbo angle.
The results of a model calculation in \cite{Cheng:2018} are shown in Table
\ref{tab:lifetime3}. The obtained lifetime pattern
 \begin{eqnarray} \label{eq:lifepattern}
\tau(\Xi_c^+)>\tau(\Lambda_c^+)>\tau(\Xi_c^0)>\tau(\Omega_c^0)
 \end{eqnarray}
seems to be in agreement with 2018 PDG  \cite{PDG2018}.
However, it is also clear from Table
\ref{tab:lifetime3} that, although the qualitative feature of the
lifetime pattern is comprehensive, the quantitative estimates of
charmed baryon lifetimes and their ratios are still rather poor.  For example, $R_1=\tau(\Xi_c^+)/\tau(\Lambda_c^+)$ is calculated to be 1.03 \cite{Cheng:2018}, while experimentally it is measured to be $2.21\pm0.15$ \cite{PDG}. Contrary to $B$ meson and bottom baryon cases where HQE in $1/m_b$ leads to the lifetime ratios in excellent agreement with experiment, the heavy quark expansion in $1/m_c$ does not work well for describing the lifetime pattern of charmed baryons.
Since the charm quark is not heavy,
it is thus natural to consider the effects stemming from the next-order $1/m_c$ expansion. This calls for the subleading $1/m_Q$ corrections to spectator effects.

\begin{table}[t]
\caption{Various contributions to the decay rates (in units of
$10^{-12}$ GeV) of singly charmed baryons to order $1/m_c^4$ in HQE to include subleading $1/m_c$ corrections to spectator effects \cite{Cheng:2018}.
}
\label{tab:lifetime4}
\begin{center}
\begin{tabular}{c c c c r r l  l  l} \hline \hline
 & $\Gamma^{\rm dec}$ & $\Gamma^{\rm ann}$ & $\Gamma^{\rm int}_-$ &
$\Gamma^{\rm int}_+$ & ~ $\Gamma^{\rm semi}$ & ~~$\Gamma^{\rm tot}$ &
~$\tau(10^{-13}s)$~ & ~ $\tau_{\rm expt}(10^{-13}s)$~ \\
\hline
 ~$\Lambda_c^+$ & ~0.886~ & ~2.179~ & ~$-0.211$~ & 0.022 & ~0.215 &
~~3.091~ & ~ 2.12~ &  ~$2.00\pm 0.06$~   \\
 ~$\Xi_c^+$ & 0.886 & 0.133 & $-0.186$ & 0.407 & ~0.437 &
~~1.677~ & ~ 3.92~ & ~$4.42\pm0.26$  \\
 ~$\Xi_c^0$ & 0.886 & 2.501 & $$ & 0.405 & ~0.435 &
~~4.228 & ~ 1.56 & ~$1.12^{+0.13}_{-0.10}$ \\
 ~$\Omega_c^0$ & 1.019 & 0.876 & & $-0.559$ & ~$-0.256$ &
~~1.079 & ~ 6.10  & ~$0.69\pm 0.12$  \\
\hline \hline
\end{tabular}
\end{center}
\end{table}

\begin{table}[t]
\caption{Various contributions to the decay rates (in units of
$10^{-12}$ GeV) of $\Omega_c^0$ to order $1/m_c^3$ and $1/m_c^4$, respectively.
}
\label{tab:lifetime:compare}
\begin{center}
\begin{tabular}{c c c c r r l  c  } \hline \hline
 & $\Gamma^{\rm dec}$ & ~~$\Gamma^{\rm ann}$~~ & $\Gamma^{\rm int}_-$ &
$\Gamma^{\rm int}_+$ & ~ $\Gamma^{\rm semi}$ & ~~$\Gamma^{\rm tot}$ &
  ~ $\tau(10^{-13}s)$~ \\
\hline
 $1/m_c^3$~~~ & 1.019 & 0.515 & $$ & 2.974 & ~1.901 &
~~6.409 & ~ 1.03  \\
 $1/m_c^4$~~~ & 1.019 & 0.876 & & $-0.559$ & ~$-0.256$ &
~~1.079 & ~ 6.10    \\
\hline \hline
\end{tabular}
\end{center}
\end{table}

In 2017 I started the task of calculating the subleading $1/m_c$ corrections to spectator effects described by dimension-7 operators \cite{Cheng:2018}. The numerical results are shown in Table \ref{tab:lifetime4}. Comparing Tables \ref{tab:lifetime4} and \ref{tab:lifetime3}, I found that while $\Gamma(\Lambda_c^+)$ is enhanced, $\Gamma(\Xi_c^+)$ is suppressed in such a way that the resulting lifetime ratio $R_1$ is enhanced from 1.03 to 1.84. This means that $1/m_c$ corrections to spectator effects described by dimension-7 operators were heading to the right direction. However, the calculated $\Omega_c$ lifetime became entirely unexpected: the shortest-lived $\Omega_c^0$ turned out out to be the longest-lived one to ${\cal O}(1/m_c^4)$.
Because this sounds very ridiculous, I was very disappointed with the calculations on $\tau(\Omega_c^0)$, although the results on $\Lambda_c^+$ and $\Xi_c^+$ were improved.

To see the issue of HQE with the $\Omega_c^0$, we write
$\Gamma_+^{\rm int}(\Omega_c)=\Gamma^{\rm int}_{+,6}(\Omega_c)+\Gamma^{\rm int}_{+,7}(\Omega_c)$ and  $\Gamma^{\rm semi}(\Omega_c)=\Gamma^{\rm semi}_{6}(\Omega_c)+\Gamma^{\rm semi}_{7}(\Omega_c)$ and display the results of $1/m_c^3$ and $1/m_c^4$ corrections separately in Table \ref{tab:lifetime:compare}. It is obvious that the dimension-7 contributions $\Gamma^{\rm int}_{+,7}(\Omega_c)$ and $\Gamma^{\rm semi}_7(\Omega_c)$ are destructive and their size are so large that they overcome the dimension-6 ones and flip the sign.
Of course, a negative $\Gamma^{\rm semi}(\Omega_c)$ does not make sense as the subleading corrections are too large to justify the validity of HQE. Hence, HQE fails to apply to $\Omega_c^0$ to order $1/m_c^4$! I was wondering why the $\Omega_c^0$ is so special. Maybe this is ascribed to its sextet nature.

Then in the early 2018, it occurred to me that why not I enforce the use of HQE to $\Omega_c^0$ and see what the prediction of $\tau(\Omega_c^0)$ will come out.
In order to allow a description of the $1/m_c^4$ corrections to $\Gamma(\Omega_c)$ within the realm of perturbation theory, I introdcued a phenomenological parameter $\alpha$ to suppress the contributions from dimension-7 operators. More precisely,
$\Gamma^{\rm int}_{+,7}(\Omega_c)$ and $\Gamma^{\rm semi}_7(\Omega_c)$ are multiplied by a factor of $(1-\alpha)$, so that
\footnote{Note that the above-mentioned suppression effect is not needed for the antitriplet baryons $\Lambda_c^+,\Xi_c^+$ and $\Xi_c^0$.}
\be
\Gamma_+^{\rm int}(\Omega_c) &=& \Gamma^{\rm int}_{+,6}(\Omega_c)+(1-\alpha)\Gamma^{\rm int}_{+,7}(\Omega_c)>0, \non \\
\Gamma^{\rm semi}(\Omega_c) &=& \Gamma^{\rm semi}_{6}(\Omega_c)+(1-\alpha)\Gamma^{\rm semi}_{7}(\Omega_c)>0.
\en
My guidelines for the parameter $\alpha$ are (i) both $\Gamma^{\rm int}_{+}(\Omega_c)$ and $\Gamma^{\rm semi}(\Omega_c)$ should be positive (at least, a negative $\Gamma^{\rm semi}(\Omega_c)$ does not make sense), and (ii)
$\Gamma^{\rm semi}(\Omega_c)$ is comparable to that of $\Lambda_c^+$ or $\Xi_c$.
Demanding a sensible HQE for $\Omega_c^0$, I have conjectured in \cite{Cheng:2018} that $\alpha$ lies in $0.16<\alpha<0.32$ and the $\Omega_c^0$ lifetime lies in the range of $(2.3\sim3.3)\times 10^{-13}s$.
\footnote{My early conjecture of $\tau(\Omega_c^0)$ of order $2.3\times 10^{-13}s$  was first presented  in the HIPEA Workshop held in March of 2018 \cite{Cheng:HIEPA} before the LHCb announcement of the new $\tau(\Omega_c^0)$ measurement.}
This means that $\Omega_c^0$ could live longer than $\Lambda_c^+$!

\begin{table}[t]
\caption{Calculated lifetimes  of charmed baryons (in units of $10^{-13}s$) in the HQE to order $1/m_c^3$ and $1/m_c^4$ including $\alpha$. New world averages are taken from Table \ref{tab:expt_lifetimes}.
}
\label{tab:lifetime:final}
\begin{center}
\begin{tabular}{c c c c } \hline \hline
 & ~~Theory ($1/m_c^3$) & ~~Theory ($1/m_c^4$ with $\alpha$)~~ & 2021 WA \\
\hline
 $\Xi_c^+$ & 3.06 & 3.92 & $4.56\pm0.05$  \\
 $\Omega_c^0$ & 1.03 & 2.3$\sim$3.3  & $2.746\pm0.124$    \\
 $\Lambda_c^+$ & 2.96 & 2.12 & $2.024\pm0.031$ \\
 $\Xi_c^0$ & 1.61 & 1.56 & $1.52\pm0.02$ \\
\hline \hline
\end{tabular}
\end{center}
\end{table}

To summarize, the lifetime hierarchies  in HQE to order $1/m_c^3$ and $1/m_c^4$ are given by
 \begin{eqnarray}
{\cal O}(1/m_c^3) &\Rightarrow & \tau(\Xi_c^+)>\tau(\Lambda_c^+)>\tau(\Xi_c^0)>{\bf \color{red}\tau(\Omega_c^0)}, \nonumber \\
{\cal O}(1/m_c^4) &\Rightarrow & {\bf \color{red}\tau(\Omega_c^0)}>\tau(\Xi_c^+)>\tau(\Lambda_c^+)>\tau(\Xi_c^0),  \\
{\cal O}(1/m_c^4)~{\rm with}~\alpha &\Rightarrow & \tau(\Xi_c^+)>{\bf \color{red}\tau(\Omega_c^0)}>\tau(\Lambda_c^+)>\tau(\Xi_c^0). \nonumber
 \end{eqnarray}
As discussed before, HQE to the order of $1/m_c^4$ is not applicable to the $\Omega_c^0$ because the destructive contributions from dimension-7 operators $\Gamma^{\rm int}_{+,7}(\Omega_c)$ and $\Gamma^{\rm semi}_7(\Omega_c)$ are too lage to justify the validity of HQE. We see from Table \ref{tab:lifetime:final} that the predicted lifetimes of $\Lambda_c^+$ and $\Xi_c^0$ after including $1/m_c$ corrections to spectator effects are in good agreement with the new averages, while the calculated $\tau(\Xi_c^+)$ is still too small by 15\% compared to experiment. Moreover, the $\Omega_c^0$ lifetime cannot be predicted precisely, as it depends on the unknown parameter $\alpha$ which describes the degree of suppression of the contributions from dimension-7 operators.
The origin of this suppression is unknown, but it could be due to the next-order $1/m_c$ correction. At any rate, this amazing story of the $\Omega_c^0$ lifetime should get QCD theorists thinking about how to refine the relevant calculations and work out the higher order terms in HQE.

In conclusion, the $\Omega_c^0$, which is naively expected to be shortest-lived in the charmed baryon system owing to the large constructive Pauli interference, could live longer than the $\Lambda_c^+$  due to the suppression from $1/m_c$ corrections arising from dimension-7 four-quark operators.

\subsection{Doubly charmed baryons}

The inclusive
nonleptonic rates of doubly charmed baryons in the valence quark
approximation and in the limit $m_s/m_c=0$ can be expressed approximately as
 \begin{eqnarray} \label{eq:lifetimes_dc}
 \Gamma_{\rm NL}(\Xi_{cc}^{++}) &=& \Gamma^{\rm
 dec}+\Gamma^{\rm
 int}_-,  \nonumber \\
 \Gamma_{\rm NL}(\Xi_{cc}^{+}) &=& \Gamma^{\rm
 dec}+\cos\theta_C^2\Gamma^{\rm ann}+\sin\theta_C^2\Gamma^{\rm int}_+,  \nonumber \\
 \Gamma_{\rm NL}(\Omega_{cc}^+) &=& \Gamma^{\rm
 dec}+\sin\theta_C^2\Gamma^{\rm ann}+\cos\theta_C^2\Gamma^{\rm int}_+.
 \end{eqnarray}
Because $\Gamma^{\rm int}_+$ is positive and $\Gamma^{\rm int}_-$ is
negative,  it is obvious that $\Xi_{cc}^{++}$ is longest-lived, whereas $\Xi_{cc}^+$ ($\Omega_{cc}^+$) is the shortest-lived if $\Gamma^{\rm int}_+>\Gamma^{\rm ann}$ ($\Gamma^{\rm int}_+<\Gamma^{\rm ann}$). We need to compute the spectator effects to see the relative weight between $\Gamma^{\rm int}_+$ and $\Gamma^{\rm ann}$.
Because the mass splitting between $\Xi_{cc}^*$ and $\Xi_{cc}$ and between $\Omega_{cc}^*$ and $\Omega_{cc}$ is less than 110 MeV (see Eq. ({\ref{eq:massrel})),
it is clear that only electromagnetic decays are allowed for spin-${3\over 2}$ $\Omega_{cc}^*$ and $\Xi_{cc}^*$.

\begin{table}[t]
\caption{Predicted lifetimes of doubly charmed baryons
in units of $10^{-13}s$. The results of \cite{Berezhnoy} are based on the calculation of using $m_c=1.73\pm0.07$ GeV and $m_s=0.35\pm0.20$ GeV from a fit to the LHCb measurement of $\tau(\Xi_{cc}^{++}).$
} \label{tab:lifetimes_dc}
\begin{center}
\begin{tabular}{l c c c } \hline \hline
& ~~$\Xi_{cc}^{++}$~~ & ~~$\Xi_{cc}^{+}$~~ & ~~$\Omega_{cc}^{+}$~~~~~\\
\hline
~~Kiselev et al. ('98) \cite{Kiselev:1999}~~ &  ~~~$4.3\pm1.1$~~~ & ~~~$1.1\pm0.3$~~~ &  \\
~~Kiselev et al. ('99) \cite{Kiselev:2002}~~ & $4.6\pm0.5$ & $1.6\pm0.5$ & $2.7\pm0.6$ \\
~~Guberina et al. ('99) \cite{Guberina}~~ & 15.5 & 2.2 & 2.5 \\
~~Chang et al. ('08) \cite{Chang}~~  & 6.7 & 2.5 & 2.1 \\
~~Karliner, Rosner ('14) \cite{Karliner:2014}~~ & 1.85 & 0.53 & \\
~~Cheng, Shi ('18) \cite{Cheng:doubly} & 2.98 & 0.44 & 0.75$\sim$1.80\\
~~Berezhnoy et al. ('18) \cite{Berezhnoy} & $2.6\pm0.3$ & $1.4\pm0.1$ & $1.8\pm0.2$ \\
\hline
~~Experiment \cite{LHCb:tauXiccpp} & $2.56\pm0.27$ & & \\
\hline
\end{tabular}
\end{center}
\end{table}

\begin{figure}[t]
\begin{center}
\subfigure[]{
\includegraphics[height=27mm]{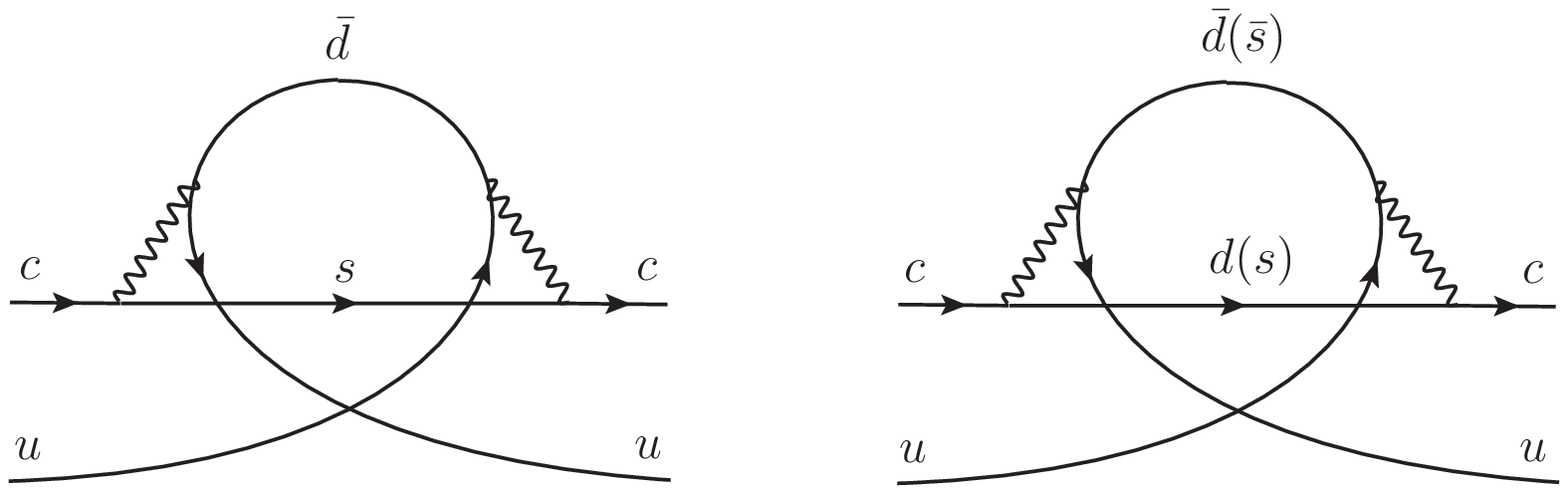}}
\subfigure[]{
\vspace{0.0cm}
\includegraphics[height=17mm]{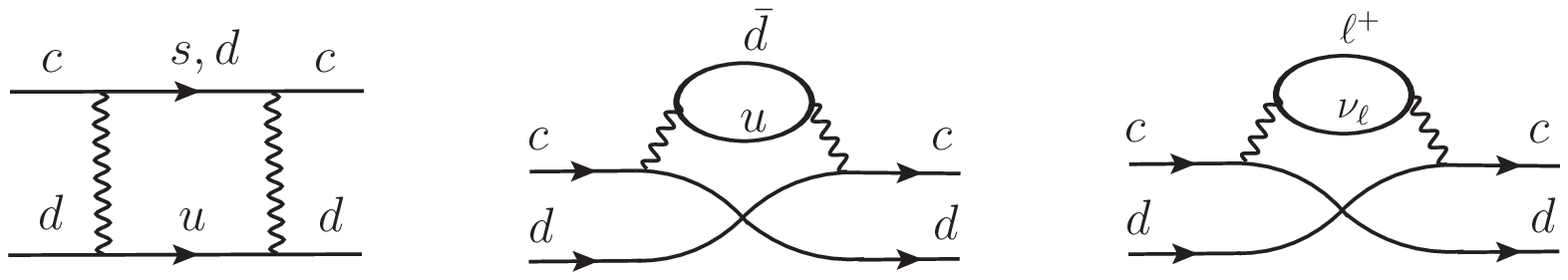}}
\subfigure[]{
\vspace{0.0cm}
\includegraphics[height=17mm]{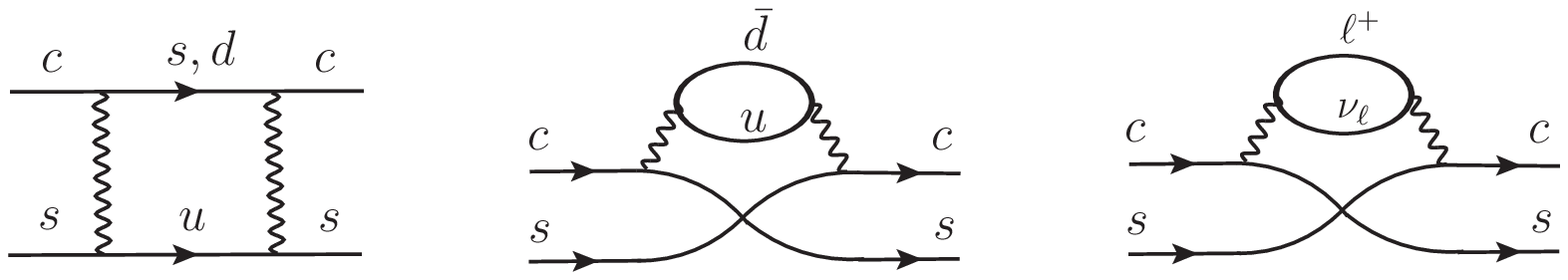}
}
\caption{Spectator effects in doubly charmed baryon decays: (a) destructive Pauli interference in $\Xi_{cc}^{++}$ decay, (b) $W$-exchange and constructive Pauli interference in $\Xi_{cc}^+$ decay, and (c) $W$-exchange and constructive Pauli interference in $\Omega_{cc}^+$ decay.}
\label{fig:spectator}
\end{center}
\end{figure}

LHCb has presented the first lifetime measurement of the charmed doubly baryon $\Xi_{cc}^{++}$ \cite{LHCb:tauXiccpp}
\begin{eqnarray} \label{eq:LHCbtauXiccpp}
\tau(\Xi_{cc}^{++})=
(2.56^{+0.24}_{-0.22}\pm0.14)\times 10^{-13}s.
\end{eqnarray}
Theoretical predictions available in the literature are collected in Table \ref{tab:lifetimes_dc} and they spread a large range, for example, $\tau(\Xi_{cc}^{++})$ ranges from 0.2 to 1.6 ps before the LHCb measurement.  The lifetime hierarchy was predicted to be of the pattern $\tau(\Xi_{cc}^{++})>\tau(\Omega_{cc}^+)>\tau(\Xi_{cc}^+)$ in \cite{Kiselev:2002,Guberina}, but $\tau(\Xi_{cc}^{++})>\tau(\Xi_{cc}^+)>\tau(\Omega_{cc}^+)$ in \cite{Chang}.

Spectator effects in the weak decays of the doubly charmed baryons $\Xi_{cc}^{++}$, $\Xi_{cc}^{+}$ and $\Omega_{cc}^+$ are depicted in Fig. \ref{fig:spectator}.
The $\Xi_{cc}^{++}$ baryon is longest-lived in the doubly charmed baryon system owing to the destructive Pauli interference absent in the $\Xi_{cc}^+$ and $\Omega_{cc}^+$. In the presence of dimension-7 contributions, its lifetime is reduced from $\sim5.2\times 10^{-13}s$ to $\sim3.0\times 10^{-13}s$ \cite{Cheng:doubly}. The $\Xi_{cc}^{+}$ baryon has the shortest lifetime of order $0.45\times 10^{-13}s$ due to a large contribution from the $W$-exchange box diagram.

It is difficult to make a precise statement on the lifetime of $\Omega_{cc}^+$. Contrary to $\Xi_{cc}$ baryons, $\tau(\Omega_{cc}^+)$ becomes longer in the presence of dimension-7 effects so that the Pauli interference $\Gamma^{\rm int}_+$ even becomes negative. This means that the subleading corrections are too large to justify the validity of the HQE. Demanding the rate $\Gamma^{\rm int}_+$ to be positive for a sensible HQE, we conjecture that the $\Omega_{cc}^+$ lifetime lies in the range of $(0.75\sim 1.80)\times 10^{-13}s$ \cite{Cheng:doubly}.
The lifetime hierarchy pattern is  given by $\tau(\Xi_{cc}^{++})>\tau(\Omega_{cc}^+)>\tau(\Xi_{cc}^+)$ and the lifetime ratio $\tau(\Xi_{cc}^{++})/\tau(\Xi_{cc}^+)$ is predicted to be of order 6.7 \cite{Cheng:doubly}.
Due to the short lifetimes of $\Xi_{cc}^+$ and $\Omega_{cc}^+$, it would be difficult to observe them in experiments \cite{LHCb:Xiccp,LHCb:Omegacc}.

\section{Hadronic weak decays}
\subsection{Nonleptonic decays of singly charmed baryons}

For a long time, both experimental and theoretical progresses in the study of hadronic decays of charmed baryons were very slow. Almost all the model calculations of two-body nonleptonic decays of charmed baryons were done before millennium and most of the experimental measurements were older ones (for a review, see \cite{Cheng:2009,Cheng:2015}). Theoretical interest in hadronic weak decays of charmed baryons peaked around the early 1990s and then faded away. Until today we still do not have a good and reliable phenomenological model to describe the complicated physics of charmed baryon decays. The situation was reversed since 2014 as there were several major breakthroughs in recent charmed-baryon experiments in regard to the weak decays of $\Lambda_c^+$ and $\Xi_c^{+,0}$ as we have briefly discussed in the Introduction and will elaborate in more detail below.

More than two decades ago,  a general formulation of the topological-diagram scheme for the nonleptonic weak decays of baryons has been proposed in \cite{Chau:1995gk}, which was then applied to all the decays of the antitriplet and sextet charmed baryons. For the weak decays $\B_{c}\to \B+M$ with $M$ being a pseudoscalar or vector meson,  the relevant topological diagrams are
the external $W$-emission $T$, the internal $W$-emission $C$, the inner $W$-emission $C'$,  and the $W$-exchange diagrams $E_1$, $E_2$ and $E_3$ as shown in Fig. \ref{fig:Bic}. Among them, $T$ and $C$ are factorizable, while $C'$ and $W$-exchange give nonfactorizable contributions.  Taking $M=P$ as an example, the amplitude for the two-body weak decay  $\mathcal{B}_i\to\mathcal{B}_f P$ reads
\begin{equation} \label{eq:generalAmp}
M(\mathcal{B}_i \to \mathcal{B}_f P)= i\bar u_f (A-B\gamma_5) u_i,
\end{equation}
where the amplitudes $A$ and $B$ correspond to the parity-violating $S$-wave and parity-conserving $P$-wave amplitudes, respectively.
In general, they receive both factorizable and nonfactorizable contributions
\be \label{eq:A&B}
& A=A^{\rm{fac}}+A^{\rm{nf}},\qquad
B=B^{\rm{fac}}+B^{\rm{nf}}.
\label{eq:amplitude}
\en
From Fig. \ref{fig:Bic} it is obvious that the topological diagram $E_3$ does not contribute to $A^{\rm{nf}}$ as weak interactions do not participate in the meson emission in this diagram.

\begin{figure}[t]
\begin{center}
\includegraphics[width=0.80\textwidth]{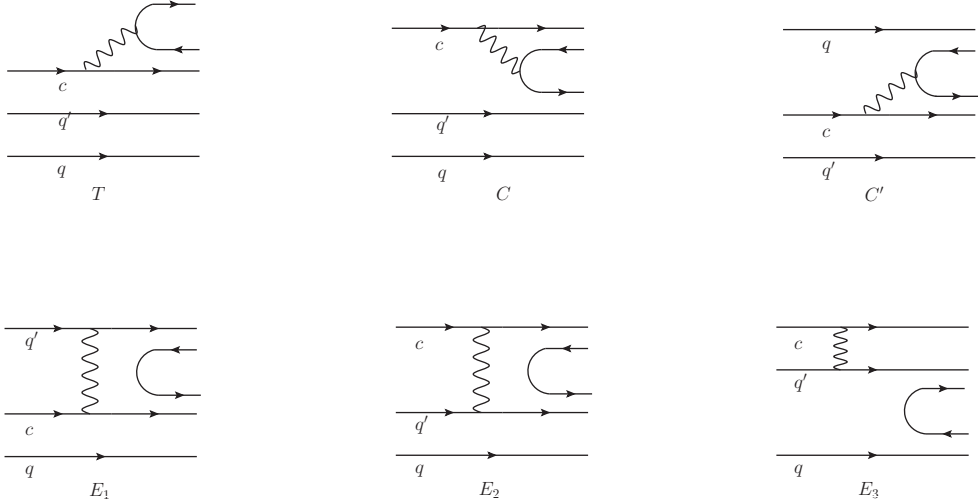}
\vspace{0.1cm}
\caption{Topological diagrams contributing to $\B_{c}\to \B+M$ decays: external $W$-emission $T$, internal $W$-emission $C$, inner $W$-emission $C'$,  $W$-exchange diagrams $E_1$, $E_2$ and $E_3$.} \label{fig:Bic}
\end{center}
\end{figure}

From the theoretical point of view, baryons being made out of three quarks, in contrast to two quarks for mesons, bring along several essential complications. First of all, the factorization approximation that the hadronic matrix element is factorized into the product of two matrix elements of single currents and that the nonfactorizable term such as the $W$-exchange contribution is negligible relative to the factorizable one is known empirically to be working reasonably well for describing the nonleptonic weak decays of heavy mesons. However, this approximation is {\it a priori} not directly applicable to the charmed baryon case as $W$-exchange there, manifested as pole diagrams, is no longer subject to helicity and color suppression. This is different from the naive color suppression of internal $W$-emission. It is known in the heavy meson case that nonfactorizable contributions will render the color suppression of internal $W$-emission ineffective. However, the $W$-exchange in baryon decays is not subject to color suppression even in the absence of nonfactorizable terms.  The
experimental measurements of the decays
$\Lambda_c^+\to\Sigma^0\pi^+,~ \Sigma^+\pi^0$ and
$\Lambda^+_c\to\Xi^0K^+$, which do not receive any factorizable
contributions,
indicate that $W$-exchange and  inner $W$-emission indeed play an essential role in charmed baryon decays.

The study of nonfactorizable effects arising from $W$-exchange and inner $W$-emission conventionally relies on the pole model. Under the pole approximation, one usually concentrates on the most important low-lying $1/2^+$ and $1/2^-$ pole states. Consider the charmed baryon decay with a pseudoscalar meson in the final state, $\B_c\to \B+P$. In general, its nonfactorizable $S$- and
$P$-wave amplitudes  are dominated by ${1\over 2}^-$ low-lying baryon resonances and ${1\over 2}^+$ ground-state baryon poles, respectively. It is known that the pole model is reduced to current algebra in the soft pseudoscalar-meson limit. The great advantage of current algebra is that the evaluation of the $S$-wave amplitude does not require the information of the troublesome negative-parity baryon resonances which are not well understood in the quark model. However, the estimation of pole amplitudes is a difficult and nontrivial
task since it involves weak baryon matrix elements and strong
coupling constants of ${1\over 2}^+$ and ${1\over 2}^-$ baryon
states.  As a consequence, the evaluation of pole diagrams is far more uncertain
than the factorizable terms. In Sec. IV.F we will discuss the recent progress in the pole model and address the limitation of the current-algebra approach.

\subsection{Discussions}
\subsubsection{$\Lambda_c^+$ decays}

There were two major breakthroughs in recent charmed-baryon experiments in regard to the hadronic weak decays of $\Lambda_c^+$. The first one is related to the absolute branching fraction of $\Lambda_c^+\to pK^-\pi^+$. Experimentally, nearly all the branching fractions of the
$\Lambda_c^+$ were measured relative to the $pK^-\pi^+$ mode.
On the basis of ARGUS and CLEO data, Particle Data Group had made a model-dependent determination of the absolute branching fraction, ${\cal B}(\Lambda_c^+\to pK^-\pi^+)=(5.0\pm1.3)\%$ \cite{PDG:2014}. In 2014 Belle reported a value of $(6.84\pm0.24^{+0.21}_{-0.27})\%$ \cite{Zupanc} from the reconstruction of $D^*p\pi$ recoiling against the $\Lambda_c^+$ production in $e^+ e^-$ annihilation. The uncertainties are much smaller, and, most importantly, this measurement is model independent! Soon after BESIII has also measured this mode directly with the result ${\cal B}(\Lambda_c^+\to pK^-\pi^+)=(5.84\pm0.27\pm0.23)\%$ \cite{BES:pKpi}. Its precision is comparable to the Belle's result. A new average of $(6.28\pm0.32)\%$ for this benchmark mode is quoted by the PDG \cite{PDG}. Another approach is to exploit a particular decay, $B^+\to p\pi^+\pi^+\overline{\Sigma}_c^{--}$, and its charge conjugate to measure
${\cal B}(\Lambda_c^+\to pK^-\pi^+)$, also in a model independent manner \cite{Contu}. For a method of determining the absolute branching fractions of $\Lambda^+_c$ baryons at hadron colliders, see \cite{Stone:2019nrd}.

Second, in 2015 BESIII has measured the absolute branching fractions for more than a dozen of  decay modes directly for the first time \cite{BES:pKpi}.  Not only are the central values  substantially different from the PDG ones (versions before 2016), but also the uncertainties are significantly improved. For example, $\B(\Lambda_c^+\to p\bar K^0)=(2.3\pm0.6)\%$  and $\B(\Lambda_c^+\to \Sigma^+\omega)=(2.7\pm1.0)\%$ quoted in 2014 PDG \cite{PDG:2014} became $(3.16\pm0.12)\%$  and $(1.74\pm0.21)\%$, respectively,  in 2016 PDG \cite{PDG} due to the new measurement of BESIII. In other words, all the PDG values before the 2016 version for the branching fractions of charmed baryon decays become obsolete. For an overview on the BESIII charmed baryon progress, the reader is referred to \cite{Li:2021iwf}.

\vskip 0.4cm
\noindent {\bf (a) Cabibbo-allowed two-body decays}
\vskip 0.2cm

The branching fractions of the Cabibbo-allowed two-body decays of $\Lambda_c^+$ are listed in Table \ref{tab:BRs}.
Many of the $\Lambda_c^+$ decay modes such as $\Sigma^0\pi^+$, $\Sigma^+\pi^0$, $\Sigma^+\phi$, $\Xi^{(*)}K^{(*)+}$ and $\Delta ^{++}K^-$do not receive any factorizable
contributions,  Experimental measurement of them implies the importance of $W$-exchange and inner $W$-emission.

\begin{table}[btp]
\caption{The measured branching fractions of the Cabibbo-allowed two-body decays of $\Lambda_c^+$ (in units of \%) taken from 2021 Particle Data Group \cite{PDG}, where
$\Sigma(1385)$ and $\Xi(1530)$ states are denoted by $\Sigma^*$ and $\Xi^*$, respectively.
} \label{tab:BRs}
\begin{center}
\begin{tabular}{lc | lc|lc}
\hline\hline ~~~Decay & $\B$ & ~~~Decay & $\B$ & ~~~Decay & $\B$  \\
\hline
~~$\Lambda^+_c\to \Lambda \pi^+$~~ & 1.30$\pm$0.07 & ~~$\Lambda^+_c\to \Lambda \rho^+$~~ & $<6$ & ~~$\Lambda^+_c\to \Delta^{++}K^-$  &  $1.08\pm0.25$\\
~~$\Lambda^+_c\to \Sigma^0 \pi^+$~~ & 1.29$\pm$0.07 & ~~$\Lambda^+_c\to \Sigma^0 \rho^+$  & & ~~$\Lambda^+_c\to \Sigma^{*0} \pi^+$ & \\
~~$\Lambda^+_c\to \Sigma^+ \pi^0$~~ & 1.25$\pm$0.10 & ~~$\Lambda^+_c\to \Sigma^+ \rho^0$~~  & $<1.7$ & ~~$\Lambda^+_c\to \Sigma^{*+} \pi^0$& \\
~~$\Lambda^+_c\to \Sigma^+ \eta$~~ & 0.44$\pm$0.20 & ~~$\Lambda^+_c\to \Sigma^+ \omega$~~  &  1.70$\pm$0.21 & ~~$\Lambda^+_c\to \Sigma^{*+}\eta$~~ & $0.91\pm0.20$ \\
~~$\Lambda^+_c\to \Sigma^+ \eta^\prime$~~ & 1.5$\pm$0.6 & ~~$\Lambda^+_c\to \Sigma^+ \phi$~~ & 0.39$\pm$0.06 & ~~$\Lambda^+_c\to \Sigma^{*+} \eta^\prime$ &\\
~~$\Lambda^+_c\to \Xi^0 K^+$~~ & 0.55$\pm$0.07 & ~~$\Lambda^+_c\to \Xi^0 K^{*+}$~~ &  & ~~$\Lambda^+_c\to \Xi^{*0}K^+$~~  & 0.43$\pm$0.09 \\
~~$\Lambda^+_c\to p K_S$~~ &  1.59$\pm$0.08 & ~~$\Lambda^+_c\to p \bar K^{*0}$~~  & 1.96$\pm$0.27 & ~~$\Lambda^+_c\to \Delta^+ K_S$~~ &  \\
\hline \hline
\end{tabular}
\end{center}
\end{table}

In the 1990s various approaches were developed to describe the nonfactorizable effects in hadronic decays of singly charmed baryons $\Lambda_c^+$, $\Xi_c^{+,0}$ and $\Omega_c^0$. These include the covariant confined quark model \cite{Korner,Ivanov98}, the pole model \cite{CT92,CT93,XK92,XK92b,Verma98,Zen},
current algebra \cite{Korner79,CT93,Uppal,Verma98} and the SU(3) flavor symmetry to be discussed below.
In Table \ref{tab:CF} we show various model calculations of branching fractions  and  decay asymmetry parameters  of Cabibbo-allowed $\Lambda_c^+\to{\cal B}+P$ decays. \footnote{
Various model calculations of the branching fractions of two-body charmed baryon decays can be found in Table VIII  of \cite{Cheng:2009} for ${\cal B}_c\to{\cal B}+V$ decays
decays and Table IX for ${\cal B}_c\to{\cal B}({3\over 2}^+)+P(V)$ ones.}

\begin{table}[t]
\scriptsize{
 \caption{Branching fractions (upper entry) in $\%$ and decay asymmetry parameters (lower entry) of Cabibbo-allowed $\Lambda_c^+\to{\cal
B}+P$ decays in various models. Model results of \cite{Korner,XK92,CT93,Zen}
have been normalized using the current world average of $\tau(\Lambda_c^+)$ \cite{PDG}. Branching fractions cited from
\cite{Verma98} are for $\phi_{\eta-\eta'}=-23^\circ$ and
$r\equiv |\psi^{{\cal B}_c}(0)|^2/|\psi^{\cal B}(0)|^2=1.4$\,.
 } \label{tab:CF}
\centering
\begin{ruledtabular}
\begin{tabular}{l r r c c c c c c c c}
~~~~Decay & K\"{o}rner           & Xu
& Cheng             & \, Ivanov \,  &  \.Zenczykowski & Sharma & Zou & Geng & Experiment \\
& CCQM & Pole & CA \quad Pole & CCQM & Pole & CA & CA & SU(3) & \\
 &  \cite{Korner} &  \cite{XK92}
& \cite{CT93} &  \cite{Ivanov98} & \cite{Zen} &  \cite{Verma98} & \cite{Zou:2019kzq} & \cite{Geng:alpha} & \cite{PDG} \\
 \hline
$\Lambda^+_c\to \Lambda \pi^+$ & input & 1.62 & 1.46 \quad 0.88 & 0.79 & 0.52 & 1.12 & 1.30 & $1.30\pm0.07$ & $1.30\pm 0.07$ \\
$\Lambda^+_c\to p \bar K^0$ & input & 1.20 & 3.64 \quad 1.26 & 2.06 & 1.71 & 1.64 & 2.11 & $3.16\pm0.16$ &  $3.18\pm0.16$ \\
$\Lambda^+_c\to \Sigma^0 \pi^+$ & 0.32 & 0.34 & 1.76 \quad 0.72 & 0.88 & 0.39 & 1.34 & 2.34 & $1.27\pm0.06$ & $1.29\pm0.07$ \\
$\Lambda^+_c\to \Sigma^+ \pi^0$ & 0.32 & 0.34 & 1.76 \quad 0.72 & 0.88 & 0.39 & 1.34 & 2.34 & $1.27\pm0.06$ & $1.25\pm0.10$ \\
$\Lambda^+_c\to \Sigma^+ \eta$ & 0.16 & & & 0.11 & 0.90 & 0.57 & 0.74 & $0.32\pm0.13$ & $0.44\pm0.20$ \\
$\Lambda^+_c\to \Sigma^+ \eta^\prime$ & 1.28 & & & 0.12& 0.11 &  0.10 &  & $1.44\pm0.56$ & $1.5\pm0.6$ & \\
$\Lambda^+_c\to \Xi^0 K^+$ & 0.26 & 0.10 & & 0.31 & 0.34 &  0.13 & 0.73 & $0.56\pm0.09$ & $0.55\pm0.07$ \\
\hline
$\Lambda^+_c\to \Lambda \pi^+$ & $-$0.70 & $-$0.67 &  $-$0.99 \quad
$-$0.95 &  $-$0.95 & $-$0.99 &  $-$0.99 & $-0.93$ & $-0.87\pm0.10$ & $-$$0.84\pm0.09$\\
$\Lambda^+_c\to p \bar K^0$ & $-$1.0 & 0.51 & $-$0.90 \quad $-$0.49 & $-$0.97 & $-$0.66 & $-$0.99 & $-0.75$ &  $-0.89^{+0.26}_{-0.11}$ & $0.2\pm0.5$ \\
$\Lambda^+_c\to \Sigma^0 \pi^+$ & 0.70 & 0.92 & ~$-$0.49~ \quad ~0.78 & 0.43 & 0.39 & $-$0.31 & $-0.76$ & $-0.35\pm0.27$ & $-0.73\pm0.18$ \\
$\Lambda^+_c\to \Sigma^+ \pi^0$ & 0.70 & 0.92 &  ~$-$0.49~ \quad ~0.78 & 0.43
& 0.39 & $-$0.31 &  $-0.76$  & $-0.35\pm0.27$ & $-$$0.55\pm0.11$   \\
$\Lambda^+_c\to \Sigma^+ \eta$ & 0.33 & & & 0.55 & 0 & $-$0.91 & $-0.95$ & $-0.40\pm0.47$ &  \\
$\Lambda^+_c\to \Sigma^+ \eta^\prime$ & $-$0.45 & & & $-$0.05 & $-$0.91 & ~~0.78 & & $1.00^{+0.00}_{-0.17}$ \\
$\Lambda^+_c\to \Xi^0 K^+$ & 0 & 0 & & 0 & 0 & ~~0  & ~~0.90 & $0.94^{+0.06}_{-0.11}$ & \\
\end{tabular}
\end{ruledtabular}
 }
\end{table}

Besides dynamical model calculations, two-body nonleptonic decays of charmed baryons have been analyzed in terms of SU(3)-irreducible-representation
amplitudes \cite{Savage,Verma}.   A general formulation of the quark-diagram scheme for charmed baryons was given in \cite{CCT} (see also \cite{Kohara}).  Analysis of  Cabibbo-suppressed decays using SU(3) flavor symmetry was first carried out in \cite{Sharma}. This approach became very popular recently \cite{Lu,Geng:Lambdac,Geng:2017mxn,Geng:2018,Hsiao:2019,He:2018}.  Although SU(3) flavor symmetry is approximate, it does provide very useful information. In Table \ref{tab:CF} we show the predictions in the $SU(3)_F$ approach  under the column denoted by ``Geng" \cite{Geng:alpha} in which the parameters for both $S$- and $P$-wave amplitudes were obtained by performing a global fit to the data. \footnote{Early studies of the $SU(3)_F$ approach have overlooked the fact that charmed baryon decays are governed by several different partial-wave amplitudes which have distinct kinematic and dynamic effects.}
It appears that the SU(3) approach gives a better
description of the measured branching fractions because it fits to the data.

It is clear from Table \ref{tab:CF} that the predictions of branching fractions and decay asymmetries based on current algebra are generally in agreement with experiment. In the following we shall discuss three selective issues in order: (i) the decays $\Lambda_c^+\to \Sigma^+\eta, \Sigma^+\eta'$, (ii) the decay asymmetry parameter $\alpha$ and (iii)
the decay $\Lambda_c^+\to \Xi^{0} K^+$.

\vskip 0.2cm
{(i) $\Lambda_c^+\to \Sigma^+\eta, \Sigma^+\eta'$}
\vskip 0.2cm

All the nonfactoizable diagrams in Fig. \ref{fig:Bic} contribute to $\Lambda_c^+\to \Sigma^+\eta^{(')}$ decays. More precisely, $C'$, $E_1$ and $E_3$ contribute to the $q\bar q$ quark component of $\eta^{(')}$, while $E_2$ to the $s\bar s$ component.  Naively, it is expected that the rates of $\Lambda_c^+\to \Sigma^+\eta$ and $\Lambda_c^+\to \Sigma^+\eta'$ are comparable or the former is larger than the latter. However, the branching fraction of $\Lambda_c^+\to \Sigma^+\eta'$  measured by BESIII \cite{BES:Sigmaeta} for the first time was found to be larger than the $\Sigma^+\eta$ mode
\begin{eqnarray}
{ {\cal B}(\Lambda_c^+\to \Sigma^+\eta')\over
{\cal B}(\Lambda_c^+\to \Sigma^+\eta)}= 3.5\pm2.1\pm0.4\,.
\end{eqnarray}
An inspection of Table \ref{tab:CF} indicates that among the early calculations in 1990s only the covariant confined quark model considered in \cite{Korner} led to the correct trend for $\Lambda_c^+\to \Sigma^+\eta^{(')}$.

\vskip 0.2cm
{(ii) Decay asymmetry parameter}
\vskip 0.2cm
A verr useful information is provided by the study of the polarization of the daughter baryon $\B_f$ in the decay $\B_i\to\B_f\pi$. Even if the parent baryon is unpolarized, the daughter baryon $\B_f$ will be longitudinally polarized by an amount of $\alpha$, ${\bf P}_{\B_f}=\alpha {\bf n}$,
where ${\bf n}$ is a unit vector along the daughter baryon in the parent baryon frame and
\be
\alpha={ 2\kappa {\rm Re}(A^*B)\over |A|^2+\kappa^2|B|^2},
\en
with $\kappa=p_c/(E_f+m_{\B_f})$. The relative sign between $S$- and $P$-wave amplitudes
is determined by the decay asymmetry parameter $\alpha$.

We see from Table \ref{tab:CF} that if the $S$-wave amplitude is evaluated in the pole model or in the covariant quark model and its variant, the decay asymmetry parameters for both $\Lambda_c^+\to \Sigma^+\pi^0$ and $\Sigma^0\pi^+$ were always predicted to be positive, while it was measured to be $-0.45\pm0.31\pm0.06$ for $\Sigma^+\pi^0$ by CLEO \cite{CLEO:alpha}. In contrast, current algebra always led to a negative decay asymmetry for aforementioned two modes: $-0.49$ in \cite{CT93}, $-0.31$ in \cite{Verma98}, $-0.76$ in \cite{Zen:1993} and $-0.47$ in \cite{Datta}.  The issue with the sign of $\alpha(\Lambda_c^+\to\Sigma^+\pi^0)$ was finally resolved by BESIII. The decay asymmetry parameters of $\Lambda_c^+\to \Lambda\pi^+,\Sigma^0\pi^+,\Sigma^+\pi^0$ and $pK_S^0$ were recently measured by BESIII \cite{Ablikim:2019zwe} (see Table \ref{tab:CF}), for example,  $\alpha(\Lambda_c^+\to\Sigma^+\pi^0)=-0.57\pm0.12$ was obtained. Hence,
the negative sign of $\alpha(\Lambda_c^+\to\Sigma^+\pi^0)$ measured by CLEO was nicely confirmed by BESIII. As for the decay asymmetry in $\Lambda_c^+\to pK_S^0$, all the models except the pole model in \cite{XK92} predict a large and negative value (see Table \ref{tab:CF}), while the current measurement is $\alpha(\Lambda_c^+\to pK_S^0)=0.18\pm0.45$ \cite{Ablikim:2019zwe}. This issue needs to be resolved in future study. As for the decay parameter in the decay  $\Lambda_c^+\to \Xi^{0} K^+$, it will be discussed below.

\vskip 0.2cm
{(iii) $\Lambda_c^+\to \Xi^{0} K^+$}
\vskip 0.2cm

The decay mode $\Lambda_c^+\to \Xi^{0} K^+$ deserves  special attention.  All the existing model predictions of its branching fraction are smaller than experiment due to a large cancellation in both $S$- and $P$-wave amplitudes (see e.g. \cite{CT93} for a detailed discussion). For example, in the framework of current algebra we have
\be \label{eq:LambdacXiK}
A^{\rm{com}}(\Lambda_c^+\to \Xi^0 K^+) &=& \frac{1}{f_{K}}(a_{\Sigma^{+} \Lambda_{c}^{+}}-a_{\Xi^{0}\Xi_{c}^{0}}), \nonumber \\
B^{\rm{ca}}(\Lambda_c^+\to \Xi^0 K^+) &=&\frac{1}{f_{K}}\Bigg(g^{A({K}^{+})}_{\Xi^{0}\Sigma^{+}}\frac{m_{\Xi^{0}}+m_{\Sigma^{+}}}
{m_{\Lambda^{+}_{c}}-m_{\Sigma^{+}}}a_{\Sigma^{+}\Lambda^{+}_{c}} + a_{\Xi^{0}\Xi^{0}_{c}}\frac{m_{\Xi_c^{0}}+m_{\Lambda_c^{+}}}
{m_{\Xi^{0}}-m_{\Xi_c^{0}}}g^{A({K}^{+})}_{\Xi_c^{0}\Lambda_c^{+}}  \nonumber \\
&+& a_{\Xi^{0}\Xi^{'0}_{c}}\frac{m_{\Xi_c^{'0}}+m_{\Lambda_c^{+}}}
{m_{\Xi^{0}}-m_{\Xi_c^{'0}}}g^{A({K}^{+})}_{\Xi_c^{'0}\Lambda_c^{+}}
\Bigg).
\en
In the SU(3) limit, the matrix elements $a_{\Sigma^{+} \Lambda_{c}^{+}}\equiv \la \Sigma^+|H|\Lambda_c^+\ra$ and $a_{\Xi^{0}\Xi_{c}^{0}}$ are identical and there is a large cancellation between the first and third terms in $B^{\rm ca}$ (no contribution from the second term due to the vanishing  $g^{A({K}^{+})}_{\Xi_c^{0}\Lambda_c^{+}}$).
Moreover, the decay asymmetry parameter was predicted to be zero in the literature owing to the vanishing $S$-wave amplitude (see Table \ref{tab:CF}). As a consequence, the predicted branching fractions of this mode were too small compared to experiment.
This is a long-standing puzzle.

\begin{figure}[t]
\begin{center}
\includegraphics[width=0.70\textwidth]{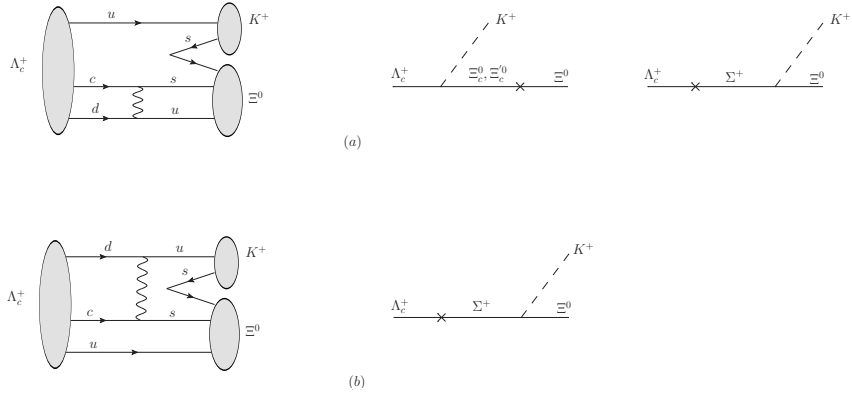}
\vspace{0.1cm}
\caption{$W$-exchange diagrams contributing to $\Lambda_c^+\to \Xi^0K^+$. The corresponding pole diagrams are also shown.} \label{fig:Lambdac_XiK} \end{center}
\end{figure}

The decay $\Lambda_c^+\to \Xi^{0} K^+$ receives contributions only from the $W$-exchange diagrams $E_1$ and $E_3$ (see Fig. \ref{fig:Bic}). It is clear that $E_3$ cannot contribute to the parity-violating $S$-wave amplitude. It can be described by two distinct pole diagrams at the hadron level shown in the right panel of the diagram Fig. \ref{fig:Lambdac_XiK}(a).
As first pointed out in \cite{Zou:2019kzq},  the conventional expression of parity-violating and -conserving amplitudes given in Eq. (\ref{eq:LambdacXiK}) is actually for the $E_3$ diagram in Fig. \ref{fig:Lambdac_XiK}(a).
As a result, non-vanishing nonfactorizable $S$- and $P$-wave amplitudes arise solely from $E_1$ as depicted in Fig. \ref{fig:Lambdac_XiK}(b).
The nonfactorizable amplitudes induced from $W$-exchange $E_1$ now read
\be \label{eq:LambdacXiK_mod}
A^{\rm{com}}(\Lambda_c^+\to \Xi^0 K^+) &=& \frac{1}{f_{K}}a_{\Sigma^{+} \Lambda_{c}^{+}}, \nonumber \\
B^{\rm{ca}}(\Lambda_c^+\to \Xi^0 K^+) &=&\frac{1}{f_{K}}\left(g^{A({K}^{+})}_{\Xi^{0}\Sigma^{+}}\frac{m_{\Xi^{0}}+m_{\Sigma^{+}}}
{m_{\Lambda^{+}_{c}}-m_{\Sigma^{+}}}a_{\Sigma^{+}\Lambda^{+}_{c}}\right).
\en
Consequently, both partial wave amplitudes are not subject to large cancellations.
Eq. (\ref{eq:LambdacXiK_mod}) leads to
${\cal B}(\Lambda_c^+\to \Xi^0 K^+)=0.71\%$, which is consistent with the data of $(0.55\pm0.07)\%$ \cite{PDG}.  Moreover, the predicted positive decay asymmetry of order $0.90$ \cite{Zou:2019kzq} is consistent with the value of $\alpha(\Lambda_c^+\to \Xi^0 K^+)=0.94^{+0.06}_{-0.11}$ obtained in the SU(3)-flavor approach \cite{Geng:alpha}. Therefore, the long-standing puzzle with the branching fraction and the decay asymmetry of $\Lambda_c^+\to\Xi^0 K^+$ is resolved and it is important to measure $\alpha(\Lambda_c^+\to \Xi^0 K^+)$ to test theory.

\vskip 0.4cm
\noindent {\bf (b) Cabibbo-suppressed two-body decays}
\vskip 0.2cm

The branching fractions of singly Cabibbo-suppressed modes are of the order of $10^{-3}\sim 10^{-4}$.
The first measured Cabibbo-suppressed mode $\Lambda_c^+\to p\phi$
is of particular interest because it receives contributions only
from the internal $W$-emission diagram $C$ and is characterized by the parameter
\be \label{eq:a2}
a_2=c_2+{c_1\over N_c^{\rm eff}},
\en
where the number of colors $N_c$ is treated as an effective parameter.
Using the current world average $\B(\Lambda_c^+\to p\phi)=(1.06\pm0.14)\times 10^{-3}$ \cite{PDG}, we obtain $|a_2|=0.45\pm0.02$ \cite{CKX}, corresponding to $N_c^{\rm eff}\approx 7$ for $c_1 = 1.346$, $c_2 = -0.636$ and $f_\phi=215$ MeV. Recall that $a_2=-0.19$ for $N_c=3$. Hence, color suppression in the factorizable amplitude of internal $W$-emission is alleviated.

\begin{table}[tbp]
\footnotesize{
 \caption{Comparison of various theoretical predictions for the branching fractions (in units of $10^{-3}$) of singly Cabibbo-suppressed decays of $\Lambda_c^+$.
 } \label{tab:SCSBF}
\centering
\begin{ruledtabular}
\begin{tabular}{ l c c c c c c c c }
 & Sharma  & Uppal   & Chen  & Lu &  Geng & Hsiao & Zou & Experiment \\
 &   \cite{Sharma} & \cite{Uppal}  &  \cite{Chen:2003} &  \cite{Lu}  & \cite{Geng:15}   &  \cite{Hsiao:2018} & \cite{Zou:2019kzq} & \cite{PDG} \\
 \hline
$\Lambda_c^+\to p\pi^0$    & 0.2 & 0.1-0.2 & 0.11-0.36 & 0.48  & $0.13\pm0.07$ & $0.08\pm0.07$ & 0.13 & $<0.08$ \footnotemark[3] \\
$\Lambda_c^+\to p\eta$ &  0.2\footnotemark[1](1.7)\footnotemark[2] & 0.3 & &  & $1.30\pm0.10$ & $1.24\pm0.21$ & 1.28 & $1.40\pm0.11$ \footnotemark[4]\\
$\Lambda_c^+\to p\eta'$ &  0.4-0.6 & 0.04-0.2 & &   &  & $0.10\pm0.03$ & &  $0.473\pm0.098$ \footnotemark[5]\\
$\Lambda_c^+\to n\pi^+$    & 0.4 & 0.8-0.9 & 0.10-0.21 & 0.97 & $0.61\pm0.20$ & &   & $0.66\pm0.13$ \footnotemark[6] \\
$\Lambda_c^+\to \Lambda K^+$ & 1.4& 1.2 & 0.18-0.39 &  & $0.64\pm0.09$ & $0.57\pm0.12$ & 1.07 & $0.61\pm0.12$ \\
$\Lambda_c^+\to \Sigma^0 K^+$ & 0.4-0.6 & 0.2-0.8 & &  & $0.57\pm0.10$ & $0.54\pm0.11$ & 0.72 & $0.52\pm0.08$ \\
$\Lambda_c^+\to \Sigma^+ K^0$ & 0.9-1.2 & 0.4-0.8 & &  & $1.14\pm0.20$ & & 1.44  &  \\
\end{tabular}
\end{ruledtabular}
\footnotetext[1]{The $P$-wave amplitude of $\Lambda_c^+\to\Xi^0K^+$ is assumed to be positive.}
\footnotetext[2]{The $P$-wave amplitude of $\Lambda_c^+\to\Xi^0K^+$ is assumed to be negative.}
\footnotetext[3]{Data taken from Ref. \cite{Belle:ppi0}.}
\footnotetext[4]{Data from Ref. \cite{Belle:ppi0} has been taken into account for the world average.}
\footnotetext[5]{Data taken from Ref. \cite{Belle:Lambdactopetap}.}
\footnotetext[6]{Data taken from Ref. \cite{BESIII:Lambdactonpi}.}
}
\end{table}

Various model predictions for the singly Cabibbo-suppressed decays $\Lambda_c^+\to\B+P$ are summarized in Table \ref{tab:SCSBF}. Except
for the dynamic calculations in \cite{Uppal,CKX,Zou:2019kzq} and the consideration of factorizable contributions in \cite{Chen:2003}, all other predictions are based on the SU(3) symmetry argument. The task of fitting to the data in terms of SU(3)-irreducible-representation amplitudes or the topological diagrams
becomes possible after combining the data of both Cabibbo-allowed and singly Cabibbo-suppressed decays and making some plausible assumption to reduce the parameters to be fitted. For example, if the SU(3) singlet contributions are not considered, then there will be three unknown parameters in $H({\bf 6}_F)$ representation and four unknown parameters in $H(\overline{\bf 15}_F)$.
Since  $c_-O_-\gg c_+O_+$, the SU(3) practitioners first tended to make the assumption of
the sextet ${\bf 6}_F$ dominance over $\overline{\bf 15}_F$. Under this hypothesis, one would have the relation \cite{Sharma,Lu}
\begin{eqnarray} \label{eq:npi}
M(\Lambda_c^+\to n\pi^+)=\sqrt{2}M(\Lambda_c^+\to p\pi^0),
\end{eqnarray}
and the sum rule
\begin{eqnarray} \label{eq:SR}
\B(\Lambda_c^+\to n\pi^+)=\sin^2\theta_C\left[ 3\B(\Lambda_c^+\to\Lambda\pi^+)+
\B(\Lambda_c^+\to\Sigma^0\pi^+)-\B(\Lambda_c^+\to p\overline{K}^0)\right].
\end{eqnarray}
The experimental values for branching fractions \cite{PDG} lead to $\B(\Lambda_c^+\to n\pi^+)\sim 0.97\times 10^{-3}$ and hence $\B(\Lambda_c^+\to p\pi^0)\sim 5\times 10^{-4}$
which exceeds the current experimental limit of $0.8\times 10^{-4}$ \cite{Belle:ppi0}.
The discrepancy between the SU(3) approach and the yet-observed $\Lambda_c^+\to p\pi^0$ is ascribed to the SU(3) relations given by Eqs. (\ref{eq:npi}) and (\ref{eq:SR}). First of all, the relation (\ref{eq:npi}) does not hold in the general quark diagram approach owing to the presence of factorizable contributions \cite{CCT}.
Since the factorizable amplitude of $\Lambda_c^+\to n\pi^+$ ($\Lambda_c^+\to p\pi^0$) is governed by the external (internal) $W$-emission, we have
\begin{eqnarray}
{M(\Lambda_c^+\to n\pi^+)^{\rm fac}\over M(\Lambda_c^+\to p\pi^0)^{\rm fac}}=-\sqrt{2}\left({a_1\over a_2}\right)\approx 2.8\sqrt{2}.
\end{eqnarray}
Hence, the factorizable amplitudes alone strongly violate the SU(3) relation (\ref{eq:npi}).
This implies that it is necessary to take into account the contributions from $H(\overline{\bf 15}_F)$. Under the pole approximation, the authors of \cite{Geng:15} showed that there was only one contribution from $H(\overline{\bf 15}_F)$ and it was factorizable. A global fit to the data led to ${\cal B}(\Lambda_c^+\to p\pi^0)=(1.3\pm0.7)\times 10^{-4}$. The current limit of $0.8\times 10^{-4}$ was set by Belle recently \cite{Belle:ppi0}.

In the dynamic model calculation of $\Lambda_c^+\to n\pi^+$ \cite{Zou:2019kzq}, we showed that the interference between factorizable and nonfactorizable contributions should be destructive. The factorizable contributions alone will lead to $\B(\Lambda_c^+\to n\pi^+)=1.5\times 10^{-3}$. However, we found that large cancellation renders the theoretical predictions of $\Lambda_c^+\to n\pi^+$ unreliable. Hence its branching fraction is not shown in Table \ref{tab:SCSBF}. In a very recent measurement BESIII obtained  $\B(\Lambda_c^+\to n\pi^+)=(6.6\pm1.2\pm0.4)\times 10^{-4}$ \cite{BESIII:Lambdactonpi}.

\subsubsection{$\Xi_c^0$ and $\Xi_c^+$ decays}

\vskip 0.2cm
\noindent {\bf (a)~ $\Xi_c^{+,0}\to{\cal
B}+P$ decays }
\vskip 0.2cm

By using a data set comprising $(772\pm11)\times 10^6$ $B\bar{B}$ pairs collected at
$\Upsilon(4S)$ resonance, Belle was able to measure the absolute branching fraction for $B^-\to \bar{\Lambda}_c^- \Xi_c^0$ and the product branching fraction $\mathcal{B}(B^-\to \bar{\Lambda}_c^- \Xi_c^0) \mathcal{B}(\Xi_c^0\to \Xi^-\pi^+)$ \cite{Belle:Xic0} and reported the first measurement of the absolute branching fraction of the $\Xi_c^0$ baryon
\footnote{However, the branching fraction of $\Xi_c^0\to \Xi^-\pi^+$ is quoted to be
$(1.43\pm 0.32)\times 10^{-2}$ by the PDG \cite{PDG}.}
\begin{equation}
\mathcal{B}(\Xi_c^0\to \Xi^-\pi^+)=(1.80\pm 0.50\pm 0.14)\times 10^{-2},
\end{equation}
so that the branching fractions of other channels can be determined from ratios of branching fractions.
Likewise, by measuring the absolute branching fraction for $\bar B^0\to \bar{\Lambda}_c^- \Xi_c^+$ and the product branching fractions $\mathcal{B}(\bar B^0\to \bar{\Lambda}_c^- \Xi_c^+) \mathcal{B}(\Xi_c^+\to \Xi^-\pi^+\pi^+)$ and $\mathcal{B}(\bar B^0\to \bar{\Lambda}_c^- \Xi_c^+) \mathcal{B}(\Xi_c^+\to pK^-\pi^+)$ \cite{Belle:Xic+},  Belle has also reported the first measurements of the absolute branching fractions of the $\Xi_c^+$ system
\be \label{eq:XicAbsolutB}
\mathcal{B}(\Xi_c^+\to \Xi^-\pi^+\pi^+) &=& (2.86\pm 1.21\pm 0.38)\%, \non \\
\mathcal{B}(\Xi_c^+\to pK^-\pi^+) &=& (0.45\pm 0.21\pm 0.07)\%.
\en
Combined with the measured ratio of $\Gamma(\Xi_c^+\to \Xi^0\pi^+)/\Gamma(\Xi_c^+\to\Xi^-\pi^+ \pi^+)=0.55\pm0.13\pm0.09$  obtained by the CLEO \cite{Edwards:1995xw}, the branching fraction of $\Xi_c^+\to \Xi^0\pi^+$ is inferred to be
\begin{equation}
\mathcal{B}(\Xi_c^+\to \Xi^0\pi^+)=
(1.57\pm0.83)\%.
\end{equation}
Later, LHCb has also measured one of the benchmark modes $\Xi_c^+\to pK^-\pi^+$ and found
\cite{LHCb:HFC}
\be \label{eq:XicAbsolutB_LHCb}
\mathcal{B}(\Xi_c^+\to pK^-\pi^+) &=& (1.135\pm 0.002\pm 0.387)\%.
\en
This is somewhat larger than the Belle measurement.

The Cabibbo-allowed decays $\Xi_c^+\to {\cal B}(3/2^+)+P$ have
been studied and were believed to be forbidden, as they do not
receive factorizable and $1/2^\pm$ pole contributions
\cite{XK92b,Korner}. However, the $\Sigma^{*+}\bar K^0$ mode was
seen earlier by FOCUS \cite{FOCUS:SigK}, and this may indicate the
importance of pole contributions beyond low-lying $1/2^\pm$
intermediate states.

\begin{table}[t]
\footnotesize{
 \caption{Branching fractions (upper entry) and decay asymmetry parameters (lower entry) of Cabibbo-allowed $\Xi_c^{+,0}\to{\cal
B}+P$ decays in various model calculations.  All the model results for branching fractions (in percent)
have been normalized using the current world averages of $\tau(\Xi_c^+)$ and $\tau(\Xi_c^0)$.
 } \label{tab:CF-his}
\centering
\begin{ruledtabular}
\begin{tabular}{l c c c c c c c c}
~~~~Decay & K\"{o}rner  {\it et al.}          & Xu  {\it et al.}
& Cheng  {\it et al.}            &  Ivanov {\it et al.}  & Zou {\it et al.} & Geng {\it et al.}   & Experiment \\
        & \cite{Korner} & \cite{XK92}
& \cite{CT93} & \cite{Ivanov98}  & \cite{Zou:2019kzq} & \cite{Geng:alpha} & \cite{PDG,Belle:XictoK} \\
& & & CA \quad~ Pole & & &  & \\
 \hline
$\Xi_c^+\to \Sigma^+ \bar K^0$ & 6.66  & 0.46 & 0.05 \quad~ 0.87 & 4.05  & 0.20 & $0.86^{+0.94}_{-0.78}$ \\
$\Xi_c^+\to \Xi^0 \pi^+ $ & 3.65 & 3.47 & 0.87 \quad~ 4.06 & 5.78  & 1.72 & $0.38\pm0.20$ & $1.6\pm0.8$ \\
$\Xi_c^0\to \Lambda \bar K^0$ & 0.17 & 0.50 & 1.36 \quad~ 0.37 & 0.55 & 1.33 & $1.05\pm0.06$  & $0.60\pm0.16$ \footnotemark[1] \\
$\Xi_c^0\to \Sigma^0 \bar K^0$ & 1.61 & 0.14 & 0.03 \quad~ 0.18 & 0.26  & 0.04 & $0.08\pm0.08$ & $0.14\pm0.05$ \\
$\Xi_c^0\to \Sigma^+ K^-$ & 0.17 &0.17 & & 0.35  &  0.78 & $0.59\pm0.11$ & $0.22\pm0.07$ \\
$\Xi_c^0\to \Xi^0 \pi^0$ & 0.05 &0.77 & 1.71 \quad~ 0.38 & 0.05  & 1.82 & $0.76\pm0.10$ & \\
$\Xi_c^0\to \Xi^- \pi^+$ & 1.42 & 2.37 &1.13 \quad~ 1.71 & 1.60  & 6.47 & $2.21\pm0.14$ & $1.43\pm 0.32$  \\
$\Xi_c^0\to \Xi^0 \eta$ & 0.32 & & & 0.37  & 2.67 & $1.03\pm0.20$  \\
$\Xi_c^0\to \Xi^0 \eta'$ & 1.16 & & & 0.41 & & $0.91\pm0.41$ \\
\hline
$\Xi_c^+\to \Sigma^+ \bar K^0$ & $-$1.0 & 0.24 & 0.43 \quad
$-$0.09 &  $-$0.99 & $-0.80$ &  $0.98^{+0.02}_{-0.16}$ & \\
$\Xi_c^+\to \Xi^0 \pi^+ $ & $-$0.78 & $-$0.81 & $-$0.77 \quad $-$0.77 & $-$0.97 & $-0.78$ & $-0.32\pm0.52$ &  \\
$\Xi_c^0\to \Lambda \bar K^0$ & $-$0.76 & 1.00 & $-$0.88 \quad $-$0.73 & $-$0.75 & $-$0.86 & $-0.68\pm0.28$ & \\
$\Xi_c^0\to \Sigma^0 \bar K^0$ & $-$0.96 & $-$0.99 &  0.85 \quad $-$0.59 & $-$0.55
& $-$0.94 & $-0.07\pm0.90$  &    \\
$\Xi_c^0\to \Sigma^+ K^-$ & 0 & 0 & & 0 & ~~0.98 & ~~$0.81\pm0.16$ &  \\
$\Xi_c^0\to \Xi^0 \pi^0$ & ~~0.92 & 0.92 &  $-$0.78 \quad $-$0.54 & ~~0.94 & $-0.77$ & $-1.00^{+0.07}_{-0.00}$ &  \\
$\Xi_c^0\to \Xi^- \pi^+$ & $-$0.38 & $-$0.38 & $-$0.47 \quad $-$0.99 & $-$0.84 & $-$0.95 & $-0.98^{+0.07}_{-0.02}$  & $-0.60\pm 0.04$ \footnotemark[2] \\
$\Xi_c^0\to \Xi^0 \eta$ & $-$0.92 &  & & $-$1.0 & ~~0.30 &  ~~$0.93^{+0.07}_{-0.19}$ &  \\
$\Xi_c^0\to \Xi^0 \eta'$ & $-$0.38 & & & $-$0.32 &  &  ~~$0.98^{+0.02}_{-0.27}$  \\
\end{tabular}
\end{ruledtabular}
\footnotetext[1]{A new Belle measurement yields $B(\Xi_c^0\to \Lambda\bar K^0)=(0.82\pm0.24)\%$ \cite{Belle:XictoK}.}
\footnotetext[2]{Dominated by the new Belle measurement of $\alpha(\Xi_c^0\to \Xi^- \pi^+)$ \cite{Belle:Xic0semi}.}
 }
\end{table}

The predicted branching fractions and decay asymmetry parameters for Cabibbo-allowed $\Xi_c^{+,0}\to{\cal B}+P$ decays in various model calculations are collected in Table \ref{tab:CF-his}. Most of them were the early studies done in 1990s except for the columns under ``Zou {\it et al.}" and ``Geng {\it et al.}" where we show the modern calculations based on current algebra and the SU(3) approach, respectively. The results for singly Cabibbo-suppressed modes are referred to Tables V and VII of \cite{Zou:2019kzq}.  We notice that there does not exist any decay mode which proceeds only through the factorizable diagram.
Among all the processes, the three modes $\Xi_c^0\to\Sigma^+ K^-, \Xi^0\pi^0, \Xi^0\eta$ in Cabibbo-favored (CF) processes
and the five singly Cabibbo-suppressed (SCS) modes $\Xi_c^+\to p\overline{K}^0$, $\Xi_c^0\to\Xi^0 K^0, p K^-, n\overline{K}^0, \Sigma^+ \pi^-$
proceed only through the nonfactorizable diagrams,
while all the other channels receive contributions from both factorizable and nonfactorizable terms.

The CF decay $\Xi_c^0\to \Sigma^+ K^-$ and the SCS modes $\Xi_c^0\to pK^-,\Sigma^+\pi^-$ are of special interest among all the $\Xi_c$ weak decays. Their naive $S$-wave amplitudes are given by
\be \label{eq:AcomXic}
A^{\rm{com}}(\Xi_c^0\to \Sigma^+ K^-) &=& \frac{1}{f_K} \left(a_{\Xi^0 \Xi_c^0}-a_{\Sigma^+\Lambda_c^+}\right),     \nonumber\\
A^{\rm{com}}(\Xi_c^0\to p K^-) &=& -\frac{1}{f_K}
\left(\frac{\sqrt{2}}{2} a_{\Sigma^0 \Xi_c^0} +\frac{\sqrt{6}}{2} a_{\Lambda \Xi_c^0}+a_{p\Lambda_c^+}\right),  \\
A^{\rm{com}}(\Xi_c^0\to \Sigma^+ \pi^-) &=& - \frac{1}{f_\pi} \left(\sqrt{2} a_{\Sigma^0 \Xi_c^0}+a_{\Sigma^+ \Xi_c^+}\right). \nonumber
\en
They all vanish in the SU(3) limit \cite{Zou:2019kzq}. Likewise, their $P$-wave amplitudes are also subject to large cancellations.
Just as the decay $\Lambda_c^+\to\Xi^0K^+$ discussed in passing, we should neglect the contributions from the  $W$-exchange diagram $E_3$ and focus on $E_1$.
The resulting amplitudes for these three modes now read
\be
A^{\rm{com}}(\Xi_c^0\to \Sigma^+ K^-) &=& \frac{1}{f_K}a_{\Xi^0 \Xi_c^0},     \nonumber\\
A^{\rm{com}}(\Xi_c^0\to p K^-) &=& -\frac{1}{f_K}
\left(\frac{\sqrt{2}}{2} a_{\Sigma^0 \Xi_c^0} +\frac{\sqrt{6}}{2} a_{\Lambda \Xi_c^0}\right),  \nonumber \\
A^{\rm{com}}(\Xi_c^0\to \Sigma^+ \pi^-) &=& - \frac{\sqrt{2}}{f_\pi} a_{\Sigma^0 \Xi_c^0}, \nonumber
\en
for $S$-wave. It was found that \cite{Zou:2019kzq}
\be
\alpha(\Xi^0_c\to \Sigma^+K^-)\approx 0.98\,, \quad \alpha(\Xi^0_c\to pK^-)\approx0.99\,, \quad \alpha(\Xi^0_c\to \Sigma^+\pi^-)\approx 0.89\,.
\en
Hence, their decay asymmetries are all positive and close to unity. It is interesting to notice that the decay asymmetries of these three modes were also predicted to be positive and large in the SU(3) approach of \cite{Geng:alpha}.

As for the two modes $\Xi_c^+\to\Xi^0\pi^+$ and $\Xi_c^0\to\Xi^-\pi^+$, we see from Table \ref{tab:CF-his} that our predictions are in good agreement with experiment for the former, but too large for the latter.
This is mainly due to the relative sign between factorizable
and nonfactorizable terms. In the absence of nonfactorizable contributions, we find ${\cal B}(
\Xi_c^+\to\Xi^0\pi^+)\approx 9.9\%$ and ${\cal B}(\Xi_c^0\to\Xi^-\pi^+)\approx 3.3\%$. Since the measured branching fractions are $(1.57\pm0.83)\%$ and $(1.43\pm0.32)\%$, respectively, this implies that there should be a large destructive interference between factorizable and nonfactorizable terms in the former and a smaller  destructive interference in the latter. However, we found that the interference between $A^{\rm f}$ and $A^{\rm nf}$ is destructive in $\Xi_c^+\to\Xi^0\pi^+$ but constructive in  $\Xi_c^0\to\Xi^-\pi^+$ (see also \cite{XK92}). As a result, the predicted branching fraction of order $6.5\%$ for the latter is too large.
We thus concluded that ${\cal B}(\Xi_c^0\to \Xi^-\pi^+)\gg \mathcal{B}(\Xi_c^+\to
\Xi^0\pi^+)$ and  that these two modes cannot be simultaneously explained within the current-algebra framework for $S$-wave amplitudes \cite{Zou:2019kzq}. On-shell corrections to the current-algebra results are probably needed to circumvent the difficulty with $\Xi_c^0\to \Xi^-\pi^+$. Table \ref{tab:CF-his} shows that all the early model calculations except for the current-algebra one \cite{CT93} led to ${\cal B}(\Xi_c^0\to \Xi^-\pi^+)\ll \mathcal{B}(\Xi_c^+\to \Xi^0\pi^+)$, whereas it is the other way around in recent studies.
Hence, more accurate measurements of them are called for in order to set the issue and see if the branching fractions of both modes are comparable as suggested by the recent data.

\vskip 0.4cm
\noindent {\bf (b)~ $\Xi_c\to{\cal B}+V$ decays}
\vskip 0.2cm

\begin{table}[tp!]
\begin{ruledtabular}
 \caption{Branching fractions (upper entry, in units of percent) and decay asymmetry parameters (lower entry) for Cabibbo-favored $\Xi_c^{0}\to \B+V$ decays.
 } \label{tab:XictoBV}
\begin{center}
\begin{tabular}
{ l c c c c c}
 Mode  & Korner & \.Zenczykowski & Hsiao & Geng & Experiment \\
 & \cite{Korner} & \cite{Zen} & \cite{Hsiao:V} & \cite{Geng:V} & \cite{Belle:XicBV} \\
 \hline
$\Xi_{c}^{0}\to\Lambda \overline{K}^{*0}$ & $1.55$   & $1.15$  & $0.46\pm0.21$ & $1.37\pm0.26$ & $0.33\pm0.03\pm0.02\pm0.10$    \\
$\Xi_{c}^{0}\to\Sigma^0 \overline{K}^{*0}$ & $0.85$   & $0.77$  & $0.27\pm0.22$ & $0.42\pm0.23$ & $1.24\pm0.05\pm0.05\pm0.36$    \\
$\Xi_{c}^{0}\to\Sigma^+ {K}^{*-}$ & $0.54$   & $0.37$ & $0.93\pm0.29$ & $0.24\pm0.17$ & $0.61\pm0.10\pm0.04\pm0.18$    \\
\hline
$\Xi_{c}^{0}\to\Lambda \overline{K}^{*0}$ & $0.58$   & $0.49$ & & $-0.67\pm0.24$ & $0.15\pm0.22\pm0.04$   \\
$\Xi_{c}^{0}\to\Sigma^0 \overline{K}^{*0}$ & $-0.87$   & $0.25$  & & $-0.42\pm0.62$ &    \\
$\Xi_{c}^{0}\to\Sigma^+ {K}^{*-}$ & $-0.60$   & $0.51$ & & $-0.76^{+0.64}_{-0.24}$ & $-0.52\pm0.30\pm0.02$   \\
\end{tabular}
\end{center}
 \end{ruledtabular}
 \end{table}

Very recently, Belle has reported the measurements of branching fractions and asymmetry parameters of Cabibbo-allowed decays $\Xi_c^0\to\Lambda \ov K^{*0},\Sigma^0\ov K^{*0}$ and $\Sigma^+K^{*-}$ for the first time \cite{Belle:XicBV}, as listed in Table \ref{tab:XictoBV}. We see that the branching fraction of $\Xi_c^0\to\Sigma^0\ov K^{*0}$ is much larger than that of $\Xi_c^0\to\Lambda \ov K^{*0}$. This contradicts all the predictions based on SU(3) flavor symmetry and dynamical models \cite{Korner,Zen,Hsiao:V,Geng:V}.

In the $\Xi_c^+$ sector, LHCb has  recently announced the observation of the doubly Cabibbo-suppressed decay $\Xi_c^+\to p\phi$ relative to the singly Cabibbo-suppressed decay $\Xi_c^+\to p K^-\pi^+$ \cite{LHCb:Xictopphi}.

\subsubsection{$\Omega_c^0$ decays}
No absolute branching fractions have been measured in $\Omega_c^0$ decays. The branching {\it ratios} were measured relative to  $\Omega_c^0\to\Omega^-\pi^+$.
A unique feature of $\Omega_c^0$ decays is that the
decay $\Omega_c^0\to\Omega^-\pi^+$ proceeds only via external
$W$-emission, whereas $\Omega_c^0\to \Xi^{*0}\bar K^0$  via internal $W$-emission diagram.  Various early model predictions of Cabibbo-allowed $\Omega_c^0\to {\cal B}(3/2^+)+P(V)$ decays, namely, $\Omega_c^0\to \Omega^-\pi^+, \Omega^-\rho^+$ and $\Omega_c^0\to \Xi^{*0}\ov K^0, \Xi^{*0}\ov K^{*0}$ are shown
in Table IV of \cite{Cheng:bottombaryon} with the unknown parameters $a_1$ and
$a_2$. From the measurement of $\Lambda_c^+\to p\phi$ we have learned that
$|a_2|=0.45\pm0.02$ \cite{CKX}. Modern studies can be found in \cite{Gutsche:Omegac} with the old value of $\tau(\Omega_c^0)$ and \cite{Hsiao:Omegac} with the new $\Omega_c^0$ lifetime which is four times longer than the old one.

The hadronic weak decays of the $\Omega_c^0$ into a ground-state octet baryon and a pseudoscalar (an axial-vector) meson were recently studied in  \cite{Xu:Omegac}  (\cite{Dhir}). The reader is referred to the original references for details. It turns out that $\Omega_c^0\to\Xi^0 \ov K^0$ is the only allowed Cabibbo-favored $\B(1/2^+)P$ decay mode. A measurement of this mode in the near future is very promising as the predicted branching fraction is of order 3.8\% \cite{Xu:Omegac}.

\subsection{Charm-flavor-conserving nonleptonic decays}
There is a special class of weak decays of charmed baryons in which the $S$-wave amplitude
can be theoretically studied in a reliable way, namely, heavy-flavor-conserving (HFC)
nonleptonic decays \cite{Cheng:HFC}. Some examples are the singly
Cabibbo-suppressed decays $\Xi_c\to\Lambda_c\pi$ and
$\Omega_c\to\Xi_c\pi$. \footnote{The decays $\Omega_Q\to \Xi'_Q\pi$ $(Q=c,b)$ are kinematically prohibited.}
Since the emitted light mesons are soft in this special class of weak decays, the nonfactorizable $S$-wave amplitude can be evaluated reliably using current algebra.
Heavy-flavor-conserving weak decays $\Xi_Q\to\Lambda_Q\pi$
with $Q=c,b$  have been studied in \cite{Sinha:HFC,Voloshin00,Faller,Gronau:2015jgh,Gronau:HFC,Cheng:HFC2016,%
Voloshin:2019,Groote:2021pxt,Niu:2021qcc,Cheng:HFC2022}.

As for the $P$-wave amplitude, if only
the light quarks inside the heavy baryon participate in weak
interactions and the heavy quark behaves as a ``spectator", then the $P$-wave amplitude will vanish in the heavy quark limit. The argument goes as follows.
The combined symmetries of heavy and light quarks severely
restrict the weak interactions allowed. In the symmetry limit and in the absence of heavy quark's participation in weak interactions, it
was found that ${\cal B}_{\bar 3}-{\cal B}_6$ and
${\cal B}^*_6-{\cal B}_6$ nonleptonic weak transitions  cannot occur \cite{Cheng:HFC}, where
${\cal B}_{\bar 3}$ and ${\cal B}_6$ are antitriplet and sextet heavy baryons, respectively,
and ${\cal B}^*_6$ the spin-3/2 heavy baryon field.
Symmetries alone permit three types of transitions:  ${\cal B}_{\bar 3}-{\cal
B}_{\bar 3}$,  ${\cal B}_{6}-{\cal B}_6$ and ${\cal B}^*_6-{\cal
B}_6$ transitions. However, in both the MIT bag and diquark
models, only ${\cal B}_{\bar 3}-{\cal B}_{\bar 3}$ transitions
have nonzero amplitudes. The general amplitude for $\B_i\to \B_f+P$ is given by Eq. (\ref{eq:generalAmp}) with $A$ and $B$ being the $S$- and $P$-wave amplitudes, respectively.
In the heavy quark limit, the diquark of the antitriplet baryon $\B_{\bar 3}$ is a scalar one with $J^P=0^+$. If only the light quarks inside the heavy baryon participate in weak
interactions, the weak diquark transition will be $0^+\to 0^+ + 0^-$ for  $\B_{\bar 3}\to \B_{\bar 3}+P$. Based on the conservation of angular momentum, it is easily seen that the parity-conserving $P$-wave amplitude vanishes in $\B_{\bar 3}\to \B_{\bar 3}+P$ decays. Hence, the $P$-wave amplitude of the $\B_{\bar 3}\to \B_{\bar 3}+P$ decay vanishes in the heavy quark limit provided that the heavy quark behaves as a spectator \cite{Cheng:HFC}.

However, both charm-flavor-conserving decays $\Xi_c^0\to \Lambda_c^+\pi^-$ and $\Xi_c^+\to \Lambda_c^+\pi^0$ receive contributions from the $W$-exchange diagrams  through the transition
$cs\to dc$. Although the charm flavor is still conserved in this process, the charm quark  does participate in  weak interactions and hence $P$-wave does not diminish even in the heavy quark limit. We will come back to this point later.

To proceed,
consider the decay $\Xi_c^0\to \Lambda_c^+\pi^-$ specifically.
The factorizable $S$-wave amplitude through external $W$-emission reads
\be
A(\Xi_c^0\to \Lambda_c^+\pi^-)^{\rm fac} &=& {G_F\over\sqrt{2}}V_{ud}^*V_{us}a_1\la \pi^-|(\bar du)|0\ra\la \Lambda_c^+|(\bar us)|\Xi_c^0\ra, \non \\
&=&
-{G_F\over\sqrt{2}}V_{ud}^*V_{us}a_1\,f_\pi (m_{\Xi_c}-m_{\Lambda_c})f_1^{\Lambda_c\Xi_c^0},
\en
where $a_1$ has the same expression as Eq. (\ref{eq:a2}) with $c_1$ and $c_2$ interchanged,
and $f_1^{\Lambda_c\Xi_c}$ is one of the $\Lambda_c -\Xi_c$ transition form factors.
Since the pion produced in the heavy-flavor-conserving decay is soft, $A^{\rm nf}$ can be evaluated using current algebra with the result
\begin{eqnarray} \label{eq:Anf}
A(\Xi_c^0\to\Lambda_c^+\pi^-)^{\rm nf}= {1\over f_\pi}\la \Lambda_c^+|{\cal H}_{\rm eff}|\Xi_c^+\ra+{1\over f_\pi}\la \Lambda_c^+|\tilde {\cal H}_{\rm eff}|\Xi_c^+\ra,
\end{eqnarray}
where the effective Hamiltonians are given by
\be
{\cal H}_{\rm eff}={G_F\over \sqrt{2}}V^*_{ud}V_{us}(c_1O_1+c_2O_2), \qquad
{\cal H}^{(c)}_{\rm eff}={G_F\over \sqrt{2}}V^*_{cd}V_{cs}(c_1\tilde O_1+c_2\tilde O_2),
\en
with $O_1=(\bar du)(\bar us)$, $O_2=(\bar ds)(\bar uu)$, $\tilde O_1=(\bar dc)(\bar cs)$, $\tilde O_2=(\bar cc)(\bar d s)$ and $(\bar q_1q_2)\equiv \bar q_1\gamma_\mu(1-\gamma_5)q_2$.
Eq. (\ref{eq:Anf}) describes  the $W$-exchange contributions through the transitions $su\to ud$
and $cs\to dc$. It is straightforward to show that
\be \label{eq:AmpHFC_2}
\left. \begin{array}{c}  A^{\rm nf}_{su\to ud} \\
A^{\rm nf}_{cs\to cd} \end{array} \right\} ={G_F\over 2\sqrt{2} f_\pi}V_{ud}^*V_{us}(c_1-c_2) \left\{ \begin{array}{c} X \\ -Y, \end{array} \right.
\en
with
\be \label{eq:Y2Z2}
X &\equiv& \la \Lambda_c^+|(\bar du)(\bar us)-(\bar uu)(\bar ds)|\Xi_c^+\ra , \non \\
Y &\equiv& \la \Lambda_c^+|(\bar dc)(\bar cs)-(\bar cc)(\bar ds)|\Xi_c^+\ra.
\en
The minus sign in front of $Y$ in Eq. (\ref{eq:AmpHFC_2}) comes the the relation for the CKM matrix elements $V_{cd}^*V_{cs}= -V_{ud}^*V_{us}$ to a very good approximation.
The final expression \cite{Cheng:HFC2016}
\be \label{eq:AmpHFC}
A(\Xi_c^0\to\Lambda_c^+\pi^-) &=& A^{\rm fac}+A^{\rm nf}_{su \to ud} + A^{\rm nf}_{sc\to cd}  \\
&=& {G_F\over\sqrt{2} f_\pi}V_{ud}^*V_{us}\left[ -a_1f_\pi^2(m_{\Xi_c}-m_{\Lambda_c})f_1^{\Lambda_c\Xi_c^0}+{1\over 2}(c_1-c_2)\left(X-Y\right)\right] \non
\en
is quite general and model independent. The remaining task is to evaluate the matrix elements $X$ and $Y$.
For given the flavor wave functions of $\Lambda_c$ and $\Xi_c$, the relative signs between various contributions are fixed. For example, we found
$f_1^{\Lambda_c\Xi_c^0}=0.985$ and $X=0.017\,{\rm GeV}^3$, so
that the amplitudes of $A^{\rm fac}$ and  $A^{\rm nf}_{su\to  ud}$ are of the opposite sign.

We now turn to the party-conserving amplitude. The factorizable $P$-wave reads
\be
B(\Xi_c^0\to \Lambda_c^+\pi^-)^{\rm fac} =
{G_F\over\sqrt{2}}V_{ud}^*V_{us}a_1\,f_\pi (m_{\Xi_c}+m_{\Lambda_c})g_1^{\Lambda_c\Xi_c^0}.
\en
However, the form factor $g_1^{\Lambda_c\Xi_c}$ vanishes in the heavy quark limit \cite{Yan}, also confirmed in the quark  model calculation. The $P$-wave amplitude arising from the $W$-exchange through $su\to ud$ transition vanishes either from the argument based on heavy quark symmetry and chiral symmetry or from explicit calculations.
Because of the additional nonspectator contribution from the $W$-exchange through $cs\to dc$, the nonfactorizable $P$-wave amplitude does not vanish in the limit of heavy quark symmetry as pointed out recently in \cite{Groote:2021pxt,Niu:2021qcc}. Explicitly,
\be \label{eq:Pwavenet}
B(\Xi_c^0\to \Lambda_c^+\pi^-)^{\rm pole} &=& - {1\over f_\pi}\left(
g_{\Lambda_c^+\Sigma_c^0}^{A(\pi^-)}{m_{\Lambda_c^+}+m_{\Sigma_c^0}\over m_{\Xi_c^0}-m_{\Sigma_c^0}}a_{\Sigma_c^0 \Xi_c^0 }
+  a_{\Lambda_c^+ {\Xi'}_c^+}{m_{\Xi_c^0}+m_{{\Xi'}_c^+}\over m_{\Lambda_c^+}-m_{{\Xi'}_c^+}}g_{{\Xi'}_c^+\Xi_c^0}^{A(\pi^-)}\right),   \\
B(\Xi_c^+\to \Lambda_c^+\pi^0)^{\rm pole} &=& - {\sqrt{2}\over f_\pi}\left(
g_{\Lambda_c^+\Sigma_c^+}^{A(\pi^0)}{m_{\Lambda_c^+}+m_{\Sigma_c^+}\over m_{\Xi_c^+}-m_{\Sigma_c^+}}a_{\Sigma_c^+ \Xi_c^+ }
+ a_{\Lambda_c^+ {\Xi'}_c^+}{m_{\Xi_c^+}+m_{{\Xi'}_c^+}\over m_{\Lambda_c^+}-m_{{\Xi'}_c^+}}g_{{\Xi'}_c^+\Xi_c^+}^{A(\pi^0)} \right), \non
\en
with
\be
a_{\Sigma_c^{0(+)} \Xi_c^{0(+)} }=\la \Sigma_c^{0(+)}|{\cal H}^{(c)}_{\rm eff}| \Xi_c^{0(+)}\ra, \qquad
 a_{\Lambda_c^+ {\Xi'}_c^+}= \la \Lambda_c^+ |{\cal H}^{(c)}_{\rm eff}|{\Xi'}_c^+ \ra.
\en
Since the masses of $\Xi_c$ and $\Sigma_c$ are very close, of order 16 MeV, it is obvious that the small mass difference between $\Xi_c$ and the $\Sigma_c$ pole gives rise to $(m_{\Lambda_c^+}+m_{\Sigma_c^+})/(m_{\Xi_c^+}-m_{\Sigma_c^+})=315$ and
$(m_{\Lambda_c^+}+m_{\Sigma_c^0})/(m_{\Xi_c^0}-m_{\Sigma_c^0})=276$,
which in turn lead to a strong enhancement of the $P$-wave pole amplitude \cite{Niu:2021qcc,Groote:2021pxt}.

For the charmed baryon $\Omega_c$, its diquark is an axial-vector one with $J^P=1^+$. Therefore, if the heavy quark $Q$ behaves as a spectator, the weak diquark transition of the HFC decay $\Omega_c\to \Xi'_c\pi$ will be $1^+\to 1^++ 0^-$. Then both $S$- and $P$-wave transitions receive factorizable contributions as pointed out in \cite{Cheng:HFC}. However, $\Omega_Q\to \Xi'_Q\pi$ are not kinematically allowed for both $Q=c$ and $b$. For $\Omega_c\to \Xi_c\pi$ decays, the diquark transition is $1^+\to 0^+ + 0^-$. This implies that contrary to $\Xi_c\to\Lambda_c\pi$ decays with the diquark weak transition $0^+\to 0^+ + 0^-$, the $S$-wave vanishes in the heavy quark limit provided that the heavy quark does not
participate in weak interactions. Just like $\Xi_c\to\Lambda_c\pi$, $\Omega_c\to \Xi_c\pi$ decays acquire additional contributions from nonspectator $W$-exchange for both $S$- and $P$-wave amplitudes. The $P$-wave transition is enhanced by the ${\Xi'}_c$ pole, though it is not so dramatic as in the case of $\Xi_c\to\Lambda_c\pi$ \cite{Cheng:HFC2022}.

The calculated $S$- and $P$-wave amplitudes, branching fractions and decay asymmetries of $\Xi_c\to \Lambda_c^+\pi$ and $\Omega_c^0\to\Xi_c\pi$ are displayed
in Table \ref{tab:HFC:AmpandBF}.
If the baryon matrix elements of 4 light-quark operators denoted by $X$ are evaluated in the bag model, the predicted branching fraction $\B(\Xi_b^-\to \Lambda_b^0\pi^-)=6.6\times 10^{-4}$ will be too small compared to the LHCb measurement given in Eq. (\ref{eq:BFXib LHCB}). Hence, we use the diquark model to estimate the baryon matrix element $X$. Nevertheless, we still use the bag model to evaluate the baryon matrix element of 4-quark oprtators involving $c$ and $\bar c$ as the diquark model is not applicable in this case.

We have confirmed that $\Xi_c\to\Lambda_c\pi$ decays are indeed dominated by the parity-conserving transition induced from  nonspectator $W$-exchange and that they receive largest contributions from the intermediate $\Sigma_c$ pole terms as stressed in \cite{Niu:2021qcc}.
We have also showed that the $S$-wave of $\Omega_b\to \Xi_b \pi$ decays vanishes in the heavy quark limit, while $\Omega_c\to \Xi_c\pi$ receive additional $W$-exchange contributions via $cs\to dc$ transition.
The asymmetry parameter $\alpha$ is found to be positive, of order $0.70$ and $0.74$ for
$\Xi_c^0\to \Lambda_c^+\pi^-$ and $\Xi_c^+\to \Lambda_c^+\pi^0$, respectively.
The predicted branching fraction is of order $5\times 10^{-4}$ for $\Omega_c^0\to\Xi_c^+\pi^-$ and $3\times 10^{-4}$ for $\Omega_c^0\to\Xi_c^0\pi^0$ both with the asymmetry parameter close to $-1$.

\begin{table}[t]
\begin{ruledtabular}
 \caption{The magnitude of $S$- and $P$-wave amplitudes (in units of $10^{-7}$), branching fractions and decay asymmetries of the HFC decays $\Xi_c\to\Lambda_c^+\pi$ and $\Omega_c^0\to\Xi_c\pi$. Uncertainties arise from the matrix element $X$ estimated in the diquark model.
 Experimental measurements are taken from Ref. \cite{LHCb:HFC}.
 } \label{tab:HFC:AmpandBF}
 \vspace{6pt}
\begin{tabular}
{ l c c c c  }
   & $\Xi_c^0\to \Lambda_c^+\pi^-$ & $\Xi_c^+\to \Lambda_c^+\pi^0$ & $\Omega_c^0\to\Xi_c^+\pi^-$ & $\Omega_c^0\to \Xi_c^0\pi^0$ \\
 \hline
 $A$ & $2.53\pm 0.75$ & $2.02\pm0.53$ & $-1.72$ & $1.21$ \\
 $B$ & $\!\!244$ & $\!\!181$ & 40.12 & $-32.19$ \\
 \hline
  $\alpha$ & $0.70^{+0.13}_{-0.17}$ & $0.74^{+0.11}_{-0.16}$ & $-0.98$ & $-0.99$ \\
 $\B$ & $(1.76^{+0.18}_{-0.12})\times 10^{-3}$ & $(3.03^{+0.29}_{-0.22})\times 10^{-3}$  & $5.1\times 10^{-4}$ & $2.8\times 10^{-4}$  \\
  $\B_{\rm expt}$ & $(5.5\pm1.8)\times 10^{-3}$ & -- & --  & -- \\
\end{tabular}
\end{ruledtabular}
\end{table}

\begin{table}[tp!]
\begin{ruledtabular}
 \caption{Branching fractions (in units of $10^{-3}$) of charm-flavor-conserving decays $\Xi_c\to\Lambda_c^+\pi$. All the model results
 have been normalized using the current world averages of lifetimes for $\Xi_c^+$ and $\Xi_c^0$ (see Table \ref{tab:expt_lifetimes}). As for the predictions of \cite{Gronau:HFC}, the first (second) entry is for destructive (constructive) interference between $A^{\rm fac}+A^{\rm nf}_{su\to ud}$ and  $A^{\rm nf}_{sc\to ud}$ amplitudes.
 } \label{tab:ComparisonBR}
\vspace{6pt}
\footnotesize{
\begin{tabular}
{ l c c c c c c c c}
 Mode  & (CLY)$^2_a$ & (CLY)$^2_b$ & Faller & Gronau & Voloshin & Niu & Cheng & Experiment \\
 & \cite{Cheng:HFC} & \cite{Cheng:HFC2016} & \cite{Faller} & \cite{Gronau:HFC} & \cite{Voloshin:2019} &  \cite{Niu:2021qcc} & \cite{Cheng:HFC2022} & \cite{LHCb:HFC} \\
 \hline
$\Xi_{c}^{0}\to\Lambda_c^+\pi^-$ & $0.39$   & $0.17$  & $<3.9$ & $0.18^{+0.23}_{-0.13}$ & $>\!0.25\pm0.15$  & $5.8\pm2.1$ & $1.76^{+0.18}_{-0.12}$  & $5.5\pm0.2\pm1.8$  \\
&    &   &  & $1.34\pm0.53$ &   &  &  &   \\
$\Xi_{c}^{+}\to\Lambda_c^+\pi^0$ & $0.69$   & $0.11$  & $<6.1$ & $<\!0.2$ & -- & $11.1\pm4.0$ & $3.03^{+0.29}_{-0.22}$ & --\\
&  &   & & $2.01\pm0.80$ & & & \\
\end{tabular}
}
 \end{ruledtabular}
 \end{table}

Various model predictions for $\Xi_c\to\Lambda_c^+\pi$ are compared in Table \ref{tab:ComparisonBR}. It is clear that all the previous predictions excluding $P$-wave contributions are too small compared to experiment. In Ref. \cite{Gronau:HFC}, Gronau and Rosner (GR) denoted $A^{\rm fac}+A^{\rm nf}_{su\to ud}$ by $A_{s\to u\bar ud}$ and made the assumption that
$A_{s\to u\bar ud}(\Xi_c^0\to\Lambda_c^+\pi^-)=A(\Xi_b^-\to\Lambda_b^0\pi^-)$.
In our case, we performed a dynamic calculation of both $A^{\rm fac}$ and $A^{\rm nf}_{s u\to ud}$. The resultant value of $A_{s\to u\bar ud}$ is consistent with that of GR (see Table I of \cite{Cheng:HFC2022}). It is the $W$-exchange term $A^{\rm nf}_{cs\to dc}$ that makes a huge difference. As the relative sign between $A(\Xi_b^-\to\Lambda_b^0\pi^-)$ and
$A^{\rm nf}_{cs\to dc}$ is not fixed, GR considered both constructive and destructive interference. In our dynamic calculation, it is evident that the interference is destructive because of the relation $V_{cd}^*V_{cs}= -V_{ud}^*V_{us}$ for CKM matrix elements. In the case of destructive interference, both this work and GR have similar branching fractions of order $10^{-4}$ from the $S$-wave alone.

In the $b$ sector, the bottom-flavor-conserving decay
$\Xi_b^-\to\Lambda_b^0\pi^-$ was first measured by the LHCb with the branching fraction
falling in the range \cite{LHCb:HFCb}
\be \label{eq:BFXib LHCB}
\B(\Xi_b^-\to\Lambda_b^0\pi^-)=(0.57\pm0.21)\sim(0.19\pm0.07)\%.
\en
Unlike the charm-flavor-conserving decay $\Xi_c^0\to\Lambda_c^+\pi^-$, it does not receive a contribution from $W$-exchange. Consequently, it is safe to neglect the $P$-wave amplitudes in the bottom-flavor-conserving decay and the calculation is greatly simplified. Explicitly,
\be
A(\Xi_b^-\to\Lambda_b^0\pi^-) = {G_F\over\sqrt{2} f_\pi}V_{ud}^*V_{us}\left[ -a_1f_\pi^2(m_{\Xi_b}-m_{\Lambda_b})f_1^{\Lambda_b\Xi_b}+{1\over 2}(c_1-c_2)X \right].
\en
We found $\B(\Xi_b^-\to\Lambda_b^0\pi^-)=6.6\times 10^{-4}$ in the bag model and $(4.7^{+2.3}_{-1.8})\times 10^{-3}$ in the diquark model estimate of the matrix element $X$ \cite{Cheng:HFC2022}.

\subsection{Nonleptonic decays of doubly charmed baryons}
Since the discovery of the doubly charmed baryon $\Xi_{cc}^{++}$ in 2017 \cite{LHCb:Xiccpp}, LHCb has proceeded to measure its lifetime \cite{LHCb:tauXiccpp} and observed the hadronic decay $\Xi_{cc}^{++}\to \Xi_c^+\pi^+$ with the result which has a
significance of 5.9 standard devitations \cite{LHCb:XiccDecay}
\begin{eqnarray} \label{eq:Xicpi}
 { {\cal B}(\Xi_{cc}^{++}\to \Xi_c^+\pi^+){\cal B}(\Xi_c^+\to pK^-\pi^+)\over
 {\cal B}(\Xi_{cc}^{++}\to \Lambda_c^+K^-\pi^+\pi^+){\cal B}(\Lambda_c^+\to pK^-\pi^+)}=0.035\pm0.009\pm0.003\,.
\end{eqnarray}
Therefore, it is important to investigate the nonleptonic weak decays of the doubly charmed baryons, $\Xi_{cc}^{++},\Xi_{cc}^+$ and $\Omega_{cc}^+$.

One unique feature of the doubly charmed baryons is that there exist certain channels which proceed only through the factorizable diagrams. Examples are $\Xi_{cc}^{++}\to \Sigma_c^{++}\overline{K}^0,\Xi_c^{\prime +}\pi^+$, $\Xi_{cc}^+\to\Sigma_c^+\eta^{(')}$, $\Omega_{cc}^+\to\Omega_c^0\pi^+,\Xi_c^{\prime +}\overline{K}^0$. Indeed, the state $\Xi_{cc}^{++}$ was discovered in the four-body channel $\Lambda_c^+K^-\pi^+\pi^+$ followed from the decay chain $\Xi_{cc}^{++}\to\Sigma_c^{++}(\to\Lambda_c^+\pi^+)\overline{K}^{*0}(\to K^-\pi^+)$ as originally suggested in \cite{Yu:Xiccpp}. Moreover, we notice that (i) the $W$-exchange contribution manifests only in $\Xi_{cc}^+$ decays, and (ii)
the inner $W$-emission amplitude $C'$ (see Fig. \ref{fig:Bic}) in $\Xi_{cc}^{++}\to \Xi_c^{'+}\pi^+$, $\Xi_{cc}^+\to \Xi_c^{'+}(\pi^0,\eta)$ and $\Omega_{cc}^+\to\Xi_c^{'+}\overline{K}^0$ should vanish
because of the Pati-Woo theorem \cite{Pati:1970fg} which results from the facts that the $(V-A)\times(V-A)$ structure of weak interactions is invariant under the Fierz transformation and that the baryon wave function is color antisymmetric. This theorem requires that the quark
pair in a baryon produced by weak interactions be antisymmetric in flavor. Since the sextet
$\Xi'_c$ is symmetric in light quark flavor, it cannot contribute to $C'$. This feature is indeed confirmed in realistic calculations \cite{Cheng:doublycharm}.

\begin{table}[!]
\begin{center}
\scriptsize{
\caption{Predicted branching fractions (in $\%$) of Cabibbo-favored doubly charmed baryon decays by different groups.
For the predictions of Dhir and Sharma \cite{Sharma:2017txj,Dhir:2018twm}, we quote the flavor-independent pole amplitudes and two different models for $\B_{cc}\to\B_c$ transition form factors: nonrelativistic quark  model (abbreviated as N) and heavy quark effective theory (H).
For the results of Gutsche {\it et al.} \cite{Gutsche:2018msz,Gutsche:2017hux,Gutsche:2019iac}, we quote the latest ones from \cite{Gutsche:2019iac}. The long-distance contributions calculated in \cite{Han:doublycharm} depend on the parameter $\eta$. We quote the results corresponding to $\eta=1.5$\,.
All the model results have been normalized using the lifetimes
$\tau(\Xi_{cc}^{++})=2.56\times10^{-13}s$, $\tau(\Xi_{cc}^{+})=0.45\times10^{-13}s$ and $\tau(\Omega_{cc}^+)=1.28\times10^{-13}s$.}\label{tab:Xicc}
 \vspace{0.3cm}
\begin{ruledtabular}
\begin{tabular}{ l c c c c c c c c }
  Mode &  Cheng  & Dhir  & Gutsche \textit{et al.} & Wang  & Gerasimov & Ke & Shi & Han \\
  & \textit{et al.} \cite{Cheng:doublycharm} & \textit{et al.} \cite{Sharma:2017txj,Dhir:2018twm} &
   \cite{Gutsche:2019iac} & \textit{et al.} \cite{Wang:doublycharm}   & \textit{et al.} \cite{Gerasimov:2019jwp} & \textit{et al.} \cite{Ke:2019lcf} & \textit{et al.} \cite{Shi:2019hbf} & \textit{et al.} \cite{Han:doublycharm} \\
\hline
$\Xi_{cc}^{++}\to \Xi_c^{+}\pi^+$  & $0.69$ & 6.64 (N) & 0.70 & 6.18  & 7.01 & $3.48\pm0.46$ & $3.1\pm0.4$ & 8.48 \\
&  & 9.19 (H) & & & &  \\
$\Xi_{cc}^{++}\to \Xi_c^{'+}\pi^+$  & 4.65 & 5.39 (N) & 3.03 & 4.33  & 5.85 & $1.96\pm0.24$ & $0.93\pm0.19$ & 4.72 \\
&  & 7.34 (H) & & &  \\
$\Xi_{cc}^{++}\to \Sigma_c^{++}\overline{K}^0$ & 1.36 & 2.39 (N) &  $1.25$ & & & & & 1.91 \\
&  & 4.69 (H) & & &  & \\
\hline
$\Xi_{cc}^{+} \to \Xi_{c}^{0}\pi^+$ &  $3.84$   & 0.59 (N) & & 1.08   & 1.23 & $0.61\pm0.08$ & $0.53\pm0.08$ & 1.47 \\
&  & 0.95 (H) & & & &  \\
$\Xi_{cc}^{+} \to \Xi_{c}^{'0}\pi^+$  & $1.55$   &  1.49 (N) &  & 0.76  & 1.04 & $0.35\pm0.04$ & $0.16\pm0.03$ & 0.91 \\
&  & 2.12 (H) & & & &  \\
$\Xi_{cc}^{+} \to \Lambda_c^+\overline{K}^0$  & $0.31$ & 0.27 (N) &  & & & & & 0.32    \\
&  & 0.37 (H) & & & &  \\
$\Xi_{cc}^{+} \to \Sigma^+_c\overline{K}^0$  & $0.38$  & 0.59 (N) &  & & & & & 1.52  \\
&  & 0.90 (H) & & & &  \\
$\Xi_{cc}^{+} \to \Xi^{+}_{c}\pi^0$  & $2.38$  &  0.50 &  & & & & & 3.28    \\
$\Xi_{cc}^{+} \to \Xi^{'+}_{c}\pi^0$   & $0.17$   & 0.054 &  & & & & & 0.32     \\
$\Xi_{cc}^{+} \to \Xi_{c}^{+} \eta$& $4.18$  &  0.063  &  & & & & & 0.11  \\
$\Xi_{cc}^{+} \to \Xi_{c}^{'+} \eta$& $0.05$  &  0.036  &  & & & & & 0.21  \\
$\Xi_{cc}^{+} \to \Sigma _c^{++}K^{-}$  & $0.13$  &  0.22 & & & & & & 0.08   \\
$\Xi_{cc}^{+} \to \Omega_c^0 K^+$  & 0.06 & 0.10 & & & & & & 0.05 \\
\hline
$\Omega_{cc}^+ \to \Omega^{0}_{c}\pi^+$ & 3.96 & 5.38 (N) & 3.08 & 3.34   & 5.30 & & $0.55\pm0.23$ & 4.33 \\
&  & 7.34 (H) & & &  \\
$\Omega_{cc}^{+}\to \Xi_c^{+}\overline{K}^0$  & 1.15 & 1.36 (N) &  1.98 & & & & &  1.64 \\
&  & 2.10 (H) & & & & & &  \\
$\Omega_{cc}^{+}\to \Xi_c^{'+}\overline{K}^0$  & 0.29 & 0.61 (N) &  0.31 & & & & &  2.40  \\
&  & 1.19 (H)& & & & & &  \\
\end{tabular}
\end{ruledtabular}}
\end{center}
\end{table}

Theoretical predictions available in the literature are collected in Table \ref{tab:Xicc}.
The channel $\Xi_{cc}^{++}\to \Xi_c^+\pi^+$ was the first two-body decay mode observed by the LHCb in the doubly charmed baryon sector. However, the prediction of $\sim 0.7\%$ in \cite{Cheng:doublycharm,Gutsche:2018msz} for its branching fraction is substantially smaller than the results of $(3\sim 9)\%$ given in the literature. This is ascribed to the destructive interference between factorizable and nonfactorizable contributions (i.e. external and internal $W$-emission diagrams) for both $S$- and $P$-wave amplitudes.
If we turn off the nonfactorizable terms, we will have a branching fraction of order $3.6\%$.
It is very interesting to notice that both calculations agree with each other
even though the estimation of nonfactorizable effects is based on entirely different approaches: current algebra and the pole model in \cite{Cheng:doublycharm} and the covariant confined quark model in \cite{Gutsche:2018msz}.
Nonfactorizable effects in $\Xi_{cc}^{++}\to \Xi_c^+\pi^+$  also have been considered in \cite{Han:doublycharm} where long-distance contributions were accounted for by final-state rescattering and partially in \cite{Sharma:2017txj} where only the nonfactorizable contribution to the $P$-wave amplitude was considered. Both found constructive interference.

Using the results $\B(\Lambda_c^+\to p K^-\pi^+)=(6.28\pm0.32)\%$ \cite{PDG} and Eq. (\ref{eq:XicAbsolutB}) for the branching fraction of $\Xi_c^+\to p K^-\pi^+$, it follows from Eq. (\ref{eq:Xicpi}) that
\footnote{The ratio of branching fractions will become smaller if the LHCb measurement of $\B(\Xi_c^+\to p K^-\pi^+)$, Eq. (\ref{eq:XicAbsolutB_LHCb}), is employed.}
\begin{equation}
{\mathcal{B}(\Xi_{cc}^{++}\to\Xi_c^+\pi^+) \over
\mathcal{B}(\Xi_{cc}^{++}\to\Lambda_c^+ K^- \pi^+\pi^+)}=0.49\pm0.27\,,
\end{equation}
where the uncertainty is dominated by the decay rate of $\Xi_c^+$ into $p K^-\pi^+$. Making the plausible assumption that $\B(\Xi_{cc}^{++}\to\Lambda_c^+ K^- \pi^+\pi^+)\approx {2\over 3}\B(
\Xi_{cc}^{++}\to\Sigma_c^{++}\overline{K}^{*0})$ and taking the latest prediction $\B(\Xi_{cc}^{++}\to\Sigma_c^{++}\overline{K}^{*0})=5.61\%$ from \cite{Gutsche:2019iac} as an example,
\footnote{The branching fraction is given by $(5.40^{+5.59}_{-3.66})\%$ in the approach of final-state rescattering \cite{Jiang:2018oak}.}
we obtain
\be
\mathcal{B}(\Xi_{cc}^{++}\to\Xi_c^+\pi^+)_{\rm expt}\approx (1.83\pm1.01)\%.
\en
Therefore, our prediction of $\mathcal{B}(\Xi_{cc}^{++}\to\Xi_c^+\pi^+)\approx 0.7\%$ is consistent with the experimental value but in the lower end. In future study, it is important to pin down the branching fraction of this mode both experimentally and theoretically.

Nonfactorizable contributions to $\Xi_{cc}^{++}\to \Xi_c^{'+}\pi^+$  were found to be very small compared to the factorizable ones in \cite{Cheng:doublycharm,Gutsche:2018msz,Han:doublycharm}. This is consistent with the Pati-Woo theorem for the inner $W$-emission amplitude as noticed in passing. Although this was also found to be true for the decay $\Omega_{cc}^+\to \Xi_c^{'+}\overline{K}^0$ in \cite{Cheng:doublycharm,Gutsche:2018msz}, the authors of \cite{Han:doublycharm} claimed that
this mode is dominated by the nonfactorizable long-distance effects (see Table IV of \cite{Han:doublycharm}). However, this will violate the Pati-Woo theorem.

\vskip 0.2cm
{\bf Search for the $\Xi_{cc}^+$ }
\vskip 0.2cm

A search for the doubly charmed baryon $\Xi_{cc}^+$ has been performed through its decay to the $\Lambda_c^+K^-\pi^+$ \cite{LHCb:Xiccp} and $\Xi_c^+\pi^-\pi^+$ \cite{LHCb:Xiccp_2} final states.
In view of the sizable branching fraction for $\Xi_{cc}^{+}\to\Xi_c^0\pi^+$ (see Table \ref{tab:Xicc}), we have proposed in \cite{Cheng:doublycharm} that experimentalists could try to search for the $\Xi_{cc}^+$ through this mode.
That is, $\Xi_{cc}^+$ is reconstructed through the $\Xi_{cc}^+\to\Xi_c^0\pi^+$ followed by the decay chain $\Xi_c^0\to \Xi^-\pi^+\to p\pi^-\pi^-\pi^+$. We have estimated in \cite{Cheng:doublycharm} that the branching fraction of $\Xi_{cc}^+\to\Lambda_c^+K^-\pi^+$ is not more than $0.8\%$. From \cite{Jiang:2018oak}, it follows that the branching fraction of $\Xi_{cc}^+\to
\Xi_c^+\pi^+\pi^-$ assuming to be dominated by the resonant contribution from $\Xi_{cc}^+\to
\Xi_c^+\rho^0$ is also small, of order $0.7\times 10^{-3}$. Therefore, it is worthwhile to search for the $\Xi_{cc}^+$ through its decay to $\Xi_c^0\pi^+$.

\subsection{Pole model versus current algebra}

As stated in Sec. IV.A, the evaluation of nonfactorizable contributions to $\B_{c}\to \B+M$ decays arising from the topological diagrams such as inner $W$-emission $C'$ and $W$-exchange
$E_1,E_2$ and $E_3$ (see Fig. \ref{fig:Bic}) often relies on the pole model in which one considers the contributions from all possible intermediate states. Among all possible pole contributions, including
resonances and continuum states, one usually focuses on the most important poles such as the low-lying ${1\over 2}^+,{1\over 2}^-$ baryon states, known as pole approximation.  More specifically,  the $S$-wave amplitude is dominated by the low-lying $1/2^-$
resonances and the $P$-wave one governed by the ground-state
$1/2^+$ poles (see Fig. \ref{fig:pole}): \footnote{Meson pole contributions are usually classified as factorizable amplitudes. Since the latter are normally calculated using the QCD-corrected effective Hamiltonian, we will not take meson-pole contributions into account to avoid the double counting problem.}
 \begin{eqnarray} \label{eq:poleamp}
 A^{\rm nf} &=& -\sum_{{\cal B}^*_n(1/2^-)}\left({g_{_{{\cal B}_f{\cal B}_{n^*}
P}}b_{_{n^*i}}\over m_i-m_{n^*}}+{b_{_{fn^*}}g_{_{{\cal
B}_{_{n^*}}{\cal B}_iP}}\over m_f-m_ {n^*}}\right)+\cdots,
\nonumber \\
 B^{\rm nf} &=& \sum_{{\cal B}_n}\left({g_{_{{\cal B}_f{\cal B}_nP}}
 a_{_{ni}}\over m_i-m_n}+{a_{_{fn}}g_{_
{{\cal B}_n{\cal B}_iP}}\over m_f-m_n}\right)+\cdots,
 \end{eqnarray}
where $A^{\rm nf}$ and $B^{\rm nf}$ are the nonfactorizable $S$-
and $P$-wave amplitudes of ${\cal B}_c\to{\cal B}P$ (c.f. Eq. (\ref{eq:A&B})), respectively,
ellipses in Eq. (\ref{eq:poleamp}) denote other pole contributions
which are negligible for our purposes, and $a_{ij}$ as well as
$b_{i^*j}$ are the baryon-baryon matrix elements defined by
 \begin{eqnarray}
\langle {\cal B}_i|{\cal H}_{_W}|{\cal B}_j\rangle =
\,\bar{u}_i(a_{ij}-b_{ij} \gamma_5 )u_j,  \qquad \langle{\cal
B}^*_i(1/2^{^-})|{\cal H}^{\rm pv}_{_W}|{\cal B}_j\rangle =
\,b_{i^*j}\, \bar{u}_iu_j,
 \end{eqnarray}
with $b_{ji^*}=-b_{i^*j}$. Evidently, the calculation of $S$-wave
amplitudes is far more difficult than the $P$-wave ones owing to the
troublesome negative-parity baryon resonances which are not well
understood in the quark model.  In \cite{CT92,CT93}, we have applied
the MIT bag model \cite{MIT} to evaluate the baryon transition matrix elements $a_{ij}$, $b_{i^*j}$ and the relevant strong coupling constants.
Unfortunately, the pole model calculations were not successful.\footnote{None of the pole model calculations in 1990s other than \cite{CT92,CT93} had evaluated the $1/2^--1/2^+$ transition matrix elements in terms of the wave functions of $1/2^-$ baryon resonances.}
From Table \ref{tab:CF} we see that the predicted branching fractions were generally too small compared to experiment. Moreover, the predicted decay asymmetry parameter $\alpha$ in $\Lambda_c^+\to \Sigma \pi$ decays was wrong in sign. This is a long-standing puzzle for the pole model. The failure of the pole-model
calculations in 1990s was ascribed to the fact that the bag model
can reproduce the qualitative behavior of the spectrum
of negative-parity baryon resonances, but is not so successful
in reproducing the quantitative features of the observed
spectroscopy  \cite{DeGrand}.

\begin{figure}[t]
\centerline{\psfig{file=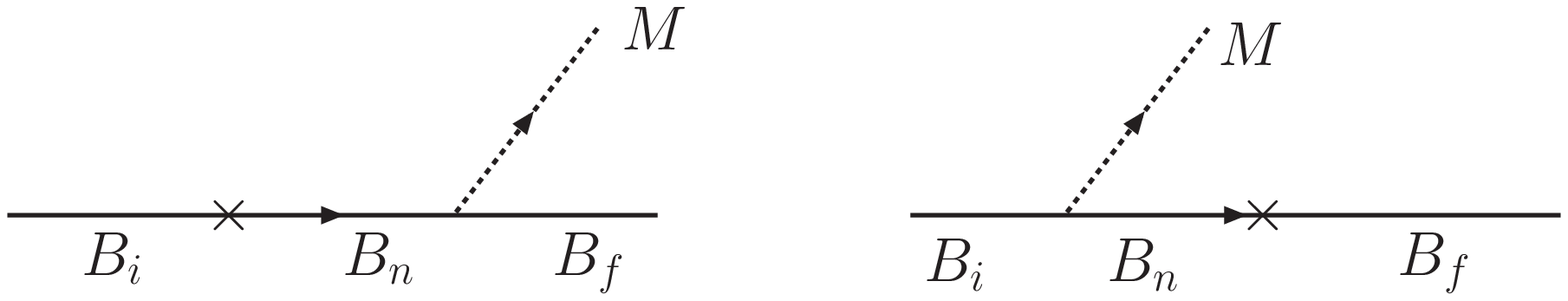,width=3.2in}}
 \caption{Pole
diagrams for charmed baryon decay ${\cal B}_i\to{\cal B}_f+M$. }
\label{fig:pole}
\end{figure}

The importance of final-state interactions (FSIs) has long been recognized in hadronic charm decays since some resonances are known to exist at energies close to
the mass of the charmed hadron. In general, nonfactorizable decay amplitudes do receive contributions from long-distance final-state rescattering. Indeed, nonfactorizable contributions modelled as final-state interactions through the one-particle-exchange process have been applied to singly and doubly charmed baryon decays \cite{Chen:FSI,Yu:Xiccpp,Jiang:2018oak,Han:doublycharm,Ke:FSI}.

The difficulty with the pole model lies in the evaluation of the $S$-wave amplitudes governed by the $1/2^-$ intermediate states. It is well known that in the soft-meson limit, the intermediate $1/2^-$ states can be summed up and reduced to a commutator term (see e.g.
\cite{CKX}); that is, $A^{\rm nf}\to A^{\rm com}$ in the soft-meson limit with
\begin{equation}
A^{\rm{com}}=-\frac{\sqrt{2}}{f_{P^a}}\la \mathcal{B}_f|[Q_5^a, {\cal H}_{\rm{eff}}^{\rm{pv}}]|\mathcal{B}_i\ra
=\frac{\sqrt{2}}{f_{P^a}}\la \mathcal{B}_f|[Q^a, {\cal H}_{\rm{eff}}^{\rm{pc}}]|\mathcal{B}_i\ra\label{eq:Apole}
\end{equation}
and
\begin{equation}
Q^a=\int d^3x \bar q\gamma^0\frac{\lambda^a}{2}q,\qquad
Q^a_5=\int d^3x \bar q\gamma^0\gamma_5\frac{\lambda^a}{2}q.
\end{equation}
The above expression for $A^{\rm com}$ is precisely the well known
soft-pion theorem in the current-algebra approach. In the
soft-pseudoscalar-meson limit, the parity-violating amplitude is reduced
to a simple commutator term expressed in terms of parity-conserving
matrix elements. Therefore, the great advantage
of current algebra is that the evaluation of the parity-violating
$S$-wave amplitude does not require the information
of the negative-parity $1/2^-$ poles.

The latest calculations of CF and SCS nonleptonic decays of $\Lambda_c^+$ using current algebra \cite{Zou:2019kzq} as shown in Tables \ref{tab:CF} and \ref{tab:SCSBF} seem to imply that current algebra is empirically working reasonably well, especially for the decay asymmetry parameter $\alpha$. However, the question is whether it is justified to apply the soft pion technique to charmed baryon decays. Notice that the pion's momentum is 864 MeV in the decay $\Lambda_c^+\to\Lambda\pi^+$, while it is only 101 MeV in $\Lambda\to p\pi^-$ decay. The produced pion is soft in hyperon decays, but not so in charmed baryon decays. Hence, {\it a priori} it is questionable to apply current algebra to $\Lambda_c$ decays.\footnote{An exception is the heavy-flavor-conserving decay where the pion is soft, 115 MeV in $\Xi_c^0\to\Lambda_c^+\pi^-$ and 99 MeV in $\Xi_b^-\to \Lambda_b^0\pi^-$.}
Why it is empirically working remains unclear. Writing
\begin{equation}
A=A^{\rm{CA}}+(A-A^{\rm{CA}}),
\end{equation}
and considering the term $(A-A^{\rm{CA}})$ as an on-shell correction to the current-algebra result, it could be that on-shell corrections happen to be small compared to the current-algebra amplitude, so that the soft-pion theorem turns out applicable even though the pion is not truly soft.

Despite of its complication in treating the $S$-wave amplitude, the pole model framework is very general and is not limited to the soft meson limit and to the pseudoscalar-meson final state. For example, current algebra is not applicable to the charmed baryon decays with a vector meson in the final state. In the past 30 years or so, progresses have been made towards the understanding of the light and heavy baryon spectroscopy. Sophisticated wave functions of the positive- and negative-parity  baryons in the quark models are available, see e.g. \cite{Chen:Sigmac,Li:LamcSL,Nagahiro:Lamc,Chua:2018lfa,Chua:2019yqh}. It is timely to
revisit the pole model for the hadronic decays of charmed baryons.

A study of hadronic weak decays of $\Lambda_c^+$ into $\Lambda\pi^+, \Sigma^0\pi^+$ and $\Sigma^+\pi^0$ in the framework of a constituent quark model
was recently presented in \cite{Niu:Lamc} in which the nonfactorizable effects of $W$-exchange  diagrams were studied in the pole model. \footnote{However, the inner $W$-emission diagram was evaluated directly in the quark model rather than through the pole model.}
The $1/2^+$ intermediate states are $\Sigma$ and $\Sigma_c(2455)$, while $1/2^-$ states are assigned to be $\Sigma^*(1620)$ and $\Sigma^*(1750)$ for the strange baryons and $\Sigma_c(2792)$ and $\Sigma_c(2806)$ for the charmed states.
The latter two states are yet to be measured and they are consistent with the quark model expectations in \cite{Ebert:2011}. It is interesting to notice that the predicted decay asymmetry parameter $\alpha(\Sigma\pi)=-0.47\pm0.19$ \cite{Niu:Lamc} in $\Lambda_c^+\to \Sigma\pi$ decays now agrees with experiment, resolving the long-standing puzzle with the sign of $\alpha(\Sigma\pi)$ in the pole model.
\footnote{It has been demonstrated recently in \cite{Ke:FSI} that the long-distance effect through the weak decay $\Lambda_c^+\to \Lambda\rho$ followed by the final-state rescattering of $\Lambda\rho\to \Sigma\pi$ will contribute destructively to the amplitude of $\Lambda_c^+\to\Sigma\pi$ and flip the sign of $\alpha$ obtained in the old pole model calculation in \cite{CT93}. However, it is clear by now that the pole model alone should be able to explain the data.
}
However, the the predicted $\alpha(\Lambda\pi^+)=-0.16\pm0.27$ differs significantly from the current average of $-0.84\pm0.09$ \cite{PDG}. This issue should be resolved in future study.
At any rate, the new pole-model calculation of $\alpha(\Sigma\pi)$  is very encouraging and
it is timely to pursue this framework further for nonleptonic charmed baryon decays and apply it to the vector meson production, a task beyond current algebra.

\subsection{Semileptonic decays}

Exclusive semileptonic decays of charmed baryons:
$\Lambda_c^+\to\Lambda e^+(\mu^+)\nu$, $\Xi_c^+\to \Xi^0
e^+\nu_e$ and $\Xi_c^0\to \Xi^-e^+\nu_e$ have been observed
experimentally. Measurement of the absolute branching fraction was first made by BESIII in 2015 for $\Lambda_c^+\to\Lambda e^+\nu_e$ \cite{BESIII:LamcSL} and subsequently  for $\Lambda_c^+\to\Lambda \mu^+\nu_\mu$ \cite{BESIII:LamcSLmu} with the results
\be
\B(\Lambda_c^+\to\Lambda e^+\nu_e)=(3.6\pm0.4)\%, \qquad
\B(\Lambda_c^+\to\Lambda \mu^+\nu_\mu)=(3.5\pm0.5)\%.
\en
Very recently, both Belle \cite{Belle:Xic0semi} and ALICE \cite{ALICE:XicSL} have reported the measurements of the branching-fraction ratios of semileptonic $\Xi_c^0\to \Xi^-\ell^+\nu_\ell$ decays:
\be \label{eq:Belle XicSL}
\B(\Xi_c^0\to \Xi^- e^+\nu_e)/\B(\Xi_c^0\to\Xi^-\pi^+) &=& 0.730\pm0.021\pm0.039, \non \\
\B(\Xi_c^0\to \Xi^- \mu^+\nu_\mu)/\B(\Xi_c^0\to\Xi^-\pi^+) &=& 0.708\pm0.033\pm0.056,
\en
by Belle and
\be \label{eq:ALICE XiCSL}
\B(\Xi_c^0\to \Xi^- e^+\nu_e)/\B(\Xi_c^0\to\Xi^-\pi^+)=1.38\pm0.14\pm0.22
\en
by ALICE. With the PDG value of $\B(\Xi_c^0\to\Xi^-\pi^+)=(1.43\pm0.32)\%$, the branching fractions of $\Xi_c^0\to\Xi^-\ell^+\mu_\ell)$ are listed in Table \ref{tab:XicSL}.

The semileptonic rates depend on the ${\cal B}_c\to{\cal B}$ transition
form factors $f_i(q^2)$ and $g_i(q^2)$ ($i=1,2,3$) defined by
\begin{eqnarray} \label{eq:FF}
 \langle {\cal B}_f(p_f)|V_\mu-A_\mu|{\cal B}_c(p_i)\rangle &=& \bar{u}_f(p_f)
\Big[f_1(q^2)\gamma_\mu+if_2(q^2)\sigma_{\mu\nu}q^\nu+f_3(q^2)q_\mu   \nonumber \\
&&
-\Big(g_1(q^2)\gamma_\mu+ig_2(q^2)\sigma_{\mu\nu}q^\nu+g_3(q^2)q_\mu\Big)\gamma_5\Big]
u_i(p_i).
\end{eqnarray}
These form factors have been evaluated using the
non-relativistic quark model (NRQM) \cite{Marcial,Singleton,CT96,Pervin},
MIT bag model \cite{Marcial},  relativistic quark model (RQM)
\cite{Ivanov96,Gutsche,Faustov:semi,Faustov:XicSL} light-front quark model (LFQM) \cite{Luo,Zhao:2018zcb,Li:LamcSL,Ke:XicSL}, QCD
sum rules (QSR) \cite{Carvalho,Huang,Azizi,Azizi2,Aliev:XicSL}, lattice QCD \cite{Meinel:LamcLam,Meinel:Lamcn,Zhang:2021oja} and SU(3) flavor symmetry \cite{Geng:semi}.

\begin{table}
\caption{Predicted branching fractions (in units of $\%$) and decay asymmetry parameters (second entry) of the semileptonic  $\Lambda_c^+\to \B\ell^+\nu_\ell$ decays in various models.  Predictions of \cite{Marcial} were obtained in the
non-relativistic quark model and the MIT bag model (in
parentheses).  There were three different scenarios for  the SU(3) predictions in \cite{Geng:semi}. We quote the results from scenario (b).
} \label{tab:LamcSL}
\begin{center}
\scriptsize{
\begin{ruledtabular}
\begin{tabular}{ l c c c c c c c c  c c  }
 Process & NRQM  & RQM & RQM & QSR & QSR & CQM & LQCD & LFQM & SU(3) & Expt \\
 & \cite{Marcial}  & \cite{Ivanov96} & \cite{Faustov:semi}
 & \cite{Carvalho} & \cite{Huang} & \cite{Gutsche}  & \cite{Meinel:LamcLam,Meinel:Lamcn} & \cite{Li:LamcSL} & \cite{Geng:semi}  &
 \cite{PDG} \\ \hline
 $\Lambda_c^+\to\Lambda^0 e^+\nu_e$  & 3.0 (2.2)  & 1.4 & 3.25  &  $2.6\pm0.4$ & $3.0\pm0.3$ & 2.78 &  $3.8\pm0.2$ & 4.04 & $3.6\pm 0.4$ & $3.6\pm 0.4$  \\
 & & $-0.812$ & & $-1$ & $-0.88\pm0.03$ & &  & & $-0.86\pm0.03$ & $-0.86\pm0.04$ \\
 $\Lambda_c^+\to\Lambda^0 \mu^+\nu_\mu$  &   &
  & 3.14 &  &  &  &  $3.7\pm0.2$ & 3.90 & $3.6\pm 0.4$ & $3.5\pm 0.5$  \\
 & & & &  & &  &  &  & $-0.86\pm0.04$ & $$ \\
 $\Lambda_c^+\to n e^+\nu_e$  & 0.22 (0.34) &
  0.26 & 0.268 & & & 0.20 &  0.41 & & $0.49\pm0.05$ & \\
   &  & &  & & & & & & $-0.89\pm0.04$ \\
\end{tabular}
\end{ruledtabular}}
\end{center}
\end{table}

\begin{table}
\footnotesize{
\caption{Same as Table \ref{tab:LamcSL} except for the semileptonic $\Xi_c^{+,0}\to \B^{0,-}\ell^+\nu_\ell$ decays.
} \label{tab:XicSL}
\centering
\begin{ruledtabular}
\begin{tabular}{ l c c c c c c c c  }
 Process & NRQM  & LFQM & RQM & LF & QSR  & LQCD & SU(3) & Experiment \\
 & \cite{Marcial}  & \cite{Zhao:2018zcb} & \cite{Faustov:XicSL} & \cite{Ke:XicSL}
 & \cite{Aliev:XicSL}   & \cite{Zhang:2021oja}  & \cite{Geng:semi}  &
 \cite{PDG} \\ \hline
 $\Xi_c^0\to\Xi^- e^+\nu_e$   & 0.40 (0.30)  & 1.35 & 2.38 & $1.72\pm0.35$ & $1.85\pm0.56$ & $2.38\pm0.45$ &   $2.4\pm0.3$ & $1.09\pm0.24$ \\
 & & & &  & &  &  $-0.83\pm0.04$ & $$ \\
 $\Xi_c^0\to\Xi^- \mu^+\nu_\mu$   &   &  & 2.31 & $$ & $1.79\pm0.54$ & $2.29\pm0.42$ &   $2.4\pm0.3$ & $1.01\pm0.25$ \\
 & & & &  & &  &  $-0.83\pm0.04$ & $$ \\
 $\Xi_c^+\to\Xi^0 e^+\nu_e$  & 1.3~(1.0) & 5.39 & 9.40 &  $5.2\pm1.0$ & $5.51\pm1.65$ & $7.18\pm1.33$  & $9.8\pm1.1$ & $7\pm4$  \\
 & & & &  & &  &   $-0.83\pm0.04$ &  \\
 $\Xi_c^+\to\Xi^0 \mu^+\nu_\mu$  &  &  & 9.11 &  $$ & $5.34\pm1.61$ & $6.91\pm1.27$  & $9.8\pm1.1$ &   \\
 & & & &  & &  &   $-0.83\pm0.04$ &  \\
\end{tabular}
\end{ruledtabular}}
\end{table}

Various model predictions of the charmed baryon semileptonic decay
rates and decay asymmetries are shown in Table \ref{tab:LamcSL} for $\Lambda_c^+$ and
Table \ref{tab:XicSL} for $\Xi_c^{+,0}$. The semileptonic decays of $\Omega_c^0$ were treated in \cite{Pervin:omegac} within the framework of a constituent quark model.
Many of the early calculations of $\B(\Lambda_c^+\to\Lambda e^+\nu)$
were slightly smaller than the experimental value of $(3.6\pm0.4)\%$ \cite{PDG}.
Lattice QCD calculations in \cite{Meinel:LamcLam} yield good agreement with experiment for both $\Lambda_c^+\to\Lambda e^+\nu_e$ and $\Lambda_c^+\to\Lambda \mu^+\nu_\mu$. As for the semileptonic $\Xi_c^0\to \Xi^-\ell^+\nu_\ell$ decays, it is clear from Table \ref{tab:XicSL} that most of recent calculations are slightly larger than the new measurements of Belle.

There have been active studies of semileptonic decays of doubly charmed baryons. The interested reader may consult  \cite{Shi,Wang:doublycharm,Albertus:2011,Faessler,Ebert:2004,Guo:doubly,Shi:2019} for further references.

Just as with the hadronic decays discussed in Sec. IV.C, there are also charm-flavor-conserving semileptonic processes such as $\Xi_c^0\to \Lambda_c^+(\Sigma_c^+)e^-\bar\nu_e$ and $\Omega_c^0\to\Xi_c^+ e^-\bar\nu_e$.
In these decays only
the light quarks inside the heavy baryon will participate in weak
interactions, while the heavy quark behaves as a spectator. This topic was recently investigated in \cite{Faller}. Owing to the severe phase-space suppression, the branching fractions are of order $10^{-6}$ in the best cases, and typically $10^{-7}$ to $10^{-8}$.

\section{Electromagnetic and weak radiative decays}

\subsection{Electromagnetic decays}
In the charmed baryon sector, only three of the radiative modes have been seen, namely, $\Xi'^0_c\to\Xi_c^0\gamma$, $\Xi'^+_c\to
\Xi^+_c\gamma$ and $\Omega_c^{*0}\to\Omega_c^0\gamma$. This is
understandable  because $m_{\Xi'_c}-m_{\Xi_c}\approx 108$ MeV and
$m_{\Omega_c^*}-m_{\Omega_c}\approx 71$ MeV. Hence, $\Xi'_c$ and
$\Omega_c^*$ are governed by the electromagnetic decays. Specifically,
the following two-body electromagnetic decays will be of interest:
 \begin{eqnarray}
\B_6 \to \B_{\overline{3}} + \gamma & : & \Sigma_c \rightarrow
\Lambda_c + \gamma,  \quad \Xi^\prime_c \rightarrow \Xi_c + \gamma , \nonumber \\
 \B^\ast_6 \rightarrow \B_{\overline{3}} + \gamma & : &
\Sigma^\ast_c \rightarrow \Lambda_c + \gamma ,  \quad \Xi^{
\ast}_c \rightarrow
\Xi_c + \gamma ,  \nonumber \\
\B^\ast_6 \rightarrow \B_6 + \gamma & : & \Sigma^\ast_c \rightarrow
\Sigma_c + \gamma , \quad \Xi^{ \ast}_c \rightarrow
\Xi^\prime_c + \gamma ,  \quad \Omega^\ast_c \rightarrow \Omega_c
+ \gamma ,
 \end{eqnarray}
where we have denoted the spin $\frac{1}{2}$ baryons as $\B_6$ and
$\B_{\overline{3}}$ for the symmetric sextet ${\bf 6}_F$  and
antisymmetric antitriplet ${\bf \bar{3}}_F$, respectively, and the spin
$\frac{3}{2}$ baryon by $\B^\ast_6$.

An ideal theoretical framework for studying the above-mentioned
electromagnetic decays is provided by the formalism in which the
heavy quark symmetry and the chiral symmetry of light quarks are
combined \cite{Yan,Wise}. When supplemented by the nonrelativistic
quark model, the formalism determines completely the low energy
dynamics of heavy hadrons.  The
chiral and electromagnetic gauge-invariant Lagrangian for heavy
baryons can be found in Eqs. (3.8) and (3.9) of \cite{Cheng93} denoted by ${\cal L}_B^{(1)}$ and ${\cal L}_B^{(2)}$, respectively.
Within the framework of heavy baryon chiral perturbation theory (HBChPT), the authors of \cite{Jiang} proceeded to construct chiral Lagrangians at the level ${\cal O}(p^2)$ and ${\cal O}(p^3)$ and then calculated the electromagnetic decay amplitudes of charmed baryons up to the next-to-leading order (NLO). This has been further generalized to the next-to-next-to-leading order (NNLO) in \cite{Wang:2018}. Besides the quark model, some of the unknown couplings (i.e. low energy coefficients) were extracted from the data of the magnetic moments and the radiative decay widths from lattice QCD. Moreover, heavy quark symmetry was utilized to reduce the number of low energy coefficients.

{\squeezetable
\begin{table}[t]
\caption{Electromagnetic decay rates (in units of keV) of $S$-wave charmed
baryons. Among the four different results listed in \cite{Dey} and \cite{Majethiya}, we quote those denoted by $\Gamma_\gamma^{(0)}$ and ``Present (ecqm)", respectively.  } \label{tab:em}
\begin{center}
\begin{tabular}{l c c c c c c c c c c }
\hline\hline ~~~~Decay & HHChPT  & HBChPT  & Dey & Ivanov & Simonis & Aliev & Wang & Bernotas & Majethiya & Hazra  \\
 & \cite{Cheng97,Cheng93}   &  \cite{Wang:2018} &  \cite{Dey} & \cite{Ivanov}  & \cite{Simonis} &  \cite{Aliev} &  \cite{ZGWang} &  \cite{Bernotas} & \cite{Majethiya} & \cite{Hazra:2021lpa}
   \\
\hline
 $\Sigma^+_c\to \Lambda_c^+\gamma$ & 91.5 & $65.6\pm2$ & 120 & $60.7\pm1.5$  & 74.1 & $50\pm17$ & & 46.1 & 60.55  & $93.5\pm0.7$  \\
 $\Sigma_c^{*+}\to\Lambda_c^+\gamma$ & 150.3 & $161.6\pm5$ & 310 & $151\pm4$ & 190 & $130\pm45$ & & 126 & 154.48  & $231\pm7$ \\
 $\Sigma_c^{*++}\to\Sigma_c^{++}\gamma$ & 1.3 & $1.20\pm0.6$ & 1.6 & & 1.96 &   $2.65\pm1.20$ & $6.36^{+6.79}_{-3.31}$ & 0.826 & 1.15 & $1.48\pm0.02$ \\
 $\Sigma_c^{*+}\to\Sigma_c^+\gamma$ & 0.002 & $0.04\pm0.03$ & 0.001 & $0.14\pm0.004$ & 0.011 &  $0.40\pm0.16$ & $0.40^{+0.43}_{-0.21}$ & 0.004 & $<10^{-4}$ & $(7\pm1)10^{-4}$ \\
 $\Sigma_c^{*0}\to\Sigma_c^0\gamma$ & 1.2 & $0.49\pm0.1$ & 1.2 &  & 1.41 & $0.08\pm0.03$ & $1.58^{+1.68}_{-0.82}$ & 1.08 & 1.12  & $1.38\pm0.02$ \\
 $\Xi'^+_c\to\Xi_c^+\gamma$ & 19.7  & $5.43\pm0.33$ & 14 & $12.7\pm1.5$ & 17.3 &  $8.5\pm2.5$ & & 10.2 &  & $21.4\pm0.3$  \\
 $\Xi^{*+}_c\to\Xi_c^+\gamma$ & 63.5  & $21.6\pm1$ & 71 & $54\pm3$ & 72.7  & $52\pm25$ & & 44.3 & 63.32   & $81.9\pm0.5$  \\
 $\Xi^{*+}_c\to\Xi_c^{'+}\gamma$ & 0.06   & $0.07\pm0.03$ & 0.10 & & 0.063 & 0.274 &  $0.96^{+1.47}_{-0.67}$ & 0.011 &  & $0.03\pm0.00$ \\
 $\Xi'^0_c\to\Xi_c^0\gamma$ & 0.4   & 0.46 &  0.33 & $0.17\pm0.02$ & 0.185 & $0.27\pm0.6$ & & 0.0015 &  & $0.34\pm0.01$ \\
 $\Xi^{*0}_c\to\Xi_c^0\gamma$ & 1.1 & 1.84 & 1.7 & $0.68\pm0.04$ & 0.745 & $0.66\pm0.32$ & & 0.908 & 0.30  & $1.32\pm0.01$  \\
 $\Xi^{*0}_c\to\Xi_c^{'0}\gamma$ & 1.0 & $0.42\pm0.16$ & 1.6 & & 1.33 & 2.14  & $1.26^{+0.80}_{-0.46}$ & 1.03 &  & $1.26\pm0.03$ \\
 $\Omega_c^{*0}\to\Omega_c^0\gamma$ & 0.9  & $0.32\pm0.20$ & 0.71 & & 1.13 & 0.932  & $1.16^{+1.12}_{-0.54}$  & 1.07& 2.02  & $1.14\pm0.13$ \\ \hline
 \hline
\end{tabular}
\end{center}
\end{table}
}

The general amplitudes of electromagnetic decays are given by
\cite{Cheng93}
 \begin{eqnarray}
 A(\B_6\to \B_{\bar 3}+\gamma) &=& i\eta_1\bar u_{\bar
3}\sigma_{\mu\nu}k^\mu \varepsilon^\nu u_6,   \nonumber \\
A(\B^*_6\to \B_{\bar 3}+\gamma) &=&
i\eta_2\epsilon_{\mu\nu\alpha\beta}
\bar u_{\bar 3}\gamma^\nu k^\alpha\varepsilon^\beta u^\mu, \ \\
A(\B^*_6\to \B_6+\gamma) &=& i\eta_3\epsilon_{\mu\nu\alpha\beta}
\bar u_6\gamma^\nu k^\alpha\varepsilon^\beta u^\mu.  \nonumber
 \end{eqnarray}
The corresponding decay rates read
 \begin{eqnarray}
\Gamma(\B_6\to \B_{\bar 3}+\gamma) &=& \eta_1^2\,{k^3\over \pi},  \nonumber \\
\Gamma(\B_6^*\to \B_{\bar 3}+\gamma) &=& \eta_2^2\,{k^3\over
3\pi}\,{ 3m_i^2+m_f^2\over 4m_i^2},    \\
\Gamma(\B_6^*\to \B_6+\gamma) &=& \eta_3^2\,{k^3\over 3\pi}\,{3m_i^2
+m_f^2\over 4m_i^2}, \nonumber
 \end{eqnarray}
where $m_i$ ($m_f$) is the mass of the parent (daughter) baryon. To the leading order,
the coupling constants $\eta_i$ can be calculated using the quark
model for $a_1$, $a_2$, and $a'_1$ \cite{Cheng93,Cheng97}:
\begin{eqnarray}
a_1=-{e\over 3}{1\over M_u}, \qquad a_2={e\over 2\sqrt{6}}{1\over M_u},
\qquad
a'_1={e\over 12}{1\over M_Q}.
\end{eqnarray}
Indeed, $\eta_1$, $\eta_2$ and $\eta_3$ are nothing but the transition magnetic moments for $\B_6\to \B_{\bar 3}$, $\B_6^*\to \B_{\bar 3}$ and $\B_6^*\to \B_6$ transitions, respectively.
The calculated results using the constituent quark masses $M_u=338$ MeV, $M_d=322$ MeV, $M_s=510$ MeV \cite{PDG}, and $M_c=1.6$ GeV, are
summarized in the second column of Table \ref{tab:em}. Some other model predictions are also listed there for comparison.

Several remarks are in order: (i) In Table \ref{tab:em} we did not quote the results of  \cite{Yang:rad}  because the calculated radiative rates are much smaller than others  even though the transition magnetic moments are comparable to each other. For example, $\Gamma(\Sigma^+_c\to \Lambda_c^+\gamma)=12.8$ keV and $\Gamma(\Sigma^{*+}_c\to \Lambda_c^+\gamma)=35.5$ keV were obtained in \cite{Yang:rad}, which are obviously too small compared to others. Recently, the decay rates of \cite{Yang:rad} were re-calculated in \cite{Hazra:2021lpa} using their own  transition moments and the results are now consistent with others (see Table XIII of \cite{Hazra:2021lpa}). (ii) Lattice calculations in  \cite{Bahtiyar,Bahtiyar:2016} yield $\Gamma(\Omega_c^*\to \Omega_c\gamma)=0.074\pm0.008$ keV, $\Gamma(\Xi'^{+}_c\to \Xi_c^+\gamma)=5.468\pm1.500$ keV and  $\Gamma(\Xi'^{0}_c\to \Xi_c^0\gamma)=0.002\pm0.004$ keV.
 (iii) The results  of \cite{Cheng97,Cheng93} and \cite{Wang:2018} shown in Table \ref{tab:em} can be regarded as the predictions of HHChPT to LO and NNLO, respectively.
It is evident  that the predicted rates of the following three modes:  $\Sigma_c^{*+}\to\Lambda_c^+\gamma$, ${\Xi'}_c^{}\to \Xi_c^{+}\gamma$ and $\Xi^{*+}_c\to\Xi_c^+\gamma$ in HHChPT to NNLO are smaller than that to LO.

\begin{table}[t]
\caption{Electromagnetic decay rates (in units of keV) of $P$-wave charmed
baryons. For the predictions of \cite{Wang:rad}, we quote the results from the $\lambda$ excitation states.
} \label{tab:empwave}
\begin{center}
\begin{tabular}{l c c c c c c}
\hline\hline ~~~~Decay &  ~~~Ivanov~~~ & ~~Tawfiq~~  & ~~Zhu~~ & ~~Wang~~  & Gamermann  & ~~Expt~~ \\
 &  \cite{Ivanov}  &  \cite{Tawfiq01}   & \cite{Zhu:2000} & \cite{Wang:rad} & \cite{Gamermann}  & \cite{Belle:rad} \\
\hline
 $1/2^-\to 1/2^+(3/2^+)\gamma$ & & & & & & \\
 $\Lambda_c(2595)^+\to\Lambda_c^+\gamma$ & $115\pm1$ &  & 36 & 0.26 & $289.1\pm51.3$ \\
 $\Lambda_c(2595)^+\to\Sigma_c^+\gamma$ & $77\pm1$ & 71 & 11 & 0.45 & $7.9\pm1.5$  \\
 $\Lambda_c(2595)^+\to\Sigma_c^{*+}\gamma$ & $6\pm0.1$ & 11 & 1 & 0.05 &  \\
 $\Xi_c(2790)^+\to\Xi_c^+\gamma$ &  & & & 4.65 & $249.6\pm41.9$ & $<350$\\
 $\Xi_c(2790)^0\to\Xi_c^0\gamma$ &  & & & 263 & $119.3\pm21.7$ & $\sim 800$\\
 $\Xi_c(2790)^+\to {\Xi'}_c^+\gamma$ &  & & & 1.43 & $0.8\pm0.5$ & \\
 $\Xi_c(2790)^0\to{\Xi'}_c^0\gamma$ &  & & & 0 & $1.3\pm0.4$ & \\
 \hline
 $3/2^-\to 1/2^+(3/2^+)\gamma$ & & & & & &  \\
 $\Lambda_c(2625)^+\to\Lambda_c^+\gamma$ & $151\pm2$ & &  48 & 0.30 & \\
 $\Lambda_c(2625)^+\to\Sigma_c^+\gamma$ & $35\pm0.5$ & 130 &  5 & 1.17 & \\
 $\Lambda_c(2625)^+\to\Sigma_c^{*+}\gamma$ & $46\pm0.6$ & 32 &  6 & 0.26 & \\
 $\Xi_c(2815)^+\to\Xi_c^+\gamma$ & $190\pm5$ & & & 2.8 & & $<80$ \\
 $\Xi_c(2815)^0\to\Xi_c^0\gamma$ & $497\pm14$ & & & 292 & & $320\pm45^{+45}_{-80}$ \\
 $\Xi_c(2815)^+\to{\Xi'}_c^+\gamma$ &  & & & 2.32 & & \\
 $\Xi_c(2815)^0\to{\Xi'}_c^0\gamma$ &  & & & 0 & &  \\
 \hline
 \hline
\end{tabular}
\end{center}
\end{table}

For the electromagnetic decays of $P$-wave charmed baryons, the
search for $\Lambda_c(2593)^+\to\Lambda_c^+\gamma$ and
$\Lambda_c(2625)^+\to\Lambda_c^+\gamma$ has not yet succeeded. Nevertheless, Belle has recently presented a successful search for the electromagnetic decays $\Xi_c(2790,2815)\to \Xi_c\gamma$ for the neutral orbitally excited $\Xi_c$ baryons \cite{Belle:rad}.  However, no signal was found for the analogous decays of charged $\Xi_c(2790)^+$ and $\Xi_c(2815)^+$ states.
The results are
\be
&& \Gamma(\Xi_c(2815)^0\to \Xi_c^0\gamma)=(320\pm45^{+45}_{-80})\,{\rm keV}, \qquad
\Gamma(\Xi_c(2815)^+\to \Xi_c^+\gamma)< 80\,{\rm keV}, \nonumber \\
&& \Gamma(\Xi_c(2790)^0\to \Xi_c^0\gamma)\sim 800\,{\rm keV},
 \qquad\qquad\qquad~~
\Gamma(\Xi_c(2790)^+\to \Xi_c^+\gamma)< 350\,{\rm keV}.
\en

Some predictions available in the literature are exhibited in Table \ref{tab:empwave}, and they are more diversified than the $S$-wave case. Nevertheless, we see that the measurement of the electromagnetic decays of the excited $\Xi_c$ state  is consistent with the model predictions for $\Xi_c(2815)^0\to \Xi_c^0\gamma$, but much higher than the theoretical expectation for $\Xi_c(2790)^0\to\Xi_c^0\gamma$. In the future it is also important to search for the electromagnetic decays of the $P$-wave $\Lambda_c^+$ states.

For the electromagnetic decays of doubly charmed baryons, see, e.g., \cite{Branz,Bernotas,Cui,Zhu:Bccrad,Simonis,Bahtiyar:2018,Lu:2017meb}. The decay widths of $\Xi_{cc}^{*++}\to\Xi_{cc}^{++}\gamma$ and $\Xi_{cc}^{*+}\to\Xi_{cc}^{+}\gamma$ are estimated to be a few or tens of keV in various models, but very small in lattice calculations \cite{Bahtiyar:2018}.

\subsection{Weak radiative decays}

Weak radiative decays of heavy baryons usually receive the short-distance (SD)
contributiuons from the so-called electromagnetic penguin diagram and long-distance (LD) contributions  from the $W$-exchange diagram accompanied by a photon emission from the
external quark, or from the same $W$-exchange diagram
but with a photon radiated from the $W$ boson. For charmed baryons,
the penguin process $c\to u\gamma$ is highly suppressed, so it
plays no role to the radiative decays of charmed hadrons. Hence, we will focus on the process such as $cu\to sd\gamma$ induced through $W$-exchange and bremsstrahlung. On the contrary, radiative bottom baryon decays are mainly governed by the SD electroweak process $b\to s\gamma$. A noticeable example is the decay $\Lambda_b^0\to\Lambda\gamma$, first studied in 1995 \cite{Cheng:rad}, was finally observed  by the LHCb in 2019 \cite{LHCb:rad}.

At the Cabibbo-allowed level, there are two modes for weak radiative decays of antitriplet charmed baryons induced from $cd\to us\gamma$, namely,
 \begin{eqnarray} \label{eq:rad}
  \Lambda_c^+\to\Sigma^+\gamma,\qquad \Xi_c^0\to\Xi^0\gamma.
 \end{eqnarray}
Singly Cabibbo-suppressed modes are $\Lambda_c^+\to p\gamma$, $\Xi_c^+\to\Sigma^+\gamma$ and $\Xi_c^0\to \Lambda(\Sigma^0)\gamma$, while   $\Xi_c^+\to p\gamma$ and $\Xi_c^0\to n\gamma$ are doubly Cabibbo-suppressed. Two-quark W-exchange
bremsstrahlung transitions are characterized by the emission of a hard photon
and the presence of a highly virtual intermediate quark between
the electromagnetic and weak vertices. It had been shown in
\cite{Cheng:rad} that these features should make it possible to
analyze these processes by perturbative QCD; that is, these
processes can be described by an effective local and gauge
invariant Lagrangian. Table \ref{tab:weakrad} shows some early predictions in different analyses. It appears that the branching fractions of the Cabibbo-allowed weak radiative decays are of order $(3.0\sim 6.0)\times 10^{-5}$. For an estimate of singly Cabibbo-suppressed ratiative modes, see \cite{Singer:rad}. For a recent SU(3) analysis of weak radiative decays, the reader is referred to \cite{Wang:radSU3}.

\begin{table}[t]
\caption{Theoretical estimates of branching fractions (upper entry, in units of $10^{-4}$) and decay asymmetry (lower entry) of Cabibbo-allowed weak radiative decays of charmed baryons.
There are two predictions in \cite{Uppal:rad} depending on the evaluation of the charmed baryon wavefunction squared at the origin $|\psi(0)|^2$. We have normalized the results to the current lifetimes of $\Lambda_c^+$ and $\Xi_c^0$.
}
\label{tab:weakrad}
\begin{center}
\begin{tabular}{l c c c } \hline \hline
 & Kamal \cite{Kamal:rad} & ~~Uppal \cite{Uppal:rad}~~ & (CLY)$^2$ \cite{Cheng:rad} \\
 &  & (I) \qquad (II) &  \\
\hline
$\Lambda_c^+\to\Sigma^+\gamma$ & 0.60 & 0.45 \qquad 2.91 & ~~0.49 \\
$\Xi_c^0\to\Xi^0\gamma$ &  & 0.30 \qquad 1.95 & ~~0.48 \\
\hline
$\Lambda_c^+\to\Sigma^+\gamma$ & 0.49 & $-0.013$~~ \quad 0.023~ & $-0.86$ \\
$\Xi_c^0\to\Xi^0\gamma$ &  & $-0.042$ \quad $-0.010$~ & $-0.86$ \\
\hline \hline
\end{tabular}
\end{center}
\end{table}

Finally, it is worth remarking that, analogous to the
heavy-flavor-conserving nonleptonic weak decays discussed in
Sec. VI.C, there is a special class of weak radiative decays in
which heavy flavor is conserved, for example, $\Xi_c \to
\Lambda_c \gamma$ and $\Omega_c \to \Xi_c \gamma$.  In these
decays, weak radiative transitions arise from the light quark sector
of the heavy baryon, whereas the heavy quark behaves as a
spectator. However, the dynamics of these radiative decays is more
complicated than that of their counterpart in nonleptonic weak decays,
e.g., $\Xi_c \to \Lambda_c \pi$. In any event, it merits an investigation.

\section{\CP violation}

The CKM matrix contains a phase which implies the existence of \CP violation, but at a very small level in the decays of charmed hadrons. This is because of the relation of the CKM matrix elements, $\lambda_s\approx -\lambda_d$ with $\lambda_p\equiv V_{cp}^*V_{up}$. As a consequence, \CP violation in the charm sector is usually governed by the CKM matrix element $\lambda_b$ which is very tiny compared to $\lambda_d$ or $\lambda_s$ in magnitude. This also indicates that the corresponding QCD penguin and electroweak penguin are rather suppressed. In $B$ physics, the interference between tree and penguin amplitudes usually will lead to sizable \CP asymmetries. Does it mean that it will be extremely difficult or even hopeless to observe \CP violation in the charm sector?

In 2019 LHCb has announced the first observation of \CP asymmetry difference between $D^0\to K^+K^-$ and $D^0\to \pi^+\pi^-$ with the result $\Delta A_{\rm CP}=(-1.54\pm0.29)\times 10^{-3}$ \cite{LHCb:CP}. In the standard-model estimate with the short-distance penguin contribution, we have the expression (see e.g. \cite{Cheng:2012a})
\be
\Delta A_{\rm CP}\approx -1.3\times 10^{-3} \left( \left|{P\over T}\right|_{_{K\!K}}\sin\theta_{_{K\!K}}+ \left|{P\over T}\right|_{_{\pi\pi}}\sin\theta_{_{\pi\pi}} \right),
\en
where the factor of $-1.3\times 10^{-3}$ comes from the imaginary part $2{\rm Im}(\lambda_d\lambda_b^*)/|\lambda_d|^2$,
$\theta{_{K\!K}}$ is the strong phase of $(P/T)_{_{K\!K}}$ and likewise for $\theta{_{\pi\pi}}$. In the most general case, one should replace $P$ by $P+P\!E+P\!A$ and $T$ by $T+E$, where $P\!E$ is the QCD-penguin exchange topology and $P\!A$ the QCD-penguin annihilation topology.
Since $|P/T|$ is na{\"i}vely expected to be of order $(\alpha_s(\mu_c)/\pi)\sim {\cal O}(0.1)$, it appears that $\Delta A_{\rm CP}$ is most likely of order $10^{-4}$ assuming strong phases close to $90^\circ$ or even less for realistic strong phases.
Does this mean that it is inevitable to introduce new physics to account for the observed \CP violation in charmed meson decays? Nowadays we learn that long-distance dynamics plays an essential role in charm physics. It has been
established sometime ago that a more suitable framework for the analysis of hadronic charmed meson decays is the so-called topological diagram approach \cite{Chau,CC86,CC87}. One of the most salient features of the topological approach is that all the topological amplitudes except the tree topology $T$ are dominated by nonfactorizable long-distance effects. It was pointed out in \cite{Cheng:2012a} that there is a resonant-like final-state rescattering which has the same topology as the QCD-penguin exchange  $P\!E$. That is, penguin annihilation receives sizable long-distance contributions through final-state interactions. It is the interference between tree and long-distance penguin exchange that pushes $\Delta A_{\rm CP}$ up to the per mille level \cite{Cheng:2012b,Cheng:2019}.  \footnote{A similar result of $\Delta A_{\rm CP}$ based on a variant of the diagrammatic approach was obtained in \cite{Li:2012}.}
And new physics is not needed to understand the the measured \CP asymmetry difference in charmed meson decays.

The theoretical study of the penguin enhancement through final-state rescattering in the charmed baryon sector is still absent. On the experimental side,
the search for \CP violation in charmed baryon decays has taken on new momentum
with the large samples of $\Lambda_c^+$ obtained by BESIII and LHCb. For two-body
decays of the $\Lambda_c^+$, \CP violation can be explored through the measurement of $C\!P$-violating decay parameter asymmetry, ${\cal A}=(\alpha+\bar\alpha)/(\alpha-\bar\alpha)$. For example,
the current result is ${\cal A}=-0.07\pm0.19\pm0.24$ for $\Lambda_c^+\to \Lambda\pi^+$ and $\bar\Lambda_c^-\to\bar\Lambda \pi^-$ \cite{Link:2005ft} and $(1.5\pm5.2\pm1.7)\%$ for
$\Xi_c^0\to\Xi^-\pi^+$ and $\bar\Xi_c^0\to \bar\Xi^+\pi^-$ \cite{Belle:Xic0semi}.
As for three-body decays,
LHCb has measured  the difference between \CP asymmetries in
$\Lambda_c^+ \to p K^+ K^-$ and $\Lambda_c^+ \to p\pi^+ \pi^-$ decay channels, in analogue to $D^0\to K^+K^-$ and $D^0\to \pi^+\pi^-$ \cite{Aaij:2017xva}.
The result  is $\Delta A_{CP}(\Lambda_c^+) = (0.30 \pm 0.91 \pm 0.61)\%$,
to be compared with a generic SM prediction of a fraction of 0.1\%
\cite{Bigi:2012ev}.
In order to probe the SM level, one has to multiply the available statistics
by at least a factor of 100. LHCb has also looked for local \CP asymmetry in $\Xi_c^+\to pK^-\pi^+$ decays and found null results \cite{LHCb:2020zkk}.

For multi-hadrons in the final state of $\Lambda_c^+$ decays such as $\Lambda_c^+\to pK^-\pi^+\pi^0$, $\Lambda_c^+\to\Lambda\pi^+\pi^+\pi^-$ and $\Lambda_c^+\to pK_S\pi^+\pi^-$, \CP violation can be exploited through several $T$-odd observables. Owing to its characters of high luminosity, broad center-of-mass energy acceptance, abundant production and clean environment, the next generation of Super Tau-Charm Facility (STCF) may provide a great platform for this kind of study.
A fast Monte Carlo simulation study in \cite{Shi:2019vus} by using the $e^+e^-$ annihilation data of 1 ab$^{-1}$ at $\sqrt{s}=4.64$ GeV, which are expected to be available at the future STCF, indicates that a sensitivity at the level of (0.25-0.5)\% is accessible for the above-mentioned three decay modes. This will be enough to measure non-zero $C\!P$-violating asymmetries as large as 1\%.

\section{Conclusions}
In this report,
we began with a brief overview of the spectroscopy of charmed baryons and discussed their possible structure and spin-parity assignments in the quark model, especially for the five new and narrow $\Omega_c$ resonances and three new excited $\Xi_c$ states. It should be stressed that the mass analysis alone is usually not sufficient to pin down the spin-parity quantum numbers of higher excited charmed baryon states, a study of their strong decays is necessary.  Meanwhile, Regge phenomenology and the mass relations for antitriplet and sextet multiplets also provide very useful guidance for the spin-parity quantum numbers.
Based on the nice property of parallelism of the Regge trajectory, we see that  the highest state $\Lambda_c(2940)^+$ does not fit if its quantum numbers are $\frac32^-$ as preferred by LHCb.  We suggest that $\Lambda_c(2940)^+$ is most likely the $\frac12^-(2P)$ state. Experimentally, it is thus important to search for the $\Lambda_c$ baryon with a mass of order 3005 MeV and verify its quantum numbers as $\frac32^-(2P)$. For the narrow $\Omega_c$ resonances, LHCb has performed the spin analysis to test their spin assignments. It turns out that the spin assignment of  $\Omega_c(3000)^0,\Omega_c(3050)^0,\Omega_c(3065)^0$ and $\Omega_c(3090)^0$ is consistent with $\lambda$-mode excitations with $J=1/2,3/2,3/2$ and $5/2$, respectively. The equal mass gaps between the last three excited $\Omega_c^0$ baryons with the resonances $\Xi_c(2923)^0$, $\Xi_c(2939)^0$ and $\Xi_c(2965)^0$, respectively, suggest that the corresponding spin quantum numbers are $3/2,3/2$ and $5/2$ for the excited $\Xi'_c$ states. The excited $5/2^-$ states identified with $\Omega_c(3090)$ and $\Xi'_c(2965)$ are further supported by the parallelism of the Regge trajectories as shown in Fig. \ref{fig:Omegac,Xicp}.

For the lifetimes of singly charmed baryons, we have witnessed the truly shocking result of the new  $\Omega_c$ lifetime measurement of the LHCb in 2018. It is nearly four times larger than that from old fixed target experiments! This was recently confirmed by the LHCb using a different measurement based on the prompt $\Omega_c$ production. Hence, a new lifetime hierarchy $\tau(\Xi_c^+)>{\tau(\Omega_c^0)}>\tau(\Lambda_c^+)>\tau(\Xi_c^0)$ is now firmly established.
On the theoretical side, it is well known that heavy quark expansion to order $1/m_c^3$ with dimension-6 four-quark operators leads to the old lifetime pattern $\tau(\Xi_c^+)>{\tau(\Lambda_c^+)}>\tau(\Xi_c^0)>\tau(\Omega_c^0)$ settled for a long time until 2018. The new LHCb measurement of the $\Omega_c^0$ lifetime
calls for the next-order $1/m_c$ corrections to spectator effects. It turns out that HQE fails to apply to $\Omega_c^0$ to $1/m_c^4$ as the subleading corrections are too large to justify the validity of HQE. By demanding a sensible HQE for $\Omega_c^0$, we have introduced a parameter $\alpha$ to suppress the contributions from dimension-7 operators.
Our study leads to the conclusion that
the $\Omega_c^0$, which is naively expected to be shortest-lived in the charmed baryon system owing to the large constructive Pauli interference, could live longer than the $\Lambda_c^+$  due to the suppression from $1/m_c$ corrections arising from dimension-7 four-quark operators.
There are many issues remained to be resolved. Why is the $\Omega_c$ so special? What is the origin of the unknown parameter $\alpha$ and how to determine its value?
More efforts towards the understanding of higher dimensional operators are thus urgently needed in the community. This is also true for the lifetimes of doubly charmed baryons.

In spite of the slow progress in the study of nonleptonic charm baryon decays for a long time both theoretically and experimentally, the situation has been dramatically changed since 2014.
Thanks to both Belle and BESIII for measuring the  absolute branching fractions of many
nonleptonic decay modes of $\Lambda_c^+$ and $\Xi_c^{+,0}$. Measurements of the  absolute branching fractions of exclusive semileptonic decays were also available from BESIII for $\Lambda_c^+\to\Lambda \ell^+\nu_\ell$  and from Belle for $\Xi_c^0\to\Xi^- \ell^+\nu_\ell$.
On the theoretical aspect, we still do not have a good and reliable phenomenological model as charm physics is essentially governed by long-distance dynamics. The main task is to evaluate the nonfactorizable amplitudes arising from $W$-exchange and inner $W$-emission diagrams. In principle, the pole model in which contributions from all possible intermediate states are considered is very general for this purpose. In practice, it requires the knowledge of wave functions for excited baryon states, which is a difficult and tedious task especially for negative-parity ones. Current algebra is very simple for estimating the nonfactorizable $S$-wave amplitude as it does not need the information of the troublesome negative-parity resonances. Although this approach is empirically working well, it cannot be fully justified as the produced pseudoscalar meson is not soft except for heavy-flavor-conserving decays. Moreover, current algebra is not applicable to the vector meson case. In view of the progress  towards the understanding of the baryon spectroscopy in the past 30 years or so, it is timely to revisit the pole model. A recent work along this direction \cite{Niu:Lamc} was very encouraging; the long-standing puzzle in the pole model with
the decay asymmetry parameter for $\Lambda_c^+\to \Sigma\pi$  was finally resolved.
By doing a global fit to the data, the approach of SU(3) flavor symmetry proves to be very useful in making predictions on yet detected modes. Nevertheless, we do need dynamical calculations which shed light on the underling mechanism for charmed baryon decays. The pole model deserves special attention in the community.

Finally, we mention a few issues with immediate interest: (i) Because of the soft pion nature in charm-flavor-conserving nonleptonic decays, the amplitude of $\Xi_c^0\to\Lambda_c^+\pi^-$ can be reliably expressed in terms of several unknown baryonic matrix elements: $X$ and $Y$, $a_{\Sigma_c^0\Xi_c^0}$, $a_{\Sigma_c^+\Xi_c^+}$ and $a_{\Lambda_c^+{\Xi'}_c^+}$  [see Eqs. (\ref{eq:AmpHFC}) and (\ref{eq:Pwavenet})]. The task is figure out a better way for estimating these unknown parameters.
(ii) Measurement of the decay asymmetry parameters in  $\Lambda_c^+\to pK_S$ and $\Lambda_c^+\to\Xi^0K^+$ is needed to test theory.  (iii) The tension between the two modes $\Xi_c^0\to\Xi^-\pi^+$ and
$\Xi_c^+\to\Xi^0\pi^+$ needs to be resolved. (iv) Measurement of $\B(\Xi_{cc}^{++}\to\Xi_c^+\pi^+)$ is vital to clarify the interference nature between external and inner $W$-emission diagrams.

\section{Acknowledgments}
This research was supported in part by the Ministry of Science and Technology of R.O.C. under Grant No. 110-2112-M-001-025.

\newcommand{\bi}{\bibitem}

\end{document}